\documentclass[11pt]{article}
\usepackage{amsmath,latexsym,amssymb,epsfig}
\usepackage{graphicx} 

\def\ltsim{\lower3pt\hbox{$\, \buildrel < \over \sim \, $}} 
\def\gtsim{\lower3pt\hbox{$\, \buildrel > \over \sim \, $}} 
\newcommand{\be}{\begin{equation}} 
\newcommand{\ee}{\end{equation}} 
\def\ga{\mathrel{\raise.3ex\hbox{$>$\kern-.75em\lower1ex\hbox{$\sim$}}}} 
\def\la{\mathrel{\raise.3ex\hbox{$<$\kern-.75em\lower1ex\hbox{$\sim$}}}}

\makeatletter 
\@addtoreset{equation}{section} 
\makeatother

\begin{document}
\baselineskip=16pt

\begin{titlepage}

\vskip 2cm
\begin{flushright}
{\bf June 2007}
\end{flushright}
\begin{center} 
\vspace{0.5cm} \Large {\sc 
Brane-Worlds Pseudo-Goldstinos}
\vspace*{5mm} 
\normalsize

{\bf
Karim~Benakli
\footnote{kbenakli@lpthe.jussieu.fr}
and
Cesar~Moura
\footnote{moura@lpthe.jussieu.fr}
}

\smallskip 
\medskip 
\it{Laboratoire de Physique Th\'eorique et Hautes
Energies}\\ 
\it{Universit\'e Pierre et Marie Curie - Paris VI â€“ Universit\'e Denis Diderot
-
Paris VII, France}

\vskip0.6in 
\end{center}

\centerline{\large\bf Abstract}

\noindent

We consider a space-time with extra dimensions  containing sectors, 
branes and bulk, that communicate only through gravitational 
interactions. In each sector, if considered separately,  supersymmetry 
could be spontaneously broken,  leading to the appearance of Goldstinos.
However, when taken all together, only certain 
combinations of the latter states turn out to be true ``would be Goldstinos'',
eaten by the gravitinos. The other (orthogonal) combinations, we call 
pseudo-Goldstinos, remain in the low energy spectrum. We discuss 
explicitly how this happen in the simplest set-up of five-dimensional 
space compactified on $S^1/\mathbb{Z}_2 $. Our results divide into two parts 
that can be considered separately. First, we build  an extension of the bulk 
five-dimensional supergravity, by a set of new auxiliary fields, that 
allows coupling it to branes where supersymmetry is spontaneously 
broken. Second, we discuss in details the super Higgs mechanism in the 
$R_\xi$ and unitary gauges, in the presence of both of a bulk 
Scherk-Schwarz mechanism and brane localized $F$-terms. This leads us to 
compute the gravitino mass and provide explicit formulae for the 
pseudo-Goldstinos spectrum.

\vspace*{2mm} 

\end{titlepage}

\section{Introduction and Conclusions}
\label{Introduction}

If supersymmetry has to play a role among the fundamental laws governing 
our world, it has to appear as spontaneously broken. For the purpose, 
the world is often described by an effective four-dimensional 
supergravity where the spontaneous breaking  corresponds to 
non-vanishing vacuum expectation values (v.e.v's) of auxiliary fields, 
called $F$-terms and $D$-terms. In the global supersymmetry limit, the 
breaking gives rise in the global limit to \textit{massless} Goldstone 
fermions, the \textit{Goldstinos} \cite{Fayet:1974jb}. Instead, in the local
version, where 
gravity is taken into account, the (would be)-Goldstinos are absorbed by 
the \textit{massive} gravitinos to become their spin $\frac{1}{2}$ 
components \cite{Volkov:1973jd,Fayet:1977vd,Casalbuoni:1988qd}.

The last decade has seen the emergence of a popular scenario for the
phenomenological
implications of the short distance description of space-time where extra
dimensions play an important role\cite{Antoniadis:1990ew}-\cite{Randall:1999ee}.
Some of the light degrees of 
freedom are confined to live on branes localized at particular points in 
a higher dimensional space. In such a set-up, supersymmetry breaking can 
happen  either on the branes or in the bulk, and it is usual to discuss 
the breaking in each sector, separately. For instance, the dynamical 
breaking of supersymmetry \cite{Witten:1981nf} is often studied in the global
limit as due to 
some non-perturbative gauge dynamics \cite{Horava:1996vs,Veneziano:1982ah} that
could happen at different 
scales on different, spatially separated, branes \cite{Intriligator:2007cp}. The
breaking of 
supersymmetry in the bulk can be instead achieved through a 
Scherk-Schwarz mechanism \cite{Scherk:1978ta}. In each of these sectors  would
be Goldstinos 
are predicted. Only certain combinations of the latter are true ones, 
eaten by the gravitinos. One asks then about the fate of the remaining 
states. This work deals with this issue.

Another problem addressed in this paper is the coupling of the bulk 
supergravity to the brane, in the presence of localized $F$-terms. In 
the global supersymmetry case, this was studied by Mirabelli and Peskin
\cite{Mirabelli:1997aj}. 
It was found that the five dimensional super Yang-Mills theory had to be 
extended off-shell  by the addition of an appropriate auxiliary field in 
order to take into  account the presence of localized $D$-terms. Such 
auxiliary fields can be integrated out, but with the price of 
introducing an explicitly singular coupling $\delta(0)$ in the scalar 
potential, which requires to be treated carefully as arising from an 
infinite sum over extra-dimensional momenta.  Here, we propose the 
adjunction of a \textit{new set of auxiliary fields} to the minimal 
five-dimensional supergravity. These fields vanish identically in the 
supersymmetric limit and allow us to keep track of the supersymmetry 
transformations in the case when boundary $F$-terms are not explicitly 
put to zero.

There have been a huge number of papers dealing with supersymmetry breaking in
extra dimensions (for a few  examples, see 
\cite{Randall:1998uk}-\cite{Gherghetta:2000qt}). We believe it useful to point
out to the reader where
    our work stands in this literature. Off-shell 
extension of minimal five-dimensional supergravity was built in
\cite{Zucker:2003qv}. This extension was further studied in 
\cite{PaccettiCorreia:2004ri}-\cite{Buchbinder:2003qu}. In particular, 
\cite{Gherghetta:2001sa,Rattazzi:2003rj} 
studied the coupling to
boundary branes and discussed in some details supersymmetry breaking through
generalized Scherk-Schwarz boundary conditions
\cite{Bagger:2001qi,vonGersdorff:2002tj} which correspond to a constant
superpotential. 
Their study uses the auxiliary
fields obtained in \cite{Zucker:2003qv}. Our approach here is different. After
tedious
computations, we build our Lagrangians \textit{from scratch} in components
fields. We identify the necessity to introduce a space-time vector and a scalar
as
auxiliary fields, and we derive transformation rules to take into account the
presence of \textit{arbitrary superpotentials} 
for breaking supersymmetry. Although we believe that our auxiliary set of fields
can be expressed as a combination of some of those of \cite{Zucker:2003qv}, 
the relation is not trivial and, for the aim of this work, we do
not find it worth to go through long computations to extract it.

The other issue discussed here is the fate of the would be Goldstinos. The
super-Higgs mechanism
in the framework of extra dimensions has been discussed in
\cite{Meissner:1999ja} for the case of a bulk Goldstino and in
\cite{DeCurtis:2003hs} for the case of 
generalized Scherk-Schwarz mechanism. The on-shell coupling of a brane Goldstino
to a bulk
supergravity was discusses by \cite{Bagger:2004rr} in the Randall-Sundrum
set-up. Our analysis
includes both bulk and boundaries would-be-Goldstinos. Finally, it might be
interesting for the reader to make the analogy with
    bosonic case of
pseudo-Goldstones. They have been
introduced by Weinberg in \cite{Weinberg:1972fn}, used for electroweak symmetry
breaking by \cite{Georgi:1974yw} 
and in the context of large
extra-dimensions, they were used as Higgs
bosons for example in \cite{Arkani-Hamed:2001nc,Hatanaka:1998yp,
Cacciapaglia:2005pa}.

Let us summarize our main results:

\begin{itemize}

\item We have introduced an extension with new auxiliary fields in order 
to keep track of supersymmetry transformation when coupling the 
five-dimensional supergravity with the branes, as mentioned above.

\item The generalized Scherk-Schwarz mechanism, as introduced by Bagger, 
Feruglio and Zwirner \cite{Bagger:2001qi},  allows  localized gravitino masses
on
boundary 
branes. We have generalize it to an arbitrary set of branes suspended in 
the bulk. The supersymmetry transformations have been derived for these 
cases. As a corollary, we obtain the 
condition for the association of the given brane-localized gravitino 
masses with non-trivial Scherk-Schwarz twist not to break supersymmetry. We 
point out explicitely the obstacles when trying to express the $F$-term
breaking 
as a generalized twist.

\item In the case of flat boundary branes and flat bulk, we study in 
details the gauge fixing for the super-Higgs mechanism. We discuss both 
the general $R_\xi$ gauge and the unitary gauge. In the latter, we 
explicitly obtain the form of the gravitino mass, and show that  from 
the four original  would-be Goldstinos (two in the bulk and one on each 
boundary), two are eaten by the $N=2$ bulk gravitinos while two 
orthogonal combinations remain. We call them pseudo-Goldstinos. We 
explicitly compute the corresponding masses for specific cases, and show 
that in the limit of infinite radius they vanish as expected when one 
decouples gravity..

\end{itemize}

Our results are obtained with particular assumptions which allow the 
computations to be carried out explicitly up to the end. The 
five-dimensional supergravity is taken with the minimal content 
on-shell. Only the corresponding states that are even under an 
appropriately defined $\mathbb{Z}_2$ action are assumed to couple on the 
branes.  In deriving the pseudo-Goldstinos spectrum, we  will take all 
the branes and the bulk to be separately flat. We also assume that we 
are working in a basis where the localized superfields providing the 
would be Goldstinos are canonically normalized. Moreover, our treatment 
is  at tree-level. We believe that the departure from these assumptions 
 should not dramatically change the qualitative 
picture presented in this first work.

The paper is organized as follows. Section \ref{secBulkBrane} displays the
Lagrangians 
corresponding to the minimal five-dimensional supergravity in the bulk 
with supersymmetry broken by non-trivial boundary conditions, as well as 
the simplest brane action for a set of chiral multiplets with a priory 
non-vanishing $F$-terms. Appendix \ref{apeConventions} summarizes the related
conventions. 
The coupling of the two sectors, bulk and branes, is performed in section 
\ref{secAuxFields} after the introduction of new auxiliary fields. The
corresponding 
supersymmetry transformations are collected in Appendix
\ref{apeTotalSusyAction}. Section \ref{secSS} 
reviews some issues of the  spontaneous breaking of supersymmetry in 
compactifications on $S^1/\mathbb{Z}_2$. In particular the interplay 
between bulk and boundary branes localized gravitino masses to break (or 
restore) supersymmetry. These results are generalized in section
\ref{secMultBranes} to 
the case with an arbitrary number of branes suspended at arbitrary 
points of $S^1/\mathbb{Z}_2$. Section \ref{secSuperHiggs} discusses in great
details the 
super Higgs mechanism when supersymmetry is spontaneously broken both in 
the bulk and on the brane. We exhibit that combination of 
would-be-Goldstinos  are not absorbed and form the remaining 
pseudo-Goldstinos. The explicit form of the latter and their masses are 
provided in section \ref{secPseudoGold}, while the general formulae are given in
Appendix \ref{apePGMassEigenstate}.

\section{Bulk with boundary branes}
\label{secBulkBrane}

Consider a five-dimensional space parametrized by coordinates $( x^{\mu} , x^{5}
)$ with 
$\mu=0, \cdots, 3$ and $x^5 \equiv y$ parametrizing the interval 
$S^1/\mathbb{Z}_2$. The latter
is constructed as an orbifold from the circle of length $2 \pi R$ ($y \sim y + 2
\pi R$) through the
identification $y \sim - y$.
Matter fields live on branes  localized for instance at particular points
$y=y_n$. We will assume here that there are only two branes sitting at the
boundaries $y_n=y_b\in \{ 0, \pi R\}$. The corresponding action can be written
as\footnote{The factor $\frac{1}{2}$ in front of ${\cal L}_{BULK}$ in equation
(\ref{Action}) comes from: 
$\int d^{5}x = \int^{\pi R}_{0} dy \int d^{4}x = \frac{1}{2} \int^{2 \pi R}_{0}
dy \int d^{4}x $.}:

\begin{eqnarray}
    S &=&  \int^{2 \pi R}_{0} dy \int d^{4}x \left\{ \frac{1}{2} {\cal L}_{BULK}
+ 
{\cal L}_{0} \delta(y)
    +  {\cal L}_{\pi} \delta(y-\pi R) \right\} .
\label{Action}
\end{eqnarray}

\subsection{On-shell supergravity action in the bulk}
\label{secBulk}

We take the theory in the bulk to be five-dimensional supergravity 
with the minimal 
on-shell content being the f\"{u}nfbein $e^{A}_{M}$, the gravitino $\Psi_{M I}$
and the 
graviphoton $B_{M}$. The on-shell Lagrangian is given by\footnote{Unless stated
otherwise, we take $\kappa$, $\bar h$ and $c$ equal to $1$. See appendix
\ref{apeConventions} for
our conventions.} \cite{Cremmer:1980gs}:
\begin{eqnarray}
    {\cal L}_{SUGRA} &=& e_{5} \bigg \lbrace -\frac{1}{2}R \left( \omega
\right) 
+ \frac{i}{2} \check{\Psi}_{M}^{I} \Gamma^{MNP} D_{N} \Psi_{P I}
- \frac{1}{4}F_{MN}F^{MN}    - \frac{1}{6\sqrt{6}}
\epsilon^{ABCDE}F_{AB}F_{CD}B_{E} 
\nonumber    \\
&& -i\frac{\sqrt{6}}{16} F_{MN} \left( 2 \check{\Psi}^{M I} \Psi^{N}_{I} 
+ \check{\Psi}_{P}^{I} \Gamma^{MNPQ} \Psi_{Q I}  \right) 
\bigg \rbrace 
\label{Lsugra}
\end{eqnarray}
and the on-shell supersymmetry transformations are 
\footnote{In this paper we make the following approximations: we drop the
four-fermions terms in the Lagrangian 
and the three and four-fermions terms in the supersymmetry transformations.}
:
\begin{eqnarray}
\delta_{\prime} e^{A}_{M}  &=& i \check{\Xi}^{I} \Gamma^{A} \Psi_{M I}
\nonumber   \\
\delta_{\prime} B_{M}  &=& i \frac{\sqrt{6}}{2} \check{\Psi}^{I}_{M} \Xi_{I} 
\nonumber   \\
\delta_{\prime} \Psi_{M I}  &=& 2 D_{M}\Xi_{I} + \frac{1}{2 \sqrt{6}} F^{NP}
\left( \Gamma_{MNP} - 4
g_{MP}\Gamma_{N} \right) \Xi_{I}
\label{SusyTransf0}
\end{eqnarray}
where $\Xi$ is the supersymmetry transformation parameter and 
$F_{MN} = \partial_{M} B_{N} -\partial_{N} B_{M} $. Note that we use here the
symbol $\delta_{\prime} $ for the variation of the fields while the usual symbol
$\delta$ will be defined later to include extra terms (in section
\ref{secAuxFields}).

The five-dimensional spinors $\Psi_{M I} $ and $\Xi_{I}$ are symplectic Majorana
spinors, 
described in appendix \ref{apeConventions}. 
In appendix \ref{apeTotalSusyAction}, we present 
the Lagrangian (\ref{Lsugra}) and the corresponding supersymmetry
transformations (\ref{SusyTransf0}) 
in two-component spinor notation. The five-dimensional gravitino $\Psi_{M I}$
will be written:
\begin{equation}
\Psi_{M 1} = 
\left(\begin{matrix}
\psi_{M 1}\\
\overline{\psi}_{M 2} 
\end{matrix}\right)  ,\qquad
\Psi_{M 2} = 
\left(\begin{matrix}
-\psi_{M 2}\\
\overline{\psi}_{M 1} 
\end{matrix}\right) 
\label{PsilNotation}
\end{equation}
using  the two-component Weyl spinors $\psi_{MI}$.

Every generic field $\varphi$ has a well defined $\mathbb{Z}_2$ transformation:
\begin{equation}
\mathbb{Z}_{2} : \quad \varphi(y) \rightarrow {\cal P}_{0} \varphi(- y)
\label{Z2Action}
\end{equation}
that allows us to define the orbifold $S^1/\mathbb{Z}_2$ from the original
five-dimensional compactification on $S^1$. Here ${\cal P}_{0}$ is the parity of
the field $\varphi$ which obeys ${\cal
P}_{0}^{2} = 1$. The Lagrangian (\ref{Lsugra}) and supersymmetry transformations
(\ref{SusyTransf0}) must be 
invariant under the action of the mapping (\ref{Z2Action}).

At the point $y = 0$, we assume the f\"{u}nfbein to transform as:
\begin{equation}
e^{a}_{\mu}(-y) = + e^{a}_{\mu}(y) , \quad 
e^{a}_{5}(-y) = - e^{a}_{5}(y) , \quad
e^{\hat{5}}_{\mu}(-y) = - e^{\hat{5}}_{\mu}(y) , \quad
e^{\hat{5}}_{5}(-y) = + e^{\hat{5}}_{5}(y).
\label{FunfbeinParities}
\end{equation}
These  assumptions and the invariance of supersymmetry transformations 
under the $\mathbb{Z}_{2}$ action  imply 
that $\psi_{M1}$ and $\psi_{M2}$ must have opposite parities. The Lagrangian
(\ref{Lsugra2}) is also invariant under an $SU(2)_{\cal R}$ R-symmetry, under
which the gravitinos $\psi_{M1}$ and $\psi_{M2}$ transform in the representation
$\textbf{2}$ of $SU(2)_{\cal R}$:
\begin{equation}
SU(2)_{\cal R}: \quad \psi_{N I} \rightarrow {U_{I}}^{J} \psi_{N J}
\label{SU2R}
\end{equation}
with $U \in SU(2)_{\cal R}$. 
A possible choice of parity assignments is 
\begin{equation}
\psi_{\mu 1}(-y) = + \psi_{\mu 1}(y) .
\label{GravitinoParity}
\end{equation}
Again, at the point $y=0$, the other
fields parity transformations are determined from equations
(\ref{FunfbeinParities}), (\ref{GravitinoParity}) and invariance 
of (\ref{SusyWeylTransf0}) under the mapping (\ref{Z2Action}), and they are
shown in table
\ref{Parities0}:
\begin{table}[htb]
      \begin{center}
         \begin{tabular}{|l|c|c|c|c|c|c|}
         \hline
         ${\cal P}_{0} = +1$ & $e^{a}_{\mu}$ & $e^{\hat{5}}_{5}$ & $B_{5}$ &
$\psi_{\mu 1}$ & $\psi_{52}$ & $\xi_{1}$ \\
         \hline
         ${\cal P}_{0} = -1$ & $e^{a}_{5}$ & $e^{\hat{5}}_{\mu}$ & $B_{\mu}$ &
$\psi_{\mu 2}$ & $\psi_{51}$ & $\xi_{2}$ \\
         \hline
         \end{tabular}
      \caption{Parity assignments for bulk fields at $y=0$.}
\label{Parities0}
      \end{center}
\end{table}

As periodicity condition, we impose the following twisted boundary conditions:
\begin{equation}
\left(\begin{matrix}
\psi_{M 1} (y + 2 \pi R) \\
\psi_{M 2} (y + 2 \pi R)
\end{matrix}\right) 
=
\left(\begin{matrix}
\cos(2 \pi \omega) & \sin(2 \pi \omega) \\
- \sin(2 \pi \omega) & \cos(2 \pi \omega)
\end{matrix}\right) 
\left(\begin{matrix}
\psi_{M 1} (y) \\
\psi_{M 2} (y)
\end{matrix}\right) 
\label{SSTwist}
\end{equation}
which correspond for $\omega \neq 0$ to implement a Scherk-Schwarz
supersymmetry 
breaking in the bulk \cite{Scherk:1978ta}. In section \ref{secSS}, it will be
shown that boundary localized masses 
for gravitinos  can be 
absorbed in a generalized Scherk-Schwarz twist.

We must assign a parity ${\cal P}_{\pi}$ for each generic field $\varphi$ at the
point
$y=\pi R$
\begin{equation}
\varphi(\pi R + y) = {\cal P}_{\pi} \varphi(\pi R - y) ,
\label{ParityPi}
\end{equation}
which keeps the Lagrangian (\ref{Lsugra2}) and 
the supersymmetry transformations (\ref{SusyWeylTransf0}) invariant.

For instance, taking  for the f\"{u}nfbein the parities,
\begin{eqnarray}
e^{a}_{\mu}(\pi R-y) = + e^{a}_{\mu}(\pi R+y) , &\quad& 
e^{a}_{5}(\pi R-y) = - e^{a}_{5}(\pi R+y)
\nonumber \\
e^{\hat{5}}_{\mu}(\pi R-y) = - e^{\hat{5}}_{\mu}(\pi R+y) , &\quad&
e^{\hat{5}}_{5}(\pi R-y) = + e^{\hat{5}}_{5}(\pi R+y),
\label{FunfbeinParitiesPi}
\end{eqnarray}
an agreement with table \ref{Parities0} and equations (\ref{SSTwist})
requires imposing:
\begin{eqnarray}
\psi_{M +}(\pi R-y) & = & \psi_{M +}(\pi R+y) 
\nonumber \\
\psi_{M -}(\pi R-y) & = & - \psi_{M -}(\pi R+y) 
\label{GravitinoParityPi}
\end{eqnarray}
where:
\begin{eqnarray}
\psi_{\mu +} & = &
\cos( \pi \omega) \psi_{\mu 1} - \sin( \pi \omega) \psi_{\mu 2} 
\nonumber \\
\psi_{\mu -} & = &
\sin( \pi \omega) \psi_{\mu 1} + \cos( \pi \omega) \psi_{\mu 2} 
\nonumber \\
\psi_{5 +} & = &
\sin( \pi \omega) \psi_{5 1} + \cos( \pi \omega) \psi_{5 2} 
\nonumber \\
\psi_{5 -} & = &
\cos( \pi \omega) \psi_{5 1} - \sin( \pi \omega) \psi_{5 2} .
\label{ParityEigenvectorsPi}
\end{eqnarray}

Invariance of the supersymmetry transformations (\ref{SusyWeylTransf0}) under 
the $\mathbb{Z}_{2}$ mapping (\ref{ParityPi}) determines the parities of all
other fields. The result is given in table
\ref{ParitiesPi}, where the following definitions have been introduced:
\begin{eqnarray}
\xi_{+} & = &
\cos( \pi \omega) \xi_{1} - \sin( \pi \omega) \xi_{2} 
\nonumber \\
\xi_{-} & = &
\sin( \pi \omega) \xi_{1} + \cos( \pi \omega) \xi_{2} 
\label{SusyparametersPi}
\end{eqnarray}
\begin{table}[htb]
      \begin{center}
         \begin{tabular}{|l|c|c|c|c|c|c|}
         \hline
         ${\cal P}_{\pi} = +1$ & $e^{a}_{\mu}$ & $e^{\hat{5}}_{5}$ & $B_{5}$ &
$\psi_{\mu +}$ & $\psi_{5 +}$ & $\xi_{+}$ \\
         \hline
         ${\cal P}_{\pi} = -1$ & $e^{a}_{5}$ & $e^{\hat{5}}_{\mu}$ & $B_{\mu}$ &
$\psi_{\mu -}$ & $\psi_{5 -}$ & $\xi_{-}$ \\
         \hline
         \end{tabular}
      \caption{Parity assignments for bulk fields at $y=\pi R$.}
\label{ParitiesPi}
      \end{center}
\end{table}

\subsection{Boundary branes actions}
\label{secBranes}

The bulk supergravity fields presented above are coupled with matter fields
living on branes. Here, we will consider the simplest case where the branes are
localized on the boundaries $y_b = 0 , \pi R$,  with the simplest  matter
content given by $N_{b}$ chiral multiplets. The case with many branes localized
on different points of $S^1/\mathbb{Z}_2$ will be discussed in section
\ref{secMultBranes}.

Each chiral multiplet contains (on-shell) a scalar  $\phi_{b}^{i}$ and a
fermionic $\chi_{b}^{i}$ fields ($i = 1 ,
\cdots , N_{b}$).
These fields are \textit{coupled to the even parity bulk fields} at the point
$y=y_b$.
For instance, the even parity bulk fields at the point $y = 0$  are 
$e^{a}_{\mu}$, $e^{\hat{5}}_{5}$, $B_{5}$, $\psi_{\mu1}$, $\psi_{52}$ and
$\xi_{1}$, 
and they appear in the Lagrangian at the brane $0$  as \cite{WessAndBagger}:
\begin{eqnarray}
{\cal L}_{0} &=& e_{4} \Bigg \lbrace - \frac{1}{2} g_{i j^*} \partial_{\mu}
\phi_{0}^{i} \partial^{\mu} \phi_{0}^{* j} 
- i \frac{1}{2} g_{i j^*} \overline{\chi}_{0}^{j}
\overline{\sigma}^{\mu}\tilde{D}_{\mu}\chi_{0}^{i}
+ \frac{1}{8}\left( {\cal G}_{0 j} \partial_{\mu} \phi_{0}^{j} - {\cal G}_{0
j^*} 
\partial_{\mu} \phi_{0}^{* j} \right) 
\epsilon^{\mu \nu \rho \lambda} \overline{\psi}_{\rho 1}
\overline{\sigma}_{\lambda} \psi_{\nu 1}
\nonumber \\ &&
- e^{{\cal G}_{0}/2} \left[ \psi_{\mu 1} \sigma^{\mu \nu} \psi_{\nu 1} 
+ i \frac{\sqrt{2}}{2} {\cal G}_{0 j^*}\overline{\chi}_{0}^{j}
\overline{\sigma}^{\mu} \psi_{\mu 1}
+ \frac{1}{2} \left( {\cal G}_{0 i j} + {\cal G}_{0 i}{\cal G}_{0 j} 
- \Gamma^{k}_{i j} {\cal G}_{0 k} \right)^{*} 
\overline{\chi}_{0}^{i} \overline{\chi}_{0}^{j} \right] 
\nonumber \\ &&
- \frac{\sqrt{2}}{2} g_{i j^*}  \partial_{\nu} \phi_{0}^{* j} \chi_{0}^{i}
\sigma^{\mu} \overline{\sigma}^{\nu} \psi_{\mu 1}
- \frac{1}{2} e^{{\cal G}_{0}} \left( g^{i j^*} {\cal G}_{0 i} {\cal G}_{0 j^*}
- 3 \right)
+ h.c. \Bigg \rbrace ,
\label{Lbrane0}
\end{eqnarray}
where
\begin{equation}
\tilde{D}_{\mu}\chi_{0}^{i} = \partial_{\mu}\chi_{0}^{i} + \frac{1}{2}
\omega_{\mu ab}\sigma^{ab}\chi_{0}^{i} 
+ \Gamma^{i}_{j k} \partial_{\mu} \phi_{0}^{j} \chi_{0}^{k}
- \frac{1}{4} \left( {\cal G}_{0 j} \partial_{\mu} \phi_{0}^{j} - {\cal G}_{0
j^*} 
\partial_{\mu} \phi_{0}^{* j} \right)  \chi_{0}^{i}
\label{CovDerivChi}
\end{equation}
and ${\cal G}_{0}(\phi_{0} , \phi_{0}^{*})$ is a hermitian function of the
fields $\phi_{0}$ and $\phi_{0}^{*}$.
Here we have used the notations:
\begin{equation}
{\cal G}_{0 j} = \frac{\partial}{\partial \phi_{0}^{j}} {\cal G}_{0} , \quad
{\cal G}_{0 j^*} = \frac{\partial}{\partial \phi_{0}^{* j}} {\cal G}_{0} , \quad
{\cal G}_{0 i j} = \frac{\partial^{2}}{\partial \phi_{0}^{i} \partial
\phi_{0}^{j}} {\cal G}_{0} .
\label{GDerivatives}
\end{equation}
We remind that the metric $g_{i j^*}$, its inverse $g^{i j^*}$ and the
Christoffel symbols
in the K\"ahler manifold are given by:
\begin{equation}
g_{i j^*} = \frac{\partial^{2}}{\partial \phi_{0}^{i} \partial \phi_{0}^{* j}}
{\cal G}_{0} , \quad
g_{i j^*} g^{j^* k} = \delta_{i}^{k} , \quad
\Gamma^{k}_{i j} = g^{k l^*} \frac{\partial}{\partial \phi_{0}^{i}} g_{j l^*}.
\label{KahlerMetric}
\end{equation}
The function ${\cal G}_{0}(\phi_{0} , \phi_{0}^{*})$ is given, in terms of the 
K\"ahler potential $K_{0}$ and superpotential $W_{0}$ at the brane $0$, by:
\begin{equation}
{\mathcal G}_{0}(\phi_{0} , \phi_{0}^{*}) = K_{0}(\phi_{0} , \phi_{0}^{*}) +
\ln\left[ W_{0}(\phi_{0})\right]
    + \ln\left[ W_{0}(\phi_{0}) \right] ^{*}.
\label{KahlerFunction}
\end{equation}

In the following, it will be useful to define the action\footnote{We take
$F^{\mu 5}=0$ on the branes, for simplicity.}:
\begin{equation}
S^{(0)}_{4d} = \int d^{4}x \left\lbrace -\frac{1}{2} e_{4}
\hat{R}(\hat{\omega}) 
+ e_{4} \epsilon^{\mu \nu \rho \lambda}\overline{\psi}_{\mu
1}\overline{\sigma}_{\nu}\hat{D}_{\rho}\psi_{\lambda1} + {\cal L}_{0}
\right\rbrace 
\label{Lsugra4d}
\end{equation}
where $\hat{R}(\hat{\omega})$ and $\hat{D}_{\rho}\psi_{\lambda1}$ are defined
in 
equations (\ref{R4dDefinition}) and (\ref{CovDer4dDefinition}), respectively. 
This action is invariant under the four-dimensional local transformations:
\begin{eqnarray}
\delta e^{a}_{\mu}  &=& i \left( \xi_{1} \sigma^{a} \overline{\psi}_{\mu 1}
\right) + h.c.
\nonumber   \\
\delta \phi_{0}^{i}  &=& \sqrt{2} \xi_{1}\chi_{0}^{i}
\nonumber   \\
\delta \chi_{0}^{i}  &=& i \sqrt{2} \sigma^{\mu} \xi_{1} \partial_{\mu}
\phi_{0}^{i}
- \sqrt{2} e^{{\cal G}_{0}/2} g^{i j^*} {\cal G}_{0 j^*} \xi_{1}
\nonumber   \\
\delta \psi_{\mu 1}  &=& 2 \hat{D}_{\mu}\xi_{1} 
+ \frac{1}{2} \left( {\cal G}_{0 j} \partial_{\mu} \phi_{0}^{j} 
- {\cal G}_{0 j^*} \partial_{\mu} \phi_{0}^{* j} \right) \xi_{1}
+ i e^{{\cal G}_{0}/2} \sigma_{\mu} \overline{\xi}_{1}.
\label{SusyTransBrane0}
\end{eqnarray}

It is important, for our concern, to note the presence of a localized mass for
the gravitino $\psi_{\mu 1}$ in the Lagrangian (\ref{Lbrane0}). 
If $\left\langle c_{i} g^{i j^*} {\cal G}_{0 j^*} \right\rangle$ is nonzero,
then the field $c_{i} \chi_{0}^{i}$ is the Goldstino associated with the 
supersymmetry breaking in the brane $0$ as indicated by its non-linear 
transformation in equations (\ref{SusyTransBrane0}).

For the $N_{\pi}$ chiral multiplets $\phi_{\pi}^{i}$, $\chi_{\pi}^{i}$, ($i = 1
, \cdots , N_{\pi}$) at 
the brane $\pi$ a similar discussion can be carried over after the following
substitutions:
\begin{equation}
brane~0 \rightarrow brane~\pi : \quad
\left \lbrace
\begin{array}{c c c c}
{\cal L}_{0} \rightarrow {\cal L}_{\pi} , \quad &
\phi_{0}^{i} \rightarrow \phi_{\pi}^{i} , \quad &
\chi_{0}^{i} \rightarrow \chi_{\pi}^{i} , \quad &
{\cal G}_{0} \rightarrow {\cal G}_{\pi} , \\
\psi_{\mu 1} \rightarrow \psi_{\mu +} , \quad &
\xi_{1} \rightarrow \xi_{+} , \quad &
K_{0} \rightarrow K_{\pi} , \quad &
W_{0} \rightarrow W_{\pi} .
\end{array}
\right .
\label{Brane0ToBranePi}
\end{equation}

\section{Coupling the branes to the bulk}
\label{secAuxFields}

Our aim is to study generic configurations where in additional to a possible
bulk Sherk-Schwarz mechanism, supersymmetry can also be spontaneously broken
through non-vanishing boundary $F$-terms for chiral multiplets.

To make the supersymmetry breaking manifestly spontaneous, we will keep the
brane action  written as above in terms of the K\"ahler functions ${\cal
G}_{b}$. The supersymmetry breaking terms can be  identified with the vacuum
expectation values of the auxiliary fields. Coupling of these vevs to the bulk
supergravity requires then to add new auxiliary fields to the on-shell
five-dimensional supergravity action written in (\ref{Lsugra}).  This local case
version is analogous to the case of spontaneous breaking of global supersymmetry
as studied by Mirabelli and Peskin \cite{Mirabelli:1997aj}, where in order to
keep the
rigid supersymmetry manifest, it was  necessary to introduce auxiliary fields
in the
bulk. Here we will present a ``partially off-shell'' extension of the bulk
supergravity with only the minimal required auxiliary fields.
These new fields vanish identically in the supersymmetric limit as their 
boundary values are proportional to the supersymmetry breaking vevs
(see equations \ref{CDFBoundaryConditions0} and \ref{CDFBoundaryConditionsPi}). 
Moreover, integrating these auxiliary fields to go on-shell leads to singular
terms  ($\delta(0)$), again as in \cite{Mirabelli:1997aj}, which will require
careful summation over the KK bulk states in order to extract sensible
results.

\subsection{The auxiliary fields action}
\label{auxfields1}

For our purpose, we introduce two auxiliary fields denoted as $u$ and $v_{M}$.
Here $u$ is a real scalar field and $v_{M}$ is a real five-dimensional vector
field.

The bulk supergravity is now written as:
\begin{equation}
{\cal L}_{BULK} = {\cal L}_{SUGRA} + {\cal L}_{AUX},
\label{Lbulk2}
\end{equation}
where ${\cal L}_{SUGRA}$ is still given in equation (\ref{Lsugra2}) and ${\cal
L}_{AUX}$ is :
\begin{equation}
{\cal L}_{AUX} = e_{5} \frac{1}{2} \left( u u + v_{M} v^{M} \right) .
\label{Laux}
\end{equation}

Taking into account the auxiliary fields and the brane supersymmetry breaking
vevs, the bulk supersymmetry transformations of the on-shell fields become:
\begin{eqnarray}
\delta e^{A}_{M} &=& \delta_{\prime} e^{A}_{M}
\nonumber \\
\delta B_{M} &=& \delta_{\prime} B_{M}
\nonumber \\
\delta \psi_{\mu 1}  &=& \delta_{\prime} \psi_{\mu 1} + i v_{\mu} \xi_{1} + i u
\sigma_{\mu} \overline{\xi}_{1}
\nonumber   \\
\delta \psi_{\mu 2}  &=& \delta_{\prime} \psi_{\mu 2} + i v_{\mu} \xi_{2} + i u
\sigma_{\mu} \overline{\xi}_{2}
\nonumber \\
\delta \psi_{5 1} &=& \delta_{\prime} \psi_{5 1} 
- 4 e^{{\cal G}_{\pi}/2} \sin(\omega \pi)\xi_{+} \delta(y-\pi R)
\nonumber \\
\delta \psi_{5 2} &=& \delta_{\prime} \psi_{5 2} - 4 e^{{\cal G}_{0}/2}\xi_{1}
\delta(y) 
- 4 e^{{\cal G}_{\pi}/2} \cos(\omega \pi)\xi_{+} \delta(y-\pi R)
\label{SusyTransfOfGravitinoModified}
\end{eqnarray}
where the supersymmetry transformations $\delta_{\prime}$ were defined in
equation (\ref{SusyWeylTransf0}).

Both $u$ and $v_{\mu}$ are taken to be even under $\mathbb{Z}_{2}$ on both
boundaries:
\begin{eqnarray}
u(-y) = u(y) , & \quad & 
u(\pi R + y) = u(\pi R - y) ,
\nonumber \\ 
v_{\mu}(-y) = v_{\mu}(y) , & \quad &
v_{\mu}(\pi R + y) = v_{\mu}(\pi R - y),
\label{PCParities}
\end{eqnarray}
They also obey the boundary conditions at $y = 0$ and $y = \pi R$:
\begin{equation}
\left. u \right|_{y = 0} = e^{{\cal G}_{0}/2} , \quad 
i \left. v_{\mu} \right|_{y = 0} = \frac{1}{2}  \left( {\cal G}_{0 j}
\partial_{\mu} \phi_{0}^{j} 
- {\cal G}_{0 j^*} \partial_{\mu} \phi_{0}^{* j} \right) , \quad
\left. F^{\mu 5} \right|_{y = 0} = 0 ,
\label{CDFBoundaryConditions0}
\end{equation}
\begin{equation}
\left. u \right|_{y = \pi R}  = e^{{\cal G}_{\pi}/2} , \quad 
i \left. v_{\mu} \right|_{y = \pi R} = \frac{1}{2} \left( {\cal G}_{\pi j}
\partial_{\mu} \phi_{\pi}^{j} 
- {\cal G}_{\pi j^*} \partial_{\mu} \phi_{\pi}^{* j} \right) , \quad
\left. F^{\mu 5} \right|_{y = \pi R} = 0 .
\label{CDFBoundaryConditionsPi}
\end{equation}
which allow matching the supersymmetry transformations
for $e^{a}_{\mu}$ and $\psi_{\mu1}$ in the brane $0$ 
(given by equation (\ref{SusyTransBrane0})), from one side, and the
supersymmetry transformations 
induced by the bulk (given by equation (\ref{SusyTransfOfGravitinoModified})
calculated at $y=0$) from 
the other side. 
A  similar result is obtained for the brane $\pi$ after taking into account the 
substitutions (\ref{Brane0ToBranePi}).

\subsection{The auxiliary fields supersymmetry transformations}
\label{secAuxFieldsSusyTransf}

We will determine here the supersymmetry transformations of the auxiliary 
fields $u$ and $v_M$ introduced above. They will be chosen such as to keep the
full 
action invariant.

On one side, under the modified transformations given in 
equation (\ref{SusyTransfOfGravitinoModified}) the bulk supergravity action
transforms as:
\begin{eqnarray}
\delta \int d^{5}x {\cal L}_{SUGRA} &=& \int d^{5}x \left\lbrace 
i \left( \frac{\partial {\cal L}_{SUGRA}}{\partial \psi_{\mu J}}
- D_{N} \left[ \frac{\partial {\cal L}_{SUGRA}}{\partial \left( D_{N}\psi_{\mu
J} \right) } \right] \right) 
\left( v_{\mu} \xi_{J} + u \sigma_{\mu} \overline{\xi}_{J} \right) 
+ h.c. \right\rbrace . 
\nonumber \\ &&
- \int d^{4}x e_{4} \left[ 
8 e^{{\cal G}_{0}/2}\xi_{1} \sigma^{\mu \nu}D_{\mu}\psi_{\nu 1} 
+ h.c. \right]_{y=0}
\nonumber \\ &&
- \int d^{4}x e_{4} \left[ 
8 e^{{\cal G}_{\pi}/2}\xi_{+} \sigma^{\mu \nu}D_{\mu}\psi_{\nu +} 
+ h.c. \right]_{y=\pi R}
\label{LsugraVariation2}
\end{eqnarray}
where the surface terms in equation (\ref{LsugraVariation2}) come from the terms
proportional to $\delta(y)$ and $\delta(y-\pi R)$ in the modified supersymmetry
transformation laws for $\psi_{51}$ and $\psi_{52}$. For the sake of keeping
compact formulae, we have not explicitly written the variation with respect to
the gravitinos.

On the other side the equations (\ref{Laux}) and supersymmetry transformations
(\ref{SusyWeylTransf0}) lead to
\begin{eqnarray}
\delta_{\prime} \int d^{5}x {\cal L}_{AUX} &=& \int d^{5}x 
e_{5} \bigg\lbrace  \frac{1}{2}
\left( u u + v_{M} v^{M}\right) 
\left( i\xi_{1} \sigma^{\mu} \overline{\psi}_{\mu1}
+ i\xi_{2} \sigma^{\mu} \overline{\psi}_{\mu2} 
+ \xi_{2}\psi_{51} - \xi_{1} \psi_{52}  + h.c. \right) 
\nonumber \\ &&
+ u \delta_{\prime} u +  v_{M} \delta_{\prime} v^{M} 
\bigg\rbrace .
\label{LauxVariation}
\end{eqnarray}

\subsubsection{Canceling the bulk terms}

We must impose transformations laws for the auxiliary fields in such a way that
the bulk 
variations in (\ref{LsugraVariation2}) and (\ref{LauxVariation}) cancel each
other. 
This is achieved by taking:
\begin{eqnarray}
\delta_{\prime} u &=& - \frac{1}{2} u \left( i\xi_{1} \sigma^{\nu}
\overline{\psi}_{\nu1}
+ i\xi_{2} \sigma^{\nu} \overline{\psi}_{\nu2} 
+ \xi_{2}\psi_{51} - \xi_{1} \psi_{52} \right) 
\nonumber \\ &&
+ \frac{i}{e_{5}} \left[
    \overline{\xi}_{J} \overline{\sigma}^{\mu} \frac{\partial {\cal
L}_{SUGRA}}{\partial \psi^{\mu}_{J}}
-  \overline{\xi}_{J} \overline{\sigma}^{\mu} D_{N} \frac{\partial {\cal
L}_{SUGRA}}{\partial \left( D_{N}\psi^{\mu}_{J} \right) } 
\right] + h.c.
\nonumber \\
\delta_{\prime} v_{\mu} &=& - \frac{1}{2} v_{\mu} \left( i\xi_{1} \sigma^{\nu}
\overline{\psi}_{\nu1}
+ i\xi_{2} \sigma^{\nu} \overline{\psi}_{\nu2} 
+ \xi_{2}\psi_{51} - \xi_{1} \psi_{52} \right) 
\nonumber \\ &&
- \frac{i}{e_{5}} \left[
    \xi_{J} \frac{\partial {\cal L}_{SUGRA}}{\partial \psi^{\mu}_{J}}
-  \xi_{J} D_{N} \frac{\partial {\cal L}_{SUGRA}}{\partial \left(
D_{N}\psi^{\mu}_{J} \right) } 
\right] + h.c.
\nonumber \\
\delta_{\prime} v_{5} &=& - \frac{1}{2} v_{5} \left( i\xi_{1} \sigma^{\nu}
\overline{\psi}_{\nu1}
+ i\xi_{2} \sigma^{\nu} \overline{\psi}_{\nu2} 
+ \xi_{2}\psi_{51} - \xi_{1} \psi_{52} \right) + h.c.
\label{SusyTransfAux}
\end{eqnarray}
Using equations (\ref{LsugraVariation2}) and (\ref{LauxVariation}) it is easy to
check that 
\begin{eqnarray}
\delta \int d^{5}x {\cal L}_{SUGRA} + \delta_{\prime} \int d^{5}x {\cal L}_{AUX}
&=&
- \int d^{4}x e_{4} \left[ 
8 e^{{\cal G}_{0}/2}\xi_{1} \sigma^{\mu \nu}D_{\mu}\psi_{\nu 1} 
+ h.c. \right]_{y=0}
\nonumber \\ &&
- \int d^{4}x e_{4} \left[ 
8 e^{{\cal G}_{\pi}/2}\xi_{+} \sigma^{\mu \nu}D_{\mu}\psi_{\nu +} 
+ h.c. \right]_{y=\pi R}
\label{LbulkVariation0}
\end{eqnarray}

\subsubsection{Canceling the boundary terms}

The above bulk variations need to completed to include the variation of the
boundary brane actions. This will determine the final modification 
$\delta_{\prime} \rightarrow \delta$ of the transformations laws for the
auxiliary fields that 
make the full action invariant.

To calculate the variation of the brane action under the supersymmetry
transformations 
one could simply plug (\ref{SusyTransfOfGravitinoModified}) in  (\ref{Lbrane0}).
This is 
straight forward but quite long and tedious. Here we exhibit a trick that
permits one to 
find the variation of the brane action in a much shorter way.
Invariance of the action (\ref{Lsugra4d}) under the supersymmetry 
transformations (\ref{SusyTransBrane0}) implies 
\begin{eqnarray}
\delta \int d^{4}x {\cal L}_{0} &=& - \delta S_{Minimal~Sugra}
\nonumber   \\
S_{Minimal~Sugra} &=& \int d^{4}x \left\lbrace -\frac{1}{2} e_{4}
\hat{R}(\hat{\omega}) 
+ e_{4} \epsilon^{\mu \nu \rho \lambda}\overline{\psi}_{\mu
1}\overline{\sigma}_{\nu}\hat{D}_{\rho}\psi_{\lambda1} \right\rbrace .
\label{L0Variation}
\end{eqnarray}
Note  that the action $S_{Minimal~Sugra}$ is invariant under the following
supersymmetry transformations:
\begin{eqnarray}
\delta_{M.S.} e^{a}_{\mu}  &=& i \left( \xi_{1} \sigma^{a} \overline{\psi}_{\mu
1} \right) + h.c.
\nonumber   \\
\delta_{M.S.} \psi_{\mu 1}  &=& 2 \hat{D}_{\mu}\xi_{1} ,
\label{SusyTransSugra4d0nShell}
\end{eqnarray}
where $\hat{D}_{\mu}\xi_{1}$ is given by equation (\ref{CovDer4dDefinition}),
which makes it easy to calculate the supersymmetry variation of the brane
action:
\begin{equation}
\delta \int d^{4}x {\cal L}_{0} = \int d^{4}x e_{4} \left\{
\frac{1}{2} \epsilon^{\mu \nu \rho \lambda}
    \left( {\cal G}_{0 j} \partial_{\mu} \phi_{0}^{j} 
- {\cal G}_{0 j^*} \partial_{\mu} \phi_{0}^{* j} \right) \overline{\xi}_{1}
\overline{\sigma}_{\nu}\hat{D}_{\rho}\psi_{\lambda1}
+ 4 e^{{\cal G}_{0}/2} \xi_{1} \sigma^{\mu \nu} \hat{D}_{\mu}\psi_{\nu 1}
+h.c. \right\} .
\label{L0Variation2}
\end{equation}

Given the parity assignments for the f\"unfbein of table \ref{Parities0}, 
equations (\ref{CovDerivWeylSpinor}) and 
(\ref{CovDer4dDefinition}) imply  $\hat{D}_{\mu}\psi = D_{\mu}\psi$
on the boundary at $y = 0$. The
    boundary conditions 
(\ref{CDFBoundaryConditions0}) leads then to:
\begin{equation}
\delta \int d^{4}x {\cal L}_{0} = \int d^{4}x \left[ e_{4} \left( 
i \epsilon^{\mu \nu \rho \lambda} 
v_{\mu} \overline{\xi}_{1}
\overline{\sigma}_{\nu} D_{\rho}\psi_{\lambda1}
+ 4 e^{{\cal G}_{0}/2} \xi_{1} \sigma^{\mu \nu} D_{\mu}\psi_{\nu 1}
+h.c. \right) \right]_{y = 0} .
\label{L0Variation3}
\end{equation}

The same analysis can be made for the brane $\pi$ through the 
substitutions (\ref{Brane0ToBranePi}) in the above formulae. The result is:
\begin{equation}
\delta \int d^{4}x {\cal L}_{\pi} = \int d^{4}x  \left[ e_{4} \left( 
i \epsilon^{\mu \nu \rho \lambda} 
v_{\mu} \overline{\xi}_{+}
\overline{\sigma}_{\nu} D_{\rho}\psi_{\lambda +}
+ 4 e^{{\cal G}_{\pi}/2} \xi_{+} \sigma^{\mu \nu} D_{\mu}\psi_{\nu +}
+ h.c. \right)  \right]_{y = \pi R} .
\label{LPiVariation}
\end{equation}

To achieve a fully invariant ``bulk plus branes action'' the auxiliary fields
transformations are modified as follow:
\begin{eqnarray}
\delta u &=& \delta_{\prime} u 
\nonumber \\
\delta v_{\mu} &=& \delta_{\prime} v_{\mu} + c_{\mu 0} \delta(y) + c_{\mu \pi}
\delta(y - \pi R)
\nonumber \\
\delta v_{5} &=& \delta_{\prime} v_{5} .
\label{SusyTransfAuxModified}
\end{eqnarray}
The coefficients $c_{\mu 0}$ and $c_{\mu \pi}$ are determined by putting
together the different pieces of the variation of the total action given in 
(\ref{LbulkVariation0}), (\ref{L0Variation3}) and (\ref{LPiVariation})
to find (at first order in $c_{\mu 0}$ and $c_{\mu \pi}$):
\begin{eqnarray}
\delta S 
&=& \int d^{4}x \bigg\{
\left[ e_{4}
i \epsilon^{\mu \nu \rho \lambda} v_{\mu} \overline{\xi}_{1}
\overline{\sigma}_{\nu} D_{\rho}\psi_{\lambda1}
+ h.c. \right]_{y=0} 
+ \left[ e_{4} 
i \epsilon^{\mu \nu \rho \lambda} v_{\mu} \overline{\xi}_{+}
\overline{\sigma}_{\nu} D_{\rho}\psi_{\lambda +}
+ h.c. \right]_{y=\pi R}
\nonumber   \\ &&
+ \frac{1}{2} \left[ e_{4} e^{\hat{5}}_{5} 
v_{\mu} c^{\mu}_{0}
\right]_{y=0}
+ \frac{1}{2} \left[ e_{4} e^{\hat{5}}_{5} 
v_{\mu} c^{\mu}_{\pi} 
\right]_{y=\pi R}
\bigg\}.
\label{TotalActionVariation}
\end{eqnarray}
It is easy to check that if we take:
\begin{eqnarray}
c^\mu_{0} &=& - 2 i e^{5}_{\hat{5}} \epsilon^{\mu \nu \rho \lambda}
\overline{\xi}_{1}
\overline{\sigma}_{\nu} D_{\rho}\psi_{\lambda 1}
    + h.c.
\nonumber \\
c^\mu_{\pi} &=& - 2 i e^{5}_{\hat{5}} \epsilon^{\mu \nu \rho \lambda}
\overline{\xi}_{+}
\overline{\sigma}_{\nu} D_{\rho}\psi_{\lambda +}
    + h.c.
\label{hdDefinition}
\end{eqnarray}
the total action variation is zero to first order in $c^{\mu}_{0}$ and
$c^{\mu}_{\pi}$.
Note that the expressions for 
$c^{\mu}_{0}$ and $c^{\mu}_{\pi}$ are quadratic in the spinor fields, so within
our
approximation, 
where the four-fermion terms in the Lagrangians are dropped, we have $\delta S =
0$.

The bulk plus brane action (\ref{Action}) is
then invariant 
if we use the  transformations (\ref{SusyTransfOfGravitinoModified}) and
(\ref{SusyTransfAuxModified}), 
the parity assignments of tables \ref{Parities0} and \ref{ParitiesPi}, as well
as
the boundary conditions (\ref{CDFBoundaryConditions0}) and
(\ref{CDFBoundaryConditionsPi}). 
These results are summarized in appendix \ref{apeTotalSusyAction} for future
reference.

\subsection{Discontinuity of spinor bulk fields at the boundaries}
\label{secSpinorsDiscintinuity}

An important consequence of the presence  gravitino masses localized on the
branes is the appearance of wave functions discontinuities
(see \cite{Bagger:2001qi}). We provide below  a straightforward
generalization for the case $\omega \neq 0$.

The equations of motion for the gravitinos $\psi_{\mu I}$ can be seen from the
Lagrangians (\ref{Lferm}) and (\ref{BraneLagrangian}) 
to take the form\footnote{we assume $e^{\hat{5}}_{5} =
1$}:
\begin{eqnarray}
- \frac{1}{2} \epsilon^{\mu \nu \rho \lambda}
     \sigma_{\nu}\partial_{\rho}\overline{\psi}_{\lambda 1}
+ \sigma^{\mu \nu}\partial_{5}\psi_{\nu 2} 
- 2 \left\langle e^{{\cal G}_{0}/2} \right\rangle \sigma^{\mu \nu} \psi_{\nu 1}
\delta(y)
\nonumber    \\
- 2 \left\langle e^{{\cal G}_{\pi}/2} \right\rangle \cos(\omega \pi) \sigma^{\mu
\nu} \psi_{\nu +} \delta(y - \pi R)
+ \cdots &=& 0
\nonumber    \\
- \frac{1}{2} \epsilon^{\mu \nu \rho \lambda}
     \sigma_{\nu}\partial_{\rho}\overline{\psi}_{\lambda 2}
- \sigma^{\mu \nu}\partial_{5}\psi_{\nu1} 
+ 2 \left\langle e^{{\cal G}_{\pi}/2} \right\rangle \sin(\omega \pi) \sigma^{\mu
\nu} \psi_{\nu +} \delta(y - \pi R)
+ \cdots &=& 0
\label{GravitinosEOM}
\end{eqnarray}
where $\cdots$ stands for terms which involve other fields that couple to the
gravitinos and that we drop for the purpose of our discussion\footnote{This
amounts to keep the quadratic terms and consider the interaction terms as
perturbations.} The four-dimensional equation of
motion for  gravitinos $\psi_{\mu I}$ of mass $m_{3/2}$:
\begin{equation}
\epsilon^{\mu \nu \rho \lambda}
\sigma_{\nu}\partial_{\rho}\overline{\psi}_{\lambda I} = 
- 2 m_{3/2} \sigma^{\mu \nu} \psi_{\nu I}
\label{MassiveGravitinoEOM}
\end{equation}
leads then to the following equations:
\begin{eqnarray}
     \partial_{5}\psi_{\mu 2} 
+ m_{3/2} \psi_{\mu 1}
&=& 2 \left\langle e^{{\cal G}_{0}/2} \right\rangle \psi_{\mu 1} \delta(y)
+ 2 \left\langle e^{{\cal G}_{\pi}/2} \right\rangle \cos(\omega \pi) \psi_{\mu
+} \delta(y - \pi R)
\nonumber    \\
     \partial_{5}\psi_{\mu 1} 
- m_{3/2} \psi_{\mu 2}
&=& 2 \left\langle e^{{\cal G}_{\pi}/2} \right\rangle \sin(\omega \pi) \psi_{\mu
+} \delta(y - \pi R).
\label{GravitinosEOM2}
\end{eqnarray}

It can be clearly seen from equations (\ref{GravitinosEOM2})  that $\psi_{\mu
1}$ is a 
continuous field near the point $y = 0$ and $\psi_{\mu +}$ is a 
continuous field near the point $y = \pi R$. In contrast, $\psi_{\mu 2}$ has a 
jump at the point $y = 0$ while $\psi_{\mu -}$ has a 
jump at the point $y = \pi R$, their first derivative being proportional to a
Dirac $\delta$ distribution - see Fig. 1.

More precisely, integration of equations (\ref{GravitinosEOM2}) near the points
$y = 0$ and $y = \pi R$, 
taking into account  the parity  assignments of tables \ref{FieldParities0} and
\ref{FieldParitiesPi},  leads to the 
following  discontinuities of the odd gravitinos  wave functions:
\begin{eqnarray}
\lim_{y \rightarrow 0, ~ y > 0}
\psi_{\mu 2}(y) = \psi_{\mu 2}(0^{+}) &=& \left\langle e^{{\cal
G}_{0}/2}\right\rangle
\psi_{\mu 1} (0)= - \psi_{\mu 2}(0^{-})
    \nonumber \\
\lim_{y \rightarrow \pi R, ~ y < \pi R}
\psi_{\mu -}(y) = \psi_{\mu -}(\pi R^{-}) &=& - \left\langle e^{{\cal
G}_{\pi}/2}\right\rangle
\psi_{\mu +} (\pi R)= - \psi_{\mu -}(\pi R^{+})
\label{GravitinoBoundaryConditions}
\end{eqnarray}

\begin{figure}[htb]
       \begin{center}
       \includegraphics[bb=88 4 376 182]{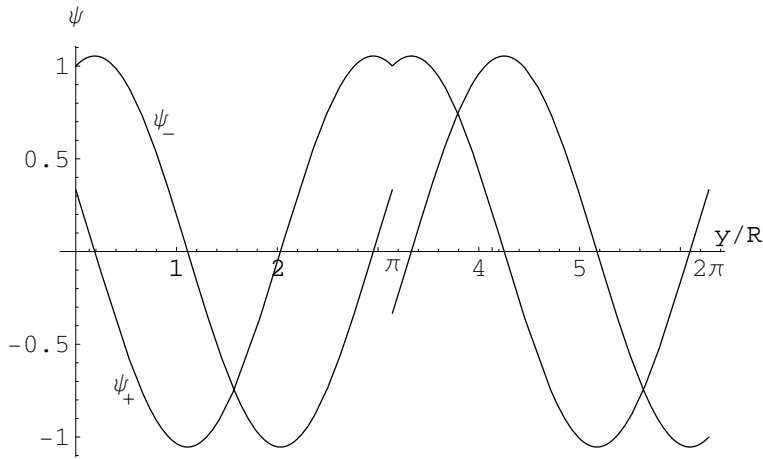}
       \end{center}
       \caption{The two bulk gravitinos wave functions along the compact
dimension. The discontinuities 
are due to the presence of brane localized masses for the other gravitino
component. In this example $\omega = 1/2 $.}
\label{fig:discont1}
\end{figure}

From the transformations of the gravitinos $\psi_{\mu I}$ in
equations (\ref{BulkFieldsSusyTransf}) it is easy to see that 
(\ref{GravitinoBoundaryConditions}) lead to the following boundary conditions
for the supersymmetry parameters:
\begin{eqnarray}
\xi_{2}(0^{+}) &=& 
\left\langle e^{{\cal G}_{0}/2}\right\rangle \xi_{1} (0)=  - \xi_{2}(0^{-})
\nonumber   \\
\xi_{-}(\pi R^{-}) &=&
- \left\langle e^{{\cal G}_{\pi}/2}\right\rangle \xi_{+} (\pi R) = - \xi_{-}(\pi
R^{+}).
\label{SusyParameterBoundaryConditions}
\end{eqnarray}
Interesting to observe is that these boundary conditions
insure that the modified  transformations of  $\psi_{5 I}$ are non
singular: the terms 
proportional to $\delta(y)$ and $\delta(y - \pi R)$ cancel with those
coming from the derivatives 
$\partial_{5} \xi_{2}$ near $y = 0$ and $\partial_{5} \xi_{-}$ near $y = \pi R$.

We end this section by a comment on the relation between the so-called orbifold
approach (used here) 
and the interval approach. We do not seem to bother
about boundary terms that arise after integration 
by parts along the fifth dimension, while it is a central issue in the interval 
approach. Here, we illustrate, through an example,  how the previous
construction can
be understood in the interval approach.

In order to perform the variation of $\int d^{5}x {\cal L}_{SUGRA}$ 
in (\ref{LsugraVariation2}) we integrated by parts in the $y$ direction. 
If the odd gravitino fields are allowed to be discontinuous in the branes the
wave
function 
$\psi_{\mu 2}(0^{+})$ and $\psi_{\mu -}(\pi R^{-})$ may be nonzero. 
So, in the interval approach, one should care about the following surface terms 
in $\delta \int d^{5}x {\cal L}_{SUGRA}$ :
\begin{equation}
\left. \delta \int d^{5}x {\cal L}_{SUGRA} \right|_{\mathrm{SurfaceTerms}} = 
\int d^{4}x \left[ i \frac{\partial {\cal L}_{SUGRA}}{\partial \left(
D_{5}\psi_{\mu J} \right) }
    \left( v_{\mu} \xi_{J} + u \sigma_{\mu} \overline{\xi}_{J} \right) 
+ h.c.\right]_{y=0^{+}}^{y=\pi R^{-}} .
\label{LsugraVariationSurfaceTerms}
\end{equation}
The Lagrangian (\ref{Lferm}) leads to
\begin{eqnarray}
\left. \delta \int d^{5}x {\cal L}_{SUGRA} \right|_{\mathrm{SurfaceTerms}} &=& i
\int d^{4}x \left[ e_{4} 
\psi_{\mu +} \sigma^{\mu \nu} \left( v_{\mu} \xi_{-} + u \sigma_{\mu}
\overline{\xi}_{-} \right)
\right. 
\nonumber \\ &&
    \left. - \psi_{\mu-} \sigma^{\mu \nu} \left( v_{\mu} \xi_{+} + u
\sigma_{\mu}
\overline{\xi}_{+} \right) 
+ h.c. \right]_{y=\pi R^{-}} 
\nonumber \\ &&
- i \int d^{4}x \left[ e_{4} 
\psi_{\mu 1} \sigma^{\mu \nu} \left( v_{\mu} \xi_{2} + u \sigma_{\mu}
\overline{\xi}_{2} \right)
\right. 
\nonumber \\ &&
    \left. - \psi_{\mu2} \sigma^{\mu \nu} \left( v_{\mu} \xi_{1} + u
\sigma_{\mu}
\overline{\xi}_{1} \right) 
+ h.c. \right]_{y=0^{+}}
\label{LsugraVariationSurfaceTerms2}
\end{eqnarray}
and the boundary conditions (\ref{GravitinoBoundaryConditions}) and 
(\ref{SusyParameterBoundaryConditions}) imply:
\begin{equation}
\left. \delta \int d^{5}x {\cal L}_{SUGRA} \right|_{\mathrm{SurfaceTerms}} = 0
\label{LsugraVariationSurfaceTerms3}
\end{equation}

\section{Inclusion of a generalized Scherk-Schwarz mechanism}
\label{secSS}

An important issue is the relation
between  bulk and brane localized gravitino masses and the twists in
Scherk-Schwarz compactifications.
This section collects a few results. Most of them, if
not all, are probably known, but  we
rederived them as they  will be useful in the rest of the paper. It also
introduces
some notations.

It is often useful to work in a basis of periodic fields $\tilde{\psi}_{M I}$ (
ie. $\tilde{\psi}_{M I}(x, y + 2 \pi R) = \tilde{\psi}_{M I}(x ,y)$) in contrast
to the multi-valued 
$\psi_{M I}$ used up to now. These are related by the rotation:
\begin{equation}
\left(\begin{matrix}
\psi_{M 1} \\
\psi_{M 2}
\end{matrix}\right) 
=
\left(\begin{matrix}
\cos[f(y) ] & \sin[ f(y) ] \\
- \sin[ f(y) ] & \cos[ f(y)]
\end{matrix}\right) 
\left(\begin{matrix}
\tilde{\psi}_{M 1} \\
\tilde{\psi}_{M 2}
\end{matrix}\right). 
\label{PsiRotation}
\end{equation}
The function $f(y)$ must obey $f(y + 2 \pi R) = f(y) + 2 \omega \pi$. Here we
follow \cite{Bagger:2001qi} and take:
\begin{equation}
f(y)= \frac{\omega_{B}}{R}y + \frac{ \Omega_{0} - \Omega_{\pi} }{2} \epsilon (y)
+ \frac{ \Omega_{0} + \Omega_{\pi} }{2} \eta (y)
\label{Fdef}
\end{equation}
with $\pi \omega_{B} + \Omega_{0} + \Omega_{\pi} = \omega \pi$. $\epsilon (y)$
is the 'sign function' on $S^{1}$:
\begin{eqnarray}
\epsilon (y) &=& + 1 , \quad 2 k \pi R < y < (2 k+1) \pi R, \quad k \in
\mathbb{Z} 
\nonumber    \\
\epsilon (y) &=& - 1 , \quad (2 k-1) \pi R < y < 2 k \pi R, \quad k \in
\mathbb{Z} 
\label{SignFct}
\end{eqnarray}
and $\eta (y)$ is the 'staircase function':
\begin{equation}
\eta(y) = 2 k + 1 , \quad  k \pi R < y < (k+1) \pi R, \quad k \in \mathbb{Z} 
\label{StaircaseFct}
\end{equation}

The supersymmetry breaking mass terms for the gravitinos is then manifest as we
perform this fields transformation in the kinetic terms of the Lagrangian
(\ref{Lferm}) to give:
\begin{eqnarray}
{\cal L}_{Kinetic} &=& e_{5} \bigg \lbrace  \frac{1}{2} \epsilon^{\mu \nu \rho
\lambda} \left( 
\tilde{\overline{\psi}}_{\mu
1}\overline{\sigma}_{\nu}D_{\rho}\tilde{\psi}_{\lambda 1}
+ \tilde{\overline{\psi}}_{\mu
2}\overline{\sigma}_{\nu}D_{\rho}\tilde{\psi}_{\lambda 2} \right)
    + e^{5}_{\hat{5}} \left(   \tilde{\psi}_{\mu 1} \sigma^{\mu
\nu}D_{5}\tilde{\psi}_{\nu 2} 
- \tilde{\psi}_{\mu2} \sigma^{\mu \nu}D_{5}\tilde{\psi}_{\nu1}  \right)
\nonumber    \\
&&
- 2 e^{5}_{\hat{5}} \left( \tilde{\psi}_{51} \sigma^{\mu
\nu}D_{\mu}\tilde{\psi}_{\nu2} 
- \tilde{\psi}_{52} \sigma^{\mu \nu}D_{\mu}\tilde{\psi}_{\nu 1} \right) 
\nonumber    \\
&&
- \left( \frac{\omega_B}{R} + 2 \Omega_{0} \delta(y) + 2 \Omega_{\pi} \delta(y -
\pi R)
\right) e^{5}_{\hat{5}} \left(  \tilde{\psi}_{\mu 1} \sigma^{\mu
\nu}\tilde{\psi}_{\nu 1} 
+  \tilde{\psi}_{\mu 2} \sigma^{\mu \nu}\tilde{\psi}_{\nu 2}  \right)
+ h.c. \bigg \rbrace .
\label{LkineticRotated}
\end{eqnarray}

The localized mass terms in (\ref{LkineticRotated}) imply discontinuities for
the gravitino wave 
functions. They are too singular to apply the variational principle without
regularization. In reference 
\cite{Bagger:2001qi} it was shown that the Lagrangian density
(\ref{LkineticRotated}) is equivalent to the action:
\begin{eqnarray}
    S_{Kinetic} &=&  \int^{2 \pi R}_{0} dy \int d^{4}x \bigg\{ \frac{1}{2}
    e_{5} \bigg[ \frac{1}{2} \epsilon^{\mu \nu \rho
\lambda} \left( 
\tilde{\overline{\psi}}_{\mu
1}\overline{\sigma}_{\nu}D_{\rho}\tilde{\psi}_{\lambda 1}
+ \tilde{\overline{\psi}}_{\mu
2}\overline{\sigma}_{\nu}D_{\rho}\tilde{\psi}_{\lambda 2} \right) 
\nonumber    \\
&&
    + e^{5}_{\hat{5}} \left(   \tilde{\psi}_{\mu 1} \sigma^{\mu
\nu}D_{5}\tilde{\psi}_{\nu 2} 
- \tilde{\psi}_{\mu2} \sigma^{\mu \nu}D_{5}\tilde{\psi}_{\nu1}  \right)
- 2 e^{5}_{\hat{5}} \left( \tilde{\psi}_{51} \sigma^{\mu
\nu}D_{\mu}\tilde{\psi}_{\nu2} 
- \tilde{\psi}_{52} \sigma^{\mu \nu}D_{\mu}\tilde{\psi}_{\nu 1} \right) 
\nonumber    \\
&&
- \left( \frac{\omega_B}{R} 
\right) e^{5}_{\hat{5}} \left(  \tilde{\psi}_{\mu 1} \sigma^{\mu
\nu}\tilde{\psi}_{\nu 1} +  \tilde{\psi}_{\mu 2} \sigma^{\mu
\nu}\tilde{\psi}_{\nu 2}  \right) \bigg]
\nonumber    \\
&&
- \left[ \tan(\Omega_{0})  \delta(y) + \tan(\Omega_{\pi}) \delta(y - \pi R)
\right] e^{5}_{\hat{5}}  \tilde{\psi}_{\mu 1} \sigma^{\mu \nu}\tilde{\psi}_{\nu
1} 
+ h.c. \bigg\} .
\label{LkineticEqiv}
\end{eqnarray}
with the fields now being piece-wise smooth.

In order to study the supersymmetry transformations of the fields
$\tilde{\psi}_{M I}$ it is convenient to 
regularize the field rotation (\ref{PsiRotation}) by introducing a regularized
function $f_{reg}(y)$ instead of the discontinuous 
function $f(y)$. The \textit{continuous} function $f_{reg}(y)$ obeys: $f_{reg}(
- \varepsilon) = - \Omega_{0}$, 
$f_{reg}( 0 ) = 0$, $f_{reg}( \varepsilon) = \Omega_{0}$, $f_{reg}( \pi R -
\varepsilon) = \Omega_{0} + \pi \omega_{B} $, 
$f_{reg}( \pi R ) = \Omega_{0} + \Omega_{\pi} + \pi \omega_{B} $, 
$f_{reg}( \pi R + \varepsilon) = \Omega_{0} + 2 \Omega_{\pi} + \pi \omega_{B}
$. 
To get the final results it suffices to take the limit $\varepsilon \rightarrow
0$ in the desired expression.

Going to the new basis requires then the following redefinition for the
supersymmetry transformation parameters,
\begin{equation}
\left(\begin{matrix}
\xi_{1} \\
\xi_{2}
\end{matrix}\right) 
=
\left(\begin{matrix}
\cos[ f_{reg}(y) ] & \sin[ f_{reg}(y) ] \\
- \sin[ f_{reg}(y) ] & \cos[ f_{reg}(y) ]
\end{matrix}\right) 
\left(\begin{matrix}
\tilde{\xi}_{1} \\
\tilde{\xi}_{2}
\end{matrix}\right), 
\label{SusyParamRedefinition}
\end{equation}
and  the supersymmetry transformations (\ref{BulkFieldsSusyTransf}) take now the
form:
\begin{eqnarray}
\delta \tilde{\psi}_{\mu 1}  &=& 2 D_{\mu}\tilde{\xi}_{1} + i v_{\mu}
\tilde{\xi}_{1} + i u \sigma_{\mu} \tilde{\overline{\xi}}_{1} + \cdots
\nonumber    \\
\delta \tilde{\psi}_{\mu 2}  &=& 2 D_{\mu}\tilde{\xi}_{2} + i v_{\mu}
\tilde{\xi}_{2} + i u \sigma_{\mu} \tilde{\overline{\xi}}_{2} + \cdots
\nonumber    \\
\delta \tilde{\psi}_{5 1}  &=& 2 D_{5}\tilde{\xi}_{1} + 2 \frac{df_{reg}}{dy}
\tilde{\xi}_{2} 
+ \cdots
\nonumber    \\
\delta \tilde{\psi}_{5 2}  &=& 2 D_{5}\tilde{\xi}_{2} - 2 \frac{df_{reg}}{dy}
\tilde{\xi}_{1} 
- 4  e^{{\cal G}_{0}/2} \, \, \tilde{\xi}_{1}  \delta(y)
- 4  e^{{\cal G}_{\pi}/2} \, \, \tilde{\xi}_{1} \delta(y-\pi R)
+ \cdots
\label{SusyTransfPsiRotated}
\end{eqnarray}
where $\cdots$ stand for terms which are proportional to $F^{MN}$.

It is important to note that the fields $\tilde{\psi}_{5 1}$ and
$\tilde{\psi}_{5 2}$ transforms non linearly 
under supersymmetry transformations: \textit{they are the  Goldstino fields
associated with the supersymmetry breaking in the bulk}, as expected.

The new brane field transformations and boundary conditions can be easily
obtained after
noticing that these redefinitions (\ref{PsiRotation}) and
(\ref{SusyParamRedefinition}) imply for the brane at  $y = \pi R$:
\begin{equation}
\psi_{\mu +}(\pi R) = \tilde{\psi}_{\mu 1} (\pi R) , \quad
    \xi_{+} (\pi R) =  \tilde{\xi}_{1} (\pi R)
, \quad \psi_{\mu -} (\pi R) = \tilde{\psi}_{\mu 2} (\pi R), \quad
\xi_{-} (\pi R) = \tilde{\xi}_{2}(\pi R).
\label{BraneSpinorsRotation}
\end{equation}

Of main interest in the generalized Sherk-Schwarz mechanism is the interplay
between bulk and brane localized gravitino mass terms in order to keep or break
supersymmetry. Given that we have explicitly obtained all the supersymmetry
transformations, it is very easy to us to answer this question by looking for
Killing spinors. Explicitly, we consider the supersymmetry 
transformations (\ref{SusyTransfPsiRotated}) evaluated with the appropriate
background and search 
for spinors $\tilde{\xi}_{I}$ which obey $\delta \tilde{\psi}_{M I}=0$. The
interesting equations arise 
from $\delta \tilde{\psi}_{5 1}$ and $\delta \tilde{\psi}_{5 2}$:
\begin{eqnarray}
    \partial_{5}\tilde{\xi}_{1} +  \frac{df_{reg}}{dy} \tilde{\xi}_{2} &=& 0
\nonumber    \\
    \partial_{5}\tilde{\xi}_{2} -  \frac{df_{reg}}{dy} \tilde{\xi}_{1} 
&=& 2 \left\langle e^{{\cal G}_{0}/2} \right\rangle \, \, \tilde{\xi}_{1} 
\delta(y)
+ 2 \left\langle e^{{\cal G}_{\pi}/2} \right\rangle \, \, \tilde{\xi}_{1}
\delta(y-\pi R) .
\label{KillingEquation}
\end{eqnarray}
The parity transformation assignments of tables \ref{FieldParities0},
\ref{FieldParitiesPi} and 
equation (\ref{SusyParamRedefinition}) imply:
\begin{eqnarray}
&
\tilde{\xi}_{1}(-y) = + \tilde{\xi}_{1}(y) , \quad \tilde{\xi}_{1}(\pi R - y) =
+ \tilde{\xi}_{1}(\pi R + y) 
\nonumber    \\ &
\tilde{\xi}_{2}(-y) = - \tilde{\xi}_{2}(y) , \quad \tilde{\xi}_{2}(\pi R - y) =
- \tilde{\xi}_{2}(\pi R + y) 
\label{TildeXiParities}
\end{eqnarray}
Integrating equations (\ref{KillingEquation}) at $y = 0$ and $y = \pi R$ and
taking into account
    (\ref{TildeXiParities}) we deduce that $\tilde{\xi}_{1}$ is a 
continuous field near the points $y = 0$ and $y = \pi R$ and that
$\tilde{\xi}_{2}$ has a jump 
at the points $y = 0$ and $y = \pi R$:
\begin{eqnarray}
\tilde{\xi}_{2}(0^{+})&=& \left\langle e^{{\cal G}_{0}/2} \right\rangle \, \, 
\tilde{\xi}_{1} (0)
\nonumber   \\
\tilde{\xi}_{2}(\pi R^{-}) &=& - \left\langle e^{{\cal G}_{\pi}/2} \right\rangle
\, \, 
\tilde{\xi}_{1} (\pi R)
\label{KillingSpinorsJunp}
\end{eqnarray}

The solutions for the Killing equations (\ref{KillingEquation}) in the interval
$0 < y < \pi R$ with the first 
boundary condition in (\ref{KillingSpinorsJunp}) are given by:
\begin{eqnarray}
\tilde{\xi}_{1}(y) &=&  \tilde{\xi}_{1}(0) \left\lbrace
    \cos [ f_{reg}(y) ] - \left\langle e^{{\cal G}_{0}/2} \right\rangle \, \, 
\sin
[ f_{reg}(y) ]  \right\rbrace
\nonumber   \\
\tilde{\xi}_{2}(y) &=& \tilde{\xi}_{1}(0) \left\lbrace
    \sin [ f_{reg}(y) ] +\, \, \left\langle e^{{\cal G}_{0}/2} \right\rangle \,
\,
    \cos
[ f_{reg}(y) ]  \right\rbrace
\label{KillingSpinorSolution}
\end{eqnarray}
The second boundary condition in (\ref{KillingSpinorsJunp}) leads to the
following relation:
\begin{eqnarray}
\frac{ \left\langle e^{{\cal G}_{0}/2} \right\rangle + \left\langle e^{{\cal
G}_{\pi}/2} \right\rangle }
{\left\langle e^{{\cal G}_{0}/2} \right\rangle\, \, \left\langle e^{{\cal
G}_{\pi}/2} \right\rangle - 1}
    & = & 
\tan [ f_{reg}(\pi R) ]
\nonumber   \\
\Rightarrow \quad
\Omega_{0} + \Omega_{\pi} + \pi \omega_{B} 
+ \arctan \left( \left\langle e^{{\cal G}_{0}/2} \right\rangle \right) 
+ \arctan \left( \left\langle e^{{\cal G}_{\pi}/2} \right\rangle \right) 
& = & n \pi , \quad
\left( n \in \mathbb{Z} \right) 
\label{SusyRestorationCondition}
\end{eqnarray}
It is sometimes useful to introduce the angles $\Theta_b$ ( $b=0,\pi$) defined
by $ \left\langle e^{{\cal G}_{b}/2} \right\rangle  = \tan {\Theta_b}$. Then,
equation (\ref{SusyRestorationCondition}) takes the simple form:
\begin{equation}
    \tan( \omega \pi + \Theta_0 +\Theta_\pi) = 0
\label{SusyRestorationCondition2}
\end{equation}
Equation (\ref{SusyRestorationCondition}) is one condition that indicates when
supersymmetry is
not spontaneously broken, other conditions are 
obtained by studding the supersymmetry 
transformations of the fields $\chi_{0}$ and $\chi_{\pi}$ in the branes. They 
imply $N_{0} + N_{\pi}$ extra conditions for the existence of Killing spinors:
\begin{equation}
\left\langle {e^{{\cal G}_{0}/2} \cal G}_{0 j}\right\rangle = 0 ,   \quad
\left\langle {e^{{\cal G}_{\pi}/2} \cal G}_{\pi j}\right\rangle = 0 .
\label{SusyBraneCondition}
\end{equation}

\section{Suspending branes in the bulk}
\label{secMultBranes}

In this section,  we generalize the previous results for the case with 
multiple branes. More precisely, we consider $N + 1$ branes placed at the points
$y = y_{n}$, $n = 0 \cdots N$ with $y_{0} = 0$, $y_{N} = \pi R$, and $y_{n} <
y_{n+1} $. The total action is given by:
\begin{equation}
    S =  \int^{2 \pi R}_{0} dy \int d^{4}x \left[ \frac{1}{2} {\cal L}_{BULK} + 
\sum_{n = 0}^{N} {\cal L}_{n} \delta(y - y_{n})  \right] .
\label{ActionN}
\end{equation}

Every ``brane $n$'' will be characterized by the choice of the bulk fields, in
particular the
gravitino, that couple to its worldvolume.
These are in fact determined as being the even fields under the $\mathbb{Z}_2$
action
at the point $y = y_{n}$:
\begin{equation}
\varphi_{even}(y_{n} + y) = {\cal P}_{n} \varphi_{even}(y_{n} - y) =
\varphi_{even}(y_{n} - y).
\label{ParityN}
\end{equation}
We adopt the parity transformations shown in table \ref{ParitiesN}, where the
following
definitions have been introduced:
\begin{eqnarray}
\psi_{\mu +}^{~n} & = &
\cos( \theta_{n} ) \psi_{\mu 1} - \sin( \theta_{n} ) \psi_{\mu 2} 
\nonumber \\
\psi_{\mu -}^{~n} & = &
\sin( \theta_{n} ) \psi_{\mu 1} + \cos( \theta_{n} ) \psi_{\mu 2} 
\nonumber \\
\psi_{5 +}^{~n} & = &
\sin( \theta_{n} ) \psi_{5 1} + \cos( \theta_{n} ) \psi_{5 2} 
\nonumber \\
\psi_{5 -}^{~n} & = &
\cos( \theta_{n} ) \psi_{5 1} - \sin( \theta_{n} ) \psi_{5 2} 
\nonumber \\
\xi_{+}^{~n} & = &
\cos( \theta_{n} ) \xi_{1} - \sin( \theta_{n} ) \xi_{2} 
\nonumber \\
\xi_{-}^{~n} & = &
\sin( \theta_{n} ) \xi_{1} + \cos( \theta_{n} ) \xi_{2} 
\label{ParityEigenvectorsN}
\end{eqnarray}
The case of boundary branes discussed in previous sections corresponds to
$\theta_{0} = 0$ and $\theta_{N} = \omega \pi$.
\begin{table}[htb]
      \begin{center}
         \begin{tabular}{|l|c|c|c|c|c|c|c|c|}
         \hline
         ${\cal P}_{n} = +1$ & $e^{a}_{\mu}$ & $e^{\hat{5}}_{5}$ & $B_{5}$ &
$\psi_{\mu +}^{~n}$
         & $\psi_{5 +}^{~n}$ & $\xi_{+}^{~n}$ & $v_{\mu}$ & $u$  \\
         \hline
         ${\cal P}_{n} = -1$ & $e^{a}_{5}$ & $e^{\hat{5}}_{\mu}$ & $B_{\mu}$ &
$\psi_{\mu -}^{~n}$
         & $\psi_{5 -}^{~n}$ & $\xi_{-}^{~n}$        & $v_{5}$ &  \\
         \hline
         \end{tabular}
      \caption{Parity assignments for bulk fields at $y=y_{n}$.}
\label{ParitiesN}
      \end{center}
\end{table}

The Lagrangian density and supersymmetry transformations for the worldvolume
fields living on the brane $n$ are given by 
equations (\ref{Lbrane0}) and (\ref{SusyTransBrane0}) after the substitutions:
\begin{equation}
brane~0 \rightarrow brane~n : \quad
\left \lbrace
\begin{array}{c c c c}
{\cal L}_{0} \rightarrow {\cal L}_{n} , \quad &
\phi_{0}^{i} \rightarrow \phi_{n}^{i} , \quad &
\chi_{0}^{i} \rightarrow \chi_{n}^{i} , \quad &
{\cal G}_{0} \rightarrow {\cal G}_{n} , \\
\psi_{\mu 1} \rightarrow \psi_{\mu +}^{~n} , \quad &
\xi_{1} \rightarrow \xi_{+}^{~n} , \quad &
K_{0} \rightarrow K_{n} , \quad &
W_{0} \rightarrow W_{n} .
\end{array}
\right .
\label{Brane0ToBraneN}
\end{equation}

The bulk Lagrangian density is given as before by equation (\ref{Lbulk3}), with
supersymmetry transformations given by:
\begin{eqnarray}
\delta e^{A}_{M} &=& \delta_{\prime} e^{A}_{M}
\nonumber \\
\delta B_{M} &=& \delta_{\prime} B_{M}
\nonumber \\
\delta \psi_{\mu 1}  &=& \delta_{\prime} \psi_{\mu 1} + i v_{\mu} \xi_{1} + i u
\sigma_{\mu} \overline{\xi}_{1}
\nonumber   \\
\delta \psi_{\mu 2}  &=& \delta_{\prime} \psi_{\mu 2} + i v_{\mu} \xi_{2} + i u
\sigma_{\mu} \overline{\xi}_{2}
\nonumber \\
\delta \psi_{5 1} &=& \delta_{\prime} \psi_{5 1} 
- 4 \sum_{n = 0}^{N} e^{{\cal G}_{n}/2} \sin(\theta_{n})\xi_{+}^{~n}
\delta(y-y_{n})
\nonumber \\
\delta \psi_{5 2} &=& \delta_{\prime} \psi_{5 2} 
- 4 \sum_{n = 0}^{N} e^{{\cal G}_{n}/2} \cos(\theta_{n})\xi_{+}^{~n}
\delta(y-y_{n}) 
\nonumber \\
\delta u &=& \delta_{\prime} u 
\nonumber \\
\delta v^{\mu} &=& \delta_{\prime} v^{\mu} - 2 i \sum_{n = 0}^{N} \left(
e^{5}_{\hat{5}} \epsilon^{\mu \nu \rho \lambda} 
\overline{\xi}_{+}^{~n} \overline{\sigma}_{\nu} D_{\rho}\psi_{\lambda +}^{~n}
+ h.c. \right) \delta(y - y_{n} )
\nonumber \\
\delta v_{5} &=& \delta_{\prime} v_{5}
\label{SusyTransfN}
\end{eqnarray}
where the transformations $\delta_{\prime}$ are given in equations
(\ref{SusyWeylTransf0}) and (\ref{SusyTransfAux}).

We also impose boundary conditions at $y = y_{n}$, these are given by the
obvious generalization of (\ref{CDFBoundaryConditions0}) and
(\ref{CDFBoundaryConditionsPi}).
With these boundary conditions and the parity assignments of table
\ref{ParitiesN} the action
(\ref{ActionN}) is invariant under the transformations (\ref{SusyTransfN}).

Consider the case of localized gravitino masses $M_n$ that include the branes
$F$-terms and generalized Scherk-Schwarz contribution  written in
\ref{LkineticEqiv}. Repeating the analysis of section
\ref{secSpinorsDiscintinuity} for the
gravitinos equations of motion
shows that the field $\psi_{\mu +}^{~n}$ is continuous at the point $y =
y_{n}$ while the 
field $\psi_{\mu -}^{~n}$ has a jump at this point:
\begin{equation}
\lim_{y \rightarrow y_{n}, ~ y > y_{n}} \psi_{\mu -}^{~n} = 
\psi_{\mu -}^{~n}(y_{n}^{+}) = M_n
\psi_{\mu +}^{~n} (y_{n}).=- \psi_{\mu -}^{~n}(y_{n}^{-})
\label{GravitinoDiscontN}
\end{equation}
The corresponding boundary conditions for the supersymmetry
transformation parameters at the point $y = y_{n}$ are:
\begin{equation}
\xi_{-}^{~n}(y_{n}^{+}) =
M_n \xi_{+}^{~n}(y_{n}) = - \xi_{-}^{~n}(y_{n}^{-})
\label{SusyParamBoundaryCondN}
\end{equation}

Supersymmetry can remain unbroken for a peculiar choice of localized and bulk 
gravitino masses, following the same lines as 
section \ref{secSS}. Again the 
equations of interest arise from requiring $\delta \psi_{5 1} = 0$ 
and $\delta \psi_{5 2} = 0$:
\begin{eqnarray}
\partial_{5} \xi_{1} 
- 2 \sum_{n = 0}^{N} M_n \sin(\theta_{n})\xi_{+}^{~n}
\delta(y-y_{n}) &=& 0
\nonumber \\
\partial_{5} \xi_{2} 
- 2 \sum_{n = 0}^{N} M_n \cos(\theta_{n})\xi_{+}^{~n}
\delta(y-y_{n}) &=& 0
\label{KillingEqMultiBrane}
\end{eqnarray}

Integrating equations (\ref{KillingEqMultiBrane}) near $y = y_{n}$, taking 
into account the parity assignments of table \ref{ParitiesN}, shows that
$\xi_{+}^{~n}$  are continuous fields  at $y = y_{n}$ while
    $\xi_{-}^{~n}$ have jumps at these points given by
(\ref{SusyParamBoundaryCondN}).

The solution of equations (\ref{KillingEqMultiBrane}) in the interval $y_{n} < y
< y_{n + 1}$  with the condition  (\ref{SusyParamBoundaryCondN}) can be written
as:
\begin{eqnarray}
\xi_{1}(y) = \left. \xi_{1} \right|_{y = y_{n}^{+}}
\nonumber \\
\xi_{2}(y) = \frac{M_n
\cos(\theta_{n}) - \sin(\theta_{n}) }
{ \cos(\theta_{n}) + M_n
\sin(\theta_{n}) } \left. \xi_{1} \right|_{y = y_{n}^{+}}
    = \tan \left[ \arctan \left( M_n
\right) - \theta_{n} \right] \left. \xi_{1} \right|_{y = y_{n}^{+}} .
\label{KillingSolutionMultiBrane}
\end{eqnarray}
then using (\ref{SusyParamBoundaryCondN}) evaluated at $y_{n+1}$, gives the
following conditions:
\begin{equation}
\theta_{n + 1} - \theta_{n} + \arctan \left( M_n\right) + \arctan \left( M_{n+1}
\right) = k \pi , \quad \left( k \in \mathbb{Z} \right).
\label{SusyRestorCondN}
\end{equation}

These N conditions generalize equation
(\ref{SusyRestorationCondition}) for the multi-brane case.
When one of the relations (\ref{SusyRestorCondN}) is not satisfied, the Killing
spinor
equations have no solution and supersymmetry is \textit{spontaneously broken in
the bulk} by a non trivial
Scherk-Schwarz twist.
The other necessary conditions for supersymmetry not to be spontaneously broken
are the direct 
generalization of \ref{SusyBraneCondition}.

\section{The super-Higgs mechanism}
\label{secSuperHiggs}

In section \ref{secSS}, we studied the supersymmetry breaking induced by
non-periodic 
boundary conditions for the gravitinos.
Here, we turn our attention to the $F$-terms of chiral multiplets living on the
branes worldvolume. More precisely, we will determine the condition for
supersymmetry breaking and study the super-Higgs effect associated.

We will perform our study in the simplest case with no branes in the bulk other
than the boundary 
ones at $y = 0$ and $y = \pi R$, as it contains all the qualitative features. 
Equations (\ref{SusyTransBrane0}) and (\ref{SusyTransfPsiRotated}) show that
four fields 
$\psi_{5 1}$, $\psi_{5 2}$, $\chi_{0}$ and $\chi_{\pi}$ transform non linearly
under supersymmetry 
transformations. These are the \textit{``local would be Goldstinos''} associated
with breaking of supersymmetry in the bulk and in the two branes respectively.
As we have two gravitinos  then two \textit{local would be Goldstinos} will be
absorbed in the super-Higgs effect to give mass to the gravitino fields
$\psi_{\mu 1}$ and $\psi_{\mu 2}$, while  two linear combinations of 
the fields $\psi_{5 1}$, $\psi_{5 2}$, $\chi_{0}$ and $\chi_{\pi}$ remain as
\textit{pseudo-Goldstinos}.

To keep the formulae explicit, we will make a number of simplifications:
\begin{itemize}
    \item We impose a zero tree level cosmological constant at each brane. This
implies that the vacuum expectation values of the bosonic fields are:
\begin{eqnarray}
& \left\langle g^{i j^*} {\cal G}_{0 i} {\cal G}_{0 j^*} \right\rangle  = 3   ,
\quad
\left\langle g^{i j^*}  {\cal G}_{0 j^*} \left( {\cal G}_{0 k i} - \Gamma_{k
i}^{l} {\cal G}_{0 l} \right)
    + {\cal G}_{0 k} \right\rangle  = 0,
\nonumber \\
& \left\langle g^{i j^*} {\cal G}_{\pi i} {\cal G}_{\pi j^*} \right\rangle = 3 
, \quad
\left\langle g^{i j^*}  {\cal G}_{\pi j^*} \left( {\cal G}_{\pi k i} - \Gamma_{k
i}^{l} {\cal G}_{\pi l} \right)
    + {\cal G}_{\pi k} \right\rangle  = 0,
\label{MinPotBoson}
\end{eqnarray}
the second and fourth equalities in equations (\ref{MinPotBoson}) come from the
extremisation 
of the scalar potential at the branes $0$ and $\pi$.
In appendix \ref{apeSimpleExample} one explicit example of K\"ahler function
which satisfies (\ref{MinPotBoson}) 
is presented.

\item We will consider that the boundaries gravitino masses arise through
explicit $F$-terms for the boundary supermultiplets as 
\begin{equation}
     M_{b} = \, \, \left\langle e^{{\cal G}_{b}/2} \right\rangle  \qquad {\rm
with}
\qquad b\in \{ 0, \pi \}
\end{equation}
while the terms $\tan(\Omega_b)$ arise from a generalized Scherk-Schwarz
mechanism and are 
absorbed by redefining the bulk twist $\omega_{B}$ by $\omega_{B~initial}
\longrightarrow \omega_{B} = 
\omega_{B~initial} + \frac{\Omega_{0}+\Omega_{\pi}}{\pi} = \omega$.

\item We adopt the following notation: 
\begin{eqnarray}
\chi_{0} &=& \frac{1}{\sqrt{3}} \left\langle {\cal G}_{0 i} \right\rangle
\chi_{0}^{i}
\nonumber \\
\chi_{\pi} &=& \frac{1}{\sqrt{3}}\left\langle {\cal G}_{\pi i} \right\rangle
\chi_{\pi}^{i} .
\label{ChiDef}
\end{eqnarray}
and we assume that the kinetic terms are canonically normalized: $g_{i j^{*}} =
\delta_{i j^{*}} + \cdots$ .

\item \textit{From now on, we drop the overscript}  $\tilde{ }$  over the 
fields defined in (\ref{PsiRotation}).

\end{itemize}

In order to study the super-Higgs effect 
we will concentrate on the bilinear terms of the fermionic fields: $\psi_{\mu
1}$, $\psi_{\mu 2}$, 
$\psi_{5 1}$, $\psi_{5 2}$, $\chi_{0}$ and $\chi_{\pi}$. These can be extracted
from equations 
(\ref{Lbrane0}) and (\ref{LkineticRotated}) and they take the form:
\begin{eqnarray}
{\cal L} &=& \frac{1}{2} \bigg \lbrace \frac{1}{2} \epsilon^{\mu \nu \rho
\lambda} \left( 
\overline{\psi}_{\mu 1}\overline{\sigma}_{\nu}\partial_{\rho}\psi_{\lambda 1}
+ \overline{\psi}_{\mu 2}\overline{\sigma}_{\nu}\partial_{\rho}\psi_{\lambda 2}
\right) 
+ \psi_{\mu 1} \sigma^{\mu \nu}\partial_{5} \psi_{\nu 2} 
- \psi_{\mu2} \sigma^{\mu \nu}\partial_{5} \psi_{\nu1} 
\nonumber    \\
&&
+ 2 \left( \psi_{52} \sigma^{\mu \nu}\partial_{\mu}\psi_{\nu 1}
    - \psi_{51} \sigma^{\mu \nu}\partial_{\mu}\psi_{\nu2} \right) 
- \frac{\omega}{R} \left( \psi_{\mu 1} \sigma^{\mu \nu} \psi_{\nu 1} 
+  \psi_{\mu 2} \sigma^{\mu \nu} \psi_{\nu 2}  \right) 
\bigg \rbrace
\nonumber    \\ &&
+ \delta (y) 
\left\lbrace  - \frac{i}{2} \overline{\chi}_{0} \overline{\sigma}^{\mu}
\partial_{\mu} \chi_{0}
- M_{0} \left[ \psi_{\mu 1} \sigma^{\mu \nu} \psi_{\nu 1} 
+ i \frac{\sqrt{6}}{2} \overline{\chi}_{0} \overline{\sigma}^{\mu} \psi_{\mu 1}
+ \chi_{0} \chi_{0} \right] \right\rbrace 
\nonumber \\ &&
+ \delta (y - \pi R) 
\left\lbrace  - \frac{i}{2} \overline{\chi}_{\pi} \overline{\sigma}^{\mu}
\partial_{\mu} \chi_{\pi}
- M_{\pi} \left[ \psi_{\mu 1} \sigma^{\mu \nu} \psi_{\nu 1} 
+ i \frac{\sqrt{6}}{2} \overline{\chi}_{\pi} \overline{\sigma}^{\mu} \psi_{\mu
1}
+ \chi_{\pi} \chi_{\pi} \right] \right\rbrace 
+ h.c. 
\label{Lkm}
\end{eqnarray}

\subsection{$R_{\xi}$ gauge}
\label{secRXiGauge}

Here we will use the analogous of $R_{\xi}$ gauges of non abelian gauge
theories. This kind of gauge fixing in supergravity theories was first discussed
in
\cite{Baulieu:1985wa}. Our discussion follows and generalizes the simpler case
of
pure
Scherk-Schwarz breaking studied in  \cite{DeCurtis:2003hs}.

Some field redefinitions allow obtaining standard kinetic terms for the fields
$\psi_{5 I}$:
\begin{eqnarray}
\psi_{\mu 1} & \rightarrow & \psi_{\mu 1} + \frac{i}{\sqrt{6}} \sigma_{\mu}
\overline{\psi}_{5 2}
\nonumber \\
\psi_{\mu 2} & \rightarrow & \psi_{\mu 2} - \frac{i}{\sqrt{6}} \sigma_{\mu}
\overline{\psi}_{5 1} 
\nonumber \\
\psi_{5 1} & \rightarrow & \frac{2}{\sqrt{6}} \psi_{5 1} 
\nonumber \\
\psi_{5 2} & \rightarrow & \frac{2}{\sqrt{6}} \psi_{5 2} .
\label{PsiRed}
\end{eqnarray}
    leading to the Lagrangian density:
\begin{eqnarray}
{\cal L} &=& \frac{1}{2} \bigg \lbrace \frac{1}{2} \epsilon^{\mu \nu \rho
\lambda} \left( 
\overline{\psi}_{\mu 1}\overline{\sigma}_{\nu}\partial_{\rho}\psi_{\lambda 1}
+ \overline{\psi}_{\mu 2}\overline{\sigma}_{\nu}\partial_{\rho}\psi_{\lambda 2}
\right) 
+ \psi_{\mu 1} \sigma^{\mu \nu}\partial_{5} \psi_{\nu 2} 
- \psi_{\mu2} \sigma^{\mu \nu}\partial_{5} \psi_{\nu1} 
\nonumber    \\
&&
- \frac{i}{2} \left( \overline{\psi}_{51} \overline{\sigma}^{\mu} \partial_{\mu}
\psi_{51} 
+  \overline{\psi}_{52} \overline{\sigma}^{\mu} \partial_{\mu} \psi_{52}
\right) 
+ \psi_{5 1} \partial_{5} \psi_{5 2} 
- \psi_{5 2} \partial_{5} \psi_{5 1} 
\nonumber    \\
&&
- \frac{\omega}{R} \left( \psi_{\mu 1} \sigma^{\mu \nu} \psi_{\nu 1} 
+  \psi_{\mu 2} \sigma^{\mu \nu} \psi_{\nu 2} 
+  \psi_{51} \psi_{51} + \psi_{52} \psi_{52} \right) 
\nonumber    \\ &&
- i\frac{\sqrt{6}}{2} \left[  \partial_{5} \overline{\psi}_{5 1}
\overline{\sigma}^{\mu} \psi_{\mu 1} 
+  \partial_{5} \overline{\psi}_{5 2} \overline{\sigma}^{\mu} \psi_{\mu 2} 
+ \frac{\omega}{R} \left( \overline{\psi}_{5 2} \overline{\sigma}^{\mu}
\psi_{\mu 1} 
- \overline{\psi}_{5 1} \overline{\sigma}^{\mu} \psi_{\mu 2}  \right) \right] 
\bigg \rbrace
\nonumber    \\ &&
+ \delta (y) 
\Bigg\lbrace   - \frac{i}{2} \overline{\chi}_{0} \overline{\sigma}^{\mu}
\partial_{\mu} \chi_{0}
- M_{0} \Bigg[ \psi_{\mu 1} \sigma^{\mu \nu} \psi_{\nu 1} 
+ i \frac{\sqrt{6}}{2} \left(  \overline{\chi}_{0} + \overline{\psi}_{5 2}
\right) \overline{\sigma}^{\mu} \psi_{\mu 1}
\nonumber \\ &&
+ \left( \chi_{0} + \psi_{5 2} \right) \left( \chi_{0} + \psi_{5 2} \right) 
\Bigg] \Bigg\rbrace 
+ \delta (y - \pi R) 
\Bigg\lbrace  - \frac{i}{2} \overline{\chi}_{\pi} \overline{\sigma}^{\mu}
\partial_{\mu} \chi_{\pi}
\nonumber \\ &&
- M_{\pi} \Bigg[ \psi_{\mu 1} \sigma^{\mu \nu} \psi_{\nu 1} 
+ i \frac{\sqrt{6}}{2} \left( \overline{\chi}_{\pi} + \overline{\psi}_{5 2}
\right) \overline{\sigma}^{\mu} \psi_{\mu 1}
+ \left( \chi_{\pi} + \psi_{5 2} \right) \left( \chi_{\pi} + \psi_{5 2} \right)
\Bigg] \Bigg\rbrace 
+ h.c. 
\label{Lkm2}
\end{eqnarray}
    instead of (\ref{Lkm}).

The gauge choice is made by the addition  to the Lagrangian density of the
$R_{\xi}$ gauge fixing term:
\begin{equation}
{\cal L}_{GF} = - \frac{i}{2 \xi} \left( \overline{h}_{1}
\overline{\sigma}^{\mu} \partial_{\mu} h_{1} 
+ \overline{h}_{2} \overline{\sigma}^{\mu} \partial_{\mu} h_{2} \right) 
\label{LGF}
\end{equation}
where 
\begin{eqnarray}
h_{1} & = & \sigma^{\mu}\overline{\psi}_{\mu 1} - \frac{\sqrt{6}}{2} \xi 
\frac{\sigma^{\mu} \partial_{\mu} }{\partial^{2}} \overline{g}_{1}
\nonumber \\
h_{2} & = & \sigma^{\mu}\overline{\psi}_{\mu 2} - \frac{\sqrt{6}}{2} \xi 
\frac{\sigma^{\mu} \partial_{\mu} }{\partial^{2}}  \overline{g}_{2}
\label{GaugeFixTerm}
\end{eqnarray}
with
\begin{eqnarray}
g_{1} & = & \partial_{5}{\psi}_{5 1} + \frac{\omega}{R} {\psi}_{5 2}
+ 2 \delta (y) M_{0} \left(  {\chi}_{0} +  {\psi}_{5 2} \right) 
+ 2 \delta (y - \pi R) M_{\pi} \left(  {\chi}_{\pi} + {\psi}_{5 2} \right)
\nonumber \\
g_{2} & = & \partial_{5}{\psi}_{5 2} - \frac{\omega}{R} {\psi}_{5 1} 
\label{GaugeFixTerm2}
\end{eqnarray}
and $\xi$ is a free constant gauge parameter.

It is straight forward to check that this gauge fixing term provides the
cancellation of mixing terms between gravitino and Goldstino fields, which is
the aim of our gauge choice :
\begin{eqnarray}
{\cal L} + {\cal L}_{GF} &=& \frac{1}{2} \bigg \lbrace  \left( 1 - \xi^{-1}
\right) 
\frac{1}{2} \epsilon^{\mu \nu \rho \lambda} \left( 
\overline{\psi}_{\mu 1}\overline{\sigma}_{\nu}\partial_{\rho}\psi_{\lambda 1}
+ \overline{\psi}_{\mu 2}\overline{\sigma}_{\nu}\partial_{\rho}\psi_{\lambda 2}
\right) 
\nonumber    \\
&&
+ \psi_{\mu 1} \sigma^{\mu \nu}\partial_{5} \psi_{\nu 2} 
- \psi_{\mu2} \sigma^{\mu \nu}\partial_{5} \psi_{\nu1} 
- \frac{i}{2} \left( \overline{\psi}_{51} \overline{\sigma}^{\mu} \partial_{\mu}
\psi_{51} 
+  \overline{\psi}_{52} \overline{\sigma}^{\mu} \partial_{\mu} \psi_{52}
\right) 
\nonumber    \\
&&
+ \psi_{5 1} \partial_{5} \psi_{5 2} 
- \psi_{5 2} \partial_{5} \psi_{5 1} 
- \frac{\omega}{R} \left( \psi_{\mu 1} \sigma^{\mu \nu} \psi_{\nu 1} 
+  \psi_{\mu 2} \sigma^{\mu \nu} \psi_{\nu 2} 
+  \psi_{51} \psi_{51} + \psi_{52} \psi_{52} \right)  \bigg \rbrace
\nonumber    \\ &&
+ \delta (y) 
\left\lbrace   - \frac{i}{2} \overline{\chi}_{0} \overline{\sigma}^{\mu}
\partial_{\mu} \chi_{0}
- M_{0} \left[ \psi_{\mu 1} \sigma^{\mu \nu} \psi_{\nu 1} 
+ \left( \chi_{0} + \psi_{5 2} \right) \left( \chi_{0} + \psi_{5 2} \right) 
\right]  \right\rbrace 
\nonumber    \\ &&
+ \delta (y - \pi R) 
\left\lbrace   - \frac{i}{2} \overline{\chi}_{\pi} \overline{\sigma}^{\mu}
\partial_{\mu} \chi_{\pi}
- M_{\pi} \left[  \psi_{\mu 1} \sigma^{\mu \nu} \psi_{\nu 1} 
+ \left( \chi_{\pi} + \psi_{5 2} \right) \left( \chi_{\pi} + \psi_{5 2} \right)
\right]  \right\rbrace 
\nonumber \\ &&
- i \frac{3}{8} \xi \left( g_{1}  \frac{\sigma^{\mu} \partial_{\mu}
}{\partial^{2}} \overline{g}_{1}
+ g_{2}  \frac{\sigma^{\mu} \partial_{\mu} }{\partial^{2}} \overline{g}_{2}
\right) 
+ h.c. 
\label{LkmGF}
\end{eqnarray}
As expected the position of the poles in the propagators of the fields $\psi_{M
I}$, $\chi_{0}$ 
and $\chi_{\pi}$ will depend on the gauge parameter $\xi$,  but of course the
gauge invariant operators and S-matrix elements 
should not depend on the parameter $\xi$.

\subsection{Unitary gauge}
\label{secUnitGauge}

The unitary gauge  can be recovered from the $R_{\xi}$ gauge 
in the limit $\xi \rightarrow \infty$. In this gauge, the gravitino 
propagators have poles at their physical mass and the unphysical degrees of
freedom 
(would-be Goldstinos) are eliminated, absorbed to provide the longitudinal
components for the gravitinos, through the 
super-Higgs  mechanism.

We first discuss the gravitino equations of motion in the bulk-branes system.
The equations 
of motion for the gravitinos $\psi_{\mu I}(y)$ in the unitary gauge 
can be extracted from the Lagrangian (\ref{LkmGF}) in the limit $\xi \rightarrow
\infty$:
\begin{eqnarray}
- \frac{1}{2} \epsilon^{\mu \nu \rho \lambda}
     \sigma_{\nu}\partial_{\rho} \overline{\psi}_{\lambda 1}
+ \sigma^{\mu \nu}\partial_{5} \psi_{\nu 2} 
- \frac{\omega}{R}\sigma^{\mu \nu} \psi_{\nu 1} 
&=&
    2 M_{0} \sigma^{\mu \nu} \psi_{\nu 1}  \delta(y)
+ 2 M_{\pi} \sigma^{\mu \nu} \psi_{\nu 1} \delta(y - \pi R) 
\nonumber    \\
- \frac{1}{2} \epsilon^{\mu \nu \rho \lambda}
     \sigma_{\nu}\partial_{\rho} \overline{\psi}_{\lambda 2}
- \sigma^{\mu \nu}\partial_{5} \psi_{\nu1} 
- \frac{\omega}{R}\sigma^{\mu \nu} \psi_{\nu 2}
    &=& 0
\label{PsiEOM}
\end{eqnarray}

Assuming the gravitinos have a \textit{four-dimensional mass} $m_{3/2}$:
\begin{equation}
\epsilon^{\mu \nu \rho \lambda}
\sigma_{\nu}\partial_{\rho} \overline{\psi}_{\lambda I} = 
- 2 m_{3/2} \sigma^{\mu \nu} \psi_{\nu I} 
\label{PsiMassEq}
\end{equation}
their equations of motion can take the form:
\begin{eqnarray}
\partial_{5} \psi_{\mu 2} 
+ \left( m_{3/2} - \frac{\omega}{R} \right) \psi_{\mu 1}
    &=&
    2 M_{0} \psi_{\mu 1} \delta(y)
+ 2 M_{\pi} \psi_{\mu 1} \delta(y - \pi R) 
\nonumber    \\
\partial_{5} \psi_{\mu1} 
- \left( m_{3/2} - \frac{\omega}{R} \right) \psi_{\mu 2}
    &=& 0
\label{PsiEOM2}
\end{eqnarray}
Integration of the equations (\ref{PsiEOM2}) near the points $y = 0$ and $y =
\pi R$, 
taking into account the parity assumptions, leads to the 
following expressions for the discontinuities of the odd gravitino fields:
\begin{eqnarray}
\psi_{\mu 2}(0^{+}) &=& M_{0} \, \, \psi_{\mu 1} (0) = -\psi_{\mu 2}(0^{-})
\nonumber   \\
\psi_{\mu 2}(\pi R^{-}) &=& - M_{\pi} \, \, \psi_{\mu 1} (\pi R) = - \psi_{\mu
2}(\pi R^{+}).
\label{PsiBC}
\end{eqnarray}

It is then straight forward to find a solution for the equations (\ref{PsiEOM2})
in the 
interval $0 < y < \pi R$ satisfying the first condition in (\ref{PsiBC}):
\begin{eqnarray}
\psi_{\mu 1}(y) &=& \left\lbrace \cos \left[ \left( m_{3/2} - \frac{\omega}{R}
\right) y\right]
+ M_{0} \sin \left[ \left( m_{3/2} - \frac{\omega}{R} \right) y \right] 
\right\rbrace \psi_{\mu 1} (0)
    \nonumber   \\
    \psi_{\mu 2}(y) &=& \left\{ M_{0} \cos \left[ \left( m_{3/2} -
\frac{\omega}{R}
\right) y\right] 
- \sin \left[ \left( m_{3/2} - \frac{\omega}{R} \right) y\right] \right\}
\psi_{\mu 1} (0) .
\label{PsiSolution}
\end{eqnarray}
The second condition in (\ref{PsiBC}) is then used to determine the gravitino
mass:
\begin{equation}
m_{3/2} = \frac{\omega}{R} + \frac{1}{\pi R} \left[ \arctan \left( M_{0}
\right) 
+ \arctan \left( M_{\pi} \right) \right] + \frac{n}{R} , 
\quad n \in \mathbb{Z}
\label{PsiMass}
\end{equation}

In remaining of the of this section we will concentrate on the would-be
Goldstino fields $\psi_{5 1}(y)$, 
$\psi_{5 2}(y)$, $\chi_{0}$ and $\chi_{\pi}$.  Note that the Lagrangian density 
(\ref{LkmGF}) shows that, in the unitary gauge $\xi \rightarrow \infty$, 
a stationary action (in order to derive of the equations of motion) is possible 
if $g_{1} = g_{2} = 0$, i.e.:
\begin{eqnarray}
\partial_{5} \psi_{5 1} + \frac{\omega}{R} \psi_{5 2}
&=& - 2 \delta (y) M_{0} \left( \chi_{0} + \psi_{5 2} \right) 
- 2 \delta (y - \pi R) M_{\pi} \left( \chi_{\pi} + \psi_{5 2} \right)
\nonumber \\
\partial_{5} \psi_{5 2} - \frac{\omega}{R} \psi_{5 1} &=& 0 .
\label{UnitGaugeCond} 
\end{eqnarray}
which imply that the fields $\psi_{5 I}(y)$, in the interval 
$0 < y < \pi R$ can be written as:
\begin{eqnarray}
\psi_{5 1}(y) &=& \frac{1}{\sqrt{\pi R}} \left[ \cos \left( \frac{\omega}{R} y +
\theta\right)  \chi_{1} +
    \sin \left( \frac{\omega}{R} y + \theta \right) \chi_{2} \right] 
\nonumber \\
\psi_{5 2}(y) &=& \frac{1}{\sqrt{\pi R}} \left[ \sin \left( \frac{\omega}{R} y +
\theta\right)  \chi_{1} - 
\cos \left( \frac{\omega}{R} y + \theta \right) \chi_{2} \right] 
\label{Psi5UG}
\end{eqnarray}
where $\chi_{1}$ and $\chi_{2}$ are $y$ independent 4d spinors and $\theta$ is a
constant which corresponds to 
a choice of basis for $\chi_{1}$ and $\chi_{2}$.

Integrating equations (\ref{UnitGaugeCond}) near $y = 0$ and $y = \pi$ we deduce
that:
\begin{eqnarray}
\psi_{5 1}(0^{+}) + M_{0} \left[ \chi_{0} + \psi_{5 2}(0) \right] &=& 0
\nonumber \\
\psi_{5 1}(\pi R^{-}) - M_{\pi} \left[ \chi_{\pi} + \psi_{5 2}(\pi R) \right]
&=& 0
\label{ChiUG}
\end{eqnarray}
which implies (for $M_\pi \neq 0$ and $M_0 \neq 0$):
\begin{eqnarray}
\chi_{\pi} &=& \frac{1}{\sqrt{\pi R}} \left[ - \sin (\omega \pi + \theta) + 
\frac{1}{M_{\pi}} \cos (\omega \pi + \theta) \right] \chi_{1} 
+ \frac{1}{\sqrt{\pi R}} \left[ \cos (\omega \pi + \theta) + 
\frac{1}{M_{\pi}} \sin (\omega \pi + \theta) \right] \chi_{2}
\nonumber \\
\chi_{0} &=& - \frac{1}{\sqrt{\pi R}} \left[ \sin (\theta) + 
\frac{1}{M_{0}} \cos (\theta) \right] \chi_{1} 
+ \frac{1}{\sqrt{\pi R}} \left[ \cos (\theta) - 
\frac{1}{M_{0}} \sin (\theta) \right] \chi_{2} .
\label{ChiUG2}
\end{eqnarray}

Here we see how the super Higgs mechanism operate, from the original two 5d and
two 4d degrees of 
freedom ($\psi_{5 1}(y)$, $\psi_{5 2}(y)$, $\chi_{0}$ and $\chi_{\pi}$), an
infinity of 
Kaluza-Klein modes is absorbed to give mass to the fields $\psi_{\mu 1}(y)$ and
$\psi_{\mu 2}(y)$ and 
only two degrees of freedom remain in the unitary gauge: the pseudo Goldstinos
$\chi_{1}$ and $\chi_{2}$.

\subsection{Comment on F-terms versus generalized Scherk-Schwarz mechanism}
\label{MbVersusTheta}

The equation (\ref{PsiMass}) raises questions about the possibility to express
 spontaneous breaking with $F$-terms ( and all the gravitinos and
pseudo-Goldstinos
masses generated)  as a
generalized Scherk-Schwarz twist, in  parallel to the case of $\Omega_b$ in
(\ref{SusyRestorationCondition}). This
is not possible, as can be seen by the following arguments.

In order to have an equivalence between the brane mass terms and a generalized
Scherk-Schwarz twist one should be able to expresses the discontinuity 
of the fields $\psi_{5 I}$ at $y = 0$ and $y = \pi R$ as an $SU(2)_{\cal R}$
rotation 
like in (\ref{PsiRotation}). This means that in order to be associated with a 
generalized Scherk-Schwarz twist the effects of the brane mass terms must 
be described by a generalized twist. So one should be able to find a rotation
such that:
\begin{equation}
\left(\begin{matrix}
\psi_{5 1} (0^{+}) \\
\psi_{5 2} (0^{+})
\end{matrix}\right) 
=
\left(\begin{matrix}
\cos(\alpha) & - \sin(\alpha) \\
\sin(\alpha) & \cos(\alpha)
\end{matrix}\right) 
\left(\begin{matrix}
\psi_{5 1} (0^{-}) \\
\psi_{5 2} (0^{-})
\end{matrix}\right). 
\label{BoundaryTwist}
\end{equation}
But equations (\ref{Psi5UG}) imply:
\begin{eqnarray}
&
\psi_{5 1}(0^{+}) = \frac{1}{\sqrt{\pi R}} \left[ \cos \left(\theta\right) 
\chi_{1} +
    \sin \left(\theta \right) \chi_{2} \right] = - \psi_{5 1}(0^{-}) 
\nonumber \\ 
&
\psi_{5 2}(0^{+}) = \frac{1}{\sqrt{\pi R}} \left[ \sin \left(\theta\right) 
\chi_{1} - 
\cos \left(\theta \right) \chi_{2} \right] = \psi_{5 2}(0^{-}) .
\label{Psi5InTheBoundary}
\end{eqnarray}
Note then that matching equations (\ref{BoundaryTwist}) and
(\ref{Psi5InTheBoundary}) 
for the coefficients of $\chi_{1}$ one finds $\alpha = 2 \theta + \pi$, while if
one matches the 
coefficients for $\chi_{2}$ in (\ref{BoundaryTwist}) 
and (\ref{Psi5InTheBoundary}) one finds $\alpha = 2 \theta$. This
incompatibility shows that $M_b$ can not be casted as $\tan(\Omega_b)$, as in
(\ref{LkineticEqiv}).

\section{The pseudo-Goldstinos spectrum}
\label{secPseudoGold}

In the previous section we have shown how some would be Goldstinos are absorbed
leading to
massive gravitinos. Here, we will discuss the spectrum of the remaining
pseudo-Goldstinos.
More precisely, we will analyze some limits or approximations which allow to
display 
compact formulae. The general case is treated in Appendix
\ref{apePGMassEigenstate}.

We will restore the explicit dependence on the (reduced) five-dimensional
Planck 
mass $ M_{5} = \kappa^{-1} $. It is related to the four-dimensional Planck 
mass $M_{4}$ by\footnote{ We recall that in our conventions the four-dimensional
Planck mass
$M_{4}$ is 
related to Newton's constant $G$ by $\sqrt{8 \pi G} = M_{4}^{-1}$.}
\begin{equation}
\pi R M_{5}^{3} = M_{4}^{2} .
\label{PlanckMasses}
\end{equation}

The lightest four-dimensional gravitino mass can be read from (\ref{PsiMass}):
\begin{eqnarray}
m_{3/2} &=& \frac{\omega}{R} + \frac{1}{\pi R} \left[ \arctan \left( \kappa
M_{0}
\right) 
+ \arctan \left(\kappa M_{\pi} \right) \right], 
\end{eqnarray}
where $M_{b} =
\left\langle e^{ \kappa^{2} {\cal G}_{b}/2} \right\rangle \kappa^{-1}
$, with $ b\in \{ 0, \pi \}$ arise from boundary $F$-terms. In the
four-dimensional limit $ \kappa  M_{b}<< 1$ the approximate  gravitino mass is:
\begin{eqnarray}
m_{3/2} &\simeq& \frac{\omega}{R} + \frac{\kappa}{\pi R} \left(  M_{0} + 
M_{\pi}
\right). 
\label{PsiMassApprox}
\end{eqnarray}

To identify the pseudo-Goldstinos mass eigenstates we shall plug 
(\ref{Psi5UG}) 
and (\ref{ChiUG2}) in the Lagrangian (\ref{LkmGF}), integrate over the $y$ 
dimension, diagonalize and canonically normalize the kinetic terms of the 
fields $\chi_{1}$ and $\chi_{2}$ and finally diagonalize their 
mass matrix. This is a tedious task, the resulting mass
eigenstates are given in the appendix \ref{apePGMassEigenstate}. Let us discuss
in
more details some
particular cases.

\subsection{Supersymmetry breaking on a single brane}
\label{subsec1Brane}

Consider the  case where the supersymmetry breaking is realized by a
combination of a Scherk-Schwarz twist $\omega$ and 
a single $F$-term, say on the brane placed at $y = 0$. This corresponds in
our generic formulae to $M_{\pi} = 0$ and $\chi_{\pi} = 0$.
Choosing a basis for 
$\chi_{1}$ and $\chi_{2}$ corresponding to $\theta = - \omega \pi$ 
equations (\ref{Psi5UG}) and (\ref{ChiUG}) imply $\chi_{1} = 0$. So, as
expected,
    \textit{only one degree of freedom} $\chi_{2}$ remains in the unitary gauge.

Substitution of (\ref{Psi5UG}) and (\ref{ChiUG2}) in (\ref{LkmGF}), integration
over $y$ and redefinition of the fields to canonically normalize their kinetic
terms
allows identifying the eigenstate, we denote as $\psi_{1}$, with mass:
\begin{equation}
m_{1} = \frac{2 M_{0} \sin(\omega \pi) \left[ \kappa M_{0} \cos(\omega \pi) +
\sin(\omega \pi) \right]}
{\kappa \pi R M_{0}^{2} + \left[ \kappa M_{0} \cos(\omega \pi) + \sin(\omega
\pi) \right]^{2} } .
\label{Mass1}
\end{equation}
The original would-be-Goldstinos are written in terms of the pseudo-Goldstino as
given by (\ref{Psi5UG}) and (\ref{ChiUG2}) in the unitary gauge, which in the
present
case reads:
\begin{eqnarray}
\psi_{5 1}(y) &=&  \frac{ \kappa M_{0} }
{\sqrt{ \kappa \pi R M_{0}^{2} + \left[ \kappa M_{0} \cos(\omega \pi) +
\sin(\omega \pi) \right]^{2} } } 
\sin \left[ \omega \left( \frac{y}{R} - \pi \right)  \right] \psi_{1} 
\nonumber \\
\psi_{5 2}(y) &=& - \frac{ \kappa M_{0} }
{\sqrt{ \kappa \pi R M_{0}^{2} + \left[ \kappa M_{0} \cos(\omega \pi) +
\sin(\omega \pi) \right]^{2} } } 
\cos \left[ \omega \left( \frac{y}{R} - \pi \right)  \right] \psi_{1} 
\nonumber \\
\chi_{0} &=& \frac{ \kappa M_{0} \cos(\omega \pi) + \sin(\omega \pi) }
{\sqrt{ \kappa \pi R M_{0}^{2} + \left[ \kappa M_{0} \cos(\omega \pi) +
\sin(\omega \pi) \right]^{2} } }  \psi_{1} 
\label{Fields1}
\end{eqnarray}

Let us discuss some particular cases which might bring to the reader some more
intuition on what is happening:
\begin{itemize}

\item Case $\omega\rightarrow 0$:

We first discuss the case of vanishing twist. The equations
(\ref{Fields1}) become: 
\begin{eqnarray}
    m_{1}&\simeq& \frac{2 \omega \pi}{\pi R + \kappa} \longrightarrow 0
\nonumber \\
\psi_{5 1}(y) &\simeq& 0 
\nonumber \\
\psi_{5 2}(y) &\simeq& -\frac {1}{\sqrt{\kappa^{-1} \pi R + 1}} \psi_{1}
\nonumber \\
\chi_{0} &\simeq& \frac {1}{\sqrt{\kappa^{-1} \pi R + 1}} \psi_{1}.
\label{1bw0}
\end{eqnarray}
There are two ways to understand these
results. First, from ``a global view'', for $\omega=0$ the
$\mathbb{Z}_2$ projected out the odd zero mode of $\psi_{51}$ ,  $\psi_{51}$
being continuous this implies $\psi_{51}= 0$. The other way is to consider ``a
local
five-dimensional description'' where the gravitino $\psi_{\mu2}$ eats the
fermion with
the same $\mathbb{Z}_2$ parity, i.e. $\psi_{51}$. The remaining gravitino
$\psi_{\mu
1}$ absorbs the linear combination $\psi_{52}(0)+\chi_0$ and reminds the
orthogonal
combination $\psi_{52}(0)-\chi_0 \sim \psi_1$ as a pseudo-Goldstino.
The only source of mass for this state is the bulk mass term.

\item Case $R \kappa^{-1} \rightarrow \infty$:

This limit gives 
\begin{eqnarray}
    m_{1}&\simeq& \frac{2  \sin(\omega \pi) \left[ \kappa M_{0} \cos(\omega \pi)
+
\sin(\omega \pi) \right]}
{\kappa \pi R M_{0} }
\nonumber \\
\psi_{5 1}(y) &\simeq&  \frac {1}{\sqrt{\kappa^{-1} \pi R}}\sin \left[ \omega
\left(
\frac{y}{R} - \pi \right)  \right] \psi_{1}
\nonumber \\
\psi_{5 2}(y) &\simeq& -\frac {1}{\sqrt{\kappa^{-1} \pi R }} \cos \left[ \omega
\left( \frac{y}{R} - \pi \right)  \right]\psi_{1}
\nonumber \\
\chi_0 &\simeq& \frac{ \kappa M_{0} \cos(\omega \pi) + \sin(\omega \pi) }
{\sqrt{ \kappa \pi R} M_{0}   }  \psi_{1}
\label{1bRfty}
\end{eqnarray}
which agrees with the fact that the absorbed Goldstino on the brane $y=0$ is
given
by $(\frac{\psi_{51}}{\kappa M_0} + \psi_{52} +\chi_0)(0^+)$ and the one eaten
at
$y=\pi$ is $\psi_{51}(\pi R^-)$.

\item Case $\omega=\frac{1}{2}$:

\begin{eqnarray}
    m_{1}&\simeq& \frac{2   M_{0} }
{\kappa \pi R M_{0}^2 +1 }
\nonumber \\
\psi_{5 1}(y) &\simeq&  - \frac {\kappa M_0}{\sqrt{\kappa \pi R M_{0}^2 +1
}}\cos
\left( \frac{y}{2R}  \right)  \psi_{1}
\nonumber \\
\psi_{5 2}(y) &\simeq& - \frac {\kappa M_0}{\sqrt{\kappa \pi R M_{0}^2 +1 }}\sin
\left( \frac{y}{2R}  \right)  \psi_{1}
\nonumber \\
\chi_0 &\simeq&  \frac {1}{\sqrt{\kappa \pi R M_{0}^2 +1 }}\psi_{1}
\label{1bRw12}
\end{eqnarray}
in which case one notes that $\psi_{52}$ decouples from the brane at $y=0$ while
the
absorbed state is $\psi_{51}(0^+)+ \kappa M_0\chi_0$.
\end{itemize}

\subsection{Hierarchical supersymmetry breaking on the boundaries}
\label{subsesBigBrane}

In this section we switch on a large  supersymmetry breaking $F$-term in on
brane 
at $M_\pi$ i.e.  $M_{\pi} > > M_{0}$. Our results assume explicitly that $\omega
\ne
0$.
They are not generically valid for  $\omega = 0$ which will be presented in
section
\ref{subsecNoSSTwixt}.
We will exhibit the first orders in an expansion in $\kappa M_{0}$ for the
pseudo-Goldstinos mass matrix  eigenvalues and eigenvectors. At the leading
order,
this is 
diagonal in the basis for $\chi_{1}$ and $\chi_{2}$ corresponding to $\theta =
0$.  The mass eigenstates are denoted as $\psi_{1}$ and  $\psi_{2}$, and have
masses
given, respectively, by:
\begin{eqnarray}
m_{1} &=& \frac{2 \sin(\omega \pi) \left[ \kappa M_{\pi} \cos(\omega \pi)
+ \sin(\omega \pi) \right] }
{ \sin(\omega \pi) \left[ \kappa M_{\pi} \cos(\omega \pi) + \sin(\omega
\pi) \right] +
\left[ \pi R + 2 \kappa - 3 \kappa \sin(\omega \pi)^{2} \right] \kappa
M_{0} M_{\pi} } M_{0}
+ O \left( \kappa^{2} M_{0}^{2} \right)
\nonumber \\
m_{2} &=& \frac{2 M_{\pi} \sin(\omega \pi) \left[ \kappa M_{\pi} \cos(\omega
\pi) + \sin(\omega \pi) \right]}
{ \kappa \pi R M_{\pi}^{2} + \left[ \kappa M_{\pi} \cos(\omega \pi) +
\sin(\omega \pi) \right]^{2} } 
+ O \left( \kappa M_{0} \right).
\label{Mass4}
\end{eqnarray}
As in the previous section, the fields $\psi_{5 1}(y)$, $\psi_{5 2}(y)$,
$\chi_{0}$
and $\chi_{\pi}$ are 
written in terms of the pseudo-Goldstinos as:
\begin{eqnarray}
\psi_{5 1}(y) &=&  \frac{ \kappa M_{\pi} }
{\sqrt{ \kappa \pi R M_{\pi}^{2} + \left[ \kappa M_{\pi} \cos(\omega \pi) +
\sin(\omega \pi) \right]^{2} } } 
\sin \left( \frac{\omega}{R} y \right) \psi_{2} + O \left( \kappa M_{0} \right)
\nonumber \\
\psi_{5 2}(y) &=& - \frac{ \kappa M_{\pi} }
{\sqrt{ \kappa \pi R M_{\pi}^{2} + \left[ \kappa M_{\pi} \cos(\omega \pi) +
\sin(\omega \pi) \right]^{2} } } 
\cos \left( \frac{\omega}{R} y \right) \psi_{2} + O \left( \kappa M_{0} \right)
\nonumber \\
\chi_{0} &=& \psi_{1} + O \left( \kappa M_{0} \right)
\nonumber \\
\chi_{\pi} &=& \frac{ \kappa M_{\pi} \cos(\omega \pi) + \sin(\omega \pi) }
{\sqrt{ \kappa \pi R M_{\pi}^{2} + \left[ \kappa M_{\pi} \cos(\omega \pi) +
\sin(\omega \pi) \right]^{2} } }
    \psi_{2} + O \left( \kappa M_{0} \right)
\label{Fields4}
\end{eqnarray}

Note that the results obtained in section \ref{subsec1Brane} can be 
derived from these formulas by taking $M_{0} = 0$ and interchanging 
the branes $0$ and $\pi$.

\subsection{The 5D or large extra dimension radius limit}
\label{subsecInfiniteRadius}

In this section we consider a very 
large extra dimensional radius, $R >> \kappa$, $R M_{0} >> 1$ and  $R M_{\pi} >>
1$,
such that the set-up is truly five-dimensional. We will compute the
pseudo-Goldstinos mass
eigenvalues and eigenvectors to leading order in a perturbation series in
$\kappa/R$.

At leading order, the mass matrix is 
diagonal in a basis for $\chi_{1}$ and $\chi_{2}$ corresponding to 
an angle $\theta$ given by:
\begin{equation} 
\tan ( 2 \theta ) = \frac{ \kappa M_{0} M_{\pi} [ 1 - \cos(2 \omega \pi) ] -
M_{0} \sin(2 \omega \pi) }
{M_{\pi} + M_{0} \cos(2 \omega \pi) - \kappa M_{0} M_{\pi} \sin(2 \omega \pi) }
\label{Angle}
\end{equation}
We choose $\theta$ in the range $- \pi/4 < \theta < \pi/4$. The 
the mass eigenstates $\psi_{1}$ and  $\psi_{2}$ have masses: 
\begin{eqnarray} 
m_{1} &=& \frac{1}{ \kappa \pi R} \left[ \frac{1}{M_{0}} + \frac{1}{M_{\pi}} +
\sqrt{\Delta} \right] 
+ O\left( \frac{\kappa^{3/2}}{R^{3/2}} \right) 
\nonumber \\
m_{2} &=& \frac{1}{ \kappa \pi R} \left[ \frac{1}{M_{0}} + \frac{1}{M_{\pi}} -
\sqrt{\Delta} \right] 
+ O\left( \frac{\kappa^{3/2}}{R^{3/2}} \right)
\label{Mass6}
\end{eqnarray}
respectively, where
\begin{equation} 
\sqrt{\Delta} = \sqrt{\frac{1}{1 + [\tan ( 2 \theta ) ]^{2} }} \left\lbrace 
\frac{1}{M_{0}} + \frac{ \cos(2 \omega \pi) }{ M_{\pi} } - \kappa \sin(2 \omega
\pi) 
+ \tan ( 2 \theta ) \left[ \kappa - \kappa \cos(2 \omega \pi) - \frac{ \sin(2
\omega \pi) }{ M_{\pi} }
\right] \right\rbrace 
\label{Delta}
\end{equation}
The original would-be-Goldstinos $\psi_{5 1}(y)$, $\psi_{5 2}(y)$, $\chi_{0}$
and $\chi_{\pi}$ are 
written in terms of the pseudo-Goldstinos $\psi_{1}$ and $\psi_{2}$ as in
(\ref{Psi5UG}) and (\ref{ChiUG2}), which read now:
\begin{eqnarray}
\psi_{5 1}(y) &=& \sqrt{\frac{\kappa}{\pi R}} \left[ \cos \left(
\frac{\omega}{R} y + \theta\right)  \psi_{1} +
    \sin \left( \frac{\omega}{R} y + \theta \right) \psi_{2} \right] + O\left(
\frac{\kappa}{R} \right)
\nonumber \\
\psi_{5 2}(y) &=& \sqrt{\frac{\kappa}{\pi R}} \left[ \sin \left(
\frac{\omega}{R} y + \theta\right)  \psi_{1} - 
\cos \left( \frac{\omega}{R} y + \theta \right) \psi_{2} \right] + O\left(
\frac{\kappa}{R} \right)
\nonumber \\
\chi_{0} &=& - \sqrt{\frac{\kappa}{\pi R}} \left[ \sin (\theta) + 
\frac{1}{ \kappa M_{0}} \cos (\theta) \right] \psi_{1} 
+ \sqrt{\frac{\kappa}{\pi R}} \left[ \cos (\theta) - 
\frac{1}{ \kappa M_{0}} \sin (\theta) \right] \psi_{2}  + O\left(
\frac{\kappa}{R} \right)
\nonumber \\
\chi_{\pi} &=& \sqrt{\frac{\kappa}{\pi R}} \left[ - \sin (\omega \pi + \theta)
+ 
\frac{1}{ \kappa M_{\pi}} \cos (\omega \pi + \theta) \right] \psi_{1} 
\nonumber \\ &&
+ \sqrt{\frac{\kappa}{\pi R}} \left[ \cos (\omega \pi + \theta) + 
\frac{1}{ \kappa M_{\pi}} \sin (\omega \pi + \theta) \right] \psi_{2} 
+ O\left( \frac{\kappa}{R} \right)
\label{Fields6}
\end{eqnarray}

\subsection{The 4D or small extra dimension radius limit}
\label{subsecSmallRadius}

In this section, we discuss the four-dimensional limit corresponding to the case
of 
a very small extra dimensional radius, $ R M_0<< 1 $ and $R M_\pi << 1$.
At the leading order, the mass eigenstates $\psi_{1}$ and  $\psi_{2}$ have
masses
given by: 
\begin{eqnarray} 
m_{1} &=& \frac{ \left( M_{0} + M_{\pi} \right) \sin (\omega \pi) 
+ 2 \kappa M_{0} M_{\pi} \cos (\omega \pi) + \sqrt{\Delta} }
{ \kappa \left( M_{0} + M_{\pi} \right) \cos (\omega \pi) - 
\left( \kappa^{2} M_{0} M_{\pi} - 1 \right) \sin (\omega \pi) }
\nonumber \\
m_{2} &=& \frac{ \left( M_{0} + M_{\pi} \right) \sin (\omega \pi) 
+ 2 \kappa M_{0} M_{\pi} \cos (\omega \pi) - \sqrt{\Delta} }
{ \kappa \left( M_{0} + M_{\pi} \right) \cos (\omega \pi) - 
\left( \kappa^{2} M_{0} M_{\pi} - 1 \right) \sin (\omega \pi) }
\label{Mass7}
\end{eqnarray}
respectively, where now $\Delta$ stands for
\begin{equation} 
\Delta = \left( M_{0} - M_{\pi} \right)^{2} \sin (\omega \pi)^{2} + 4 \left(
\kappa M_{0} M_{\pi} \right)^{2} 
\label{Delta2}
\end{equation}
Again, at the leading order, the four initial would-be-Goldstinos are expressed
in
terms of the 
pseudo-Goldstinos $\psi_{1}$ and $\psi_{2}$ as:
\begin{eqnarray}
\psi_{5 1}(y) &=&  \frac{M_{0} \left[  \sin \left( \frac{\omega}{R} y - \omega
\pi \right) 
- \kappa M_{\pi} \cos \left( \frac{\omega}{R} y - \omega \pi \right) \right]
\chi_{0}
+ M_{\pi} \left[  \sin \left( \frac{\omega}{R} y \right) 
+ \kappa M_{0} \cos \left( \frac{\omega}{R} y \right) \right] \chi_{\pi}}
{ \left( M_{0} + M_{\pi} \right) \cos (\omega \pi) - \left( \kappa M_{0} M_{\pi}
- \kappa^{-1} \right) \sin (\omega \pi) } 
\nonumber \\
\psi_{5 2}(y) &=& \frac{ - M_{0} \left[  \cos \left( \frac{\omega}{R} y - \omega
\pi \right) 
+ \kappa M_{\pi} \sin \left( \frac{\omega}{R} y - \omega \pi \right) \right]
\chi_{0}
- M_{\pi} \left[  \cos \left( \frac{\omega}{R} y \right) 
- \kappa M_{0} \sin \left( \frac{\omega}{R} y \right) \right] \chi_{\pi}}
{ \left( M_{0} + M_{\pi} \right) \cos (\omega \pi) - \left( \kappa M_{0} M_{\pi}
- \kappa^{-1} \right) \sin (\omega \pi) } 
\nonumber \\
\chi_{0} &=& \frac{\left[ \left( M_{0} - M_{\pi} \right) \sin (\omega \pi) 
+ \sqrt{\Delta} \right] \psi_{1} 
+ 2 \kappa M_{0} M_{\pi}  \psi_{2} }
{\sqrt{ 2 \left[ \Delta + \left( M_{0} - M_{\pi} \right) \sin (\omega \pi)
\sqrt{\Delta} \right] }} 
\nonumber \\
\chi_{\pi} &=& \frac{ - 2 \kappa M_{0} M_{\pi}  \psi_{1} 
+ \left[ \left( M_{0} - M_{\pi} \right) \sin (\omega \pi) 
+ \sqrt{\Delta} \right] \psi_{2} }
{\sqrt{ 2 \left[ \Delta + \left( M_{0} - M_{\pi} \right) \sin (\omega \pi)
\sqrt{\Delta} \right] }} 
\label{Fields7}
\end{eqnarray}

\subsection{No Scherk-Schwarz twist}
\label{subsecNoSSTwixt}

Another simple case corresponds to having the localized $F$-terms in the branes
as the 
only source of supersymmetry breaking, i.e. to consider a vanishing 
Scherk-Schwarz twist, $\omega = 0$.

The mass eigenstates $\psi_{1}$ and  $\psi_{2}$ have masses respectively given
by: 
\begin{eqnarray} 
m_{1} &=& 0
\nonumber \\
m_{2} &=& \frac{2 M_{0} M_{\pi} \left( M_{0} + M_{\pi} \right) (\pi R + 2
\kappa) }
{ \kappa \left( \pi R M_{0} M_{\pi} \right)^{2} + \pi R \left( 2 \kappa^{2}
M_{0}^{2} M_{\pi}^{2} + 
M_{0}^{2} + M_{\pi}^{2} \right) + \kappa \left( M_{0} + M_{\pi} \right)^{2} } .
\label{Mass8}
\end{eqnarray}
The would-be-Goldstinos $\psi_{5 1}(y)$, $\psi_{5 2}(y)$,
$\chi_{0}$ and $\chi_{\pi}$ are 
related to the pseudo-Goldstinos $\psi_{1}$ and $\psi_{2}$ through:
\begin{eqnarray}
\psi_{5 1}(y) &=& - \frac{ \kappa \sqrt{2 \kappa + \pi R} M_{0} M_{\pi}
    }
{\sqrt{ \lambda} }\psi_{2}
\nonumber \\
\psi_{5 2}(y) &=&
\frac{1}{ \sqrt{2 + \kappa^{-1} \pi R  } } \left[  - \psi_{1}
+ \frac{ \sqrt{\kappa} \left( M_{\pi} - M_{0}  \right)  }
{\sqrt{ \lambda} } \psi_{2}\right]
\nonumber \\
\chi_{0} &=&
\frac{1}{ \sqrt{2 + \kappa^{-1} \pi R  } }  \left[  \psi_{1}
+ \frac{ \sqrt{\kappa} \left( M_{0} + M_{\pi} + \kappa^{-1} \pi R M_{\pi}
\right)  }
{\sqrt{ \lambda} } \psi_{2}\right]
\nonumber \\
\chi_{\pi} &=&
\frac{1}{ \sqrt{2 + \kappa^{-1} \pi R  } }  \left[  \psi_{1}
- \frac{ \sqrt{\kappa} \left( M_{0} + M_{\pi} + \kappa^{-1} \pi R M_{0}
\right)  }
{\sqrt{ \lambda} } \psi_{2} \right]
\label{Fields8}
\end{eqnarray}
where
\begin{equation}
\lambda = \kappa \left( \pi R M_{0} M_{\pi} \right)^{2} + \pi R \left( 2
\kappa^{2} M_{0}^{2} M_{\pi}^{2} +
M_{0}^{2} + M_{\pi}^{2} \right) + \kappa \left( M_{0} + M_{\pi} \right)^{2}
\label{Delta3}
\end{equation}
Note that $\psi_{5 1}(y)$ is proportional to $M_{0} M_{\pi}$. This is expected
as
$\psi_{5 1}(y)$ is odd 
at both boundaries and for $\omega=0$ would vanish if there were not both
discontinuities at 
$y=0$ and $y=\pi R$ due to $M_{0}$ and $M_\pi$ respectively.  Note that one of
the
pseudo-goldstinos is massless. This can be understood from the following
arguments.
Generically, the pseudo-Goldstinos get masses from boundaries and bulk. 
The brane masses are for the combination $\chi_0\, +\, \psi_{52}(0)$ at $y=0$
and
$\chi_\pi\, +\, \psi_{52}(\pi R)$ at $y=\pi R$, as seen from equation 
(\ref{LkmGF}). In this case of $\omega=0$, both these combinations are
proportional
to $\psi_{5 1}(0^+) = \psi_{5 1}(\pi R^-) \sim  \psi_2$ as seen from the unitary
gauge condition (\ref{ChiUG}). The orthogonal combination, $\psi_1$, would have
received a mass from the bulk, but this vanishes now as $\omega=0$.

Let us discuss some particular limits that connect this case to the previous
ones:

\begin{itemize}
\item Case $M_\pi >> M_0$:

In subsection \ref{subsesBigBrane} we provided results for $M_{\pi} > > M_{0}$
assuming $\omega\neq 0$ and warned the reader that they are not always valid
when
$\omega = 0$. In fact, in this case the masses and the 
respective eigenstates are given instead by
\footnote{Here it is also assumed $R M_{0} << 1$.}
:
\begin{eqnarray}
m_{1} & = & 0
\nonumber \\
m_{2} & \simeq & \frac{2 (\pi R + 2\kappa) }{\pi R + \kappa}  M_{0}
\nonumber \\
\nonumber \\
\psi_{5 1}(y) &\simeq& -\frac { M_{0} \kappa \sqrt{ \kappa^{-1} \pi R + 2 }
}{\sqrt{\kappa^{-1} \pi R + 1}} \psi_{2}
\nonumber \\
\psi_{5 2}(y) &\simeq&
\frac{1}{ \sqrt{2 + \kappa^{-1} \pi R  } } \left[  - \psi_{1}
+ \frac{ 1 } {\sqrt{1 + \kappa^{-1} \pi R } } \psi_{2} \right]
\nonumber \\
\chi_{0} &\simeq&
\frac{1}{ \sqrt{2 + \kappa^{-1} \pi R  } } \left[ \psi_{1}
+ \sqrt{1 + \kappa^{-1} \pi R } \psi_{2} \right]
\nonumber \\
\chi_{\pi} &\simeq&
\frac{1}{ \sqrt{2 + \kappa^{-1} \pi R  } } \left[ \psi_{1}
- \frac{ 1 } {\sqrt{1 + \kappa^{-1} \pi R } } \psi_{2} \right]
\label{5bMpfty}
\end{eqnarray}

Note that if we take in these expression the large radius limit i.e. with
$\omega=0$,  $M_\pi >> M_0$ and $R \kappa^{-1} >>1$, the result is:
\begin{eqnarray}
m_{1} & = & 0
\nonumber \\
m_{2} & \simeq & 2  \, \, M_{0}
\nonumber \\
\nonumber \\
\psi_{5 1}(y) &\simeq& - M_{0} \, \, \kappa  \, \, \psi_{2}
\nonumber \\
\psi_{5 2}(y) &\simeq&
\frac{1}{ \sqrt{ \kappa^{-1} \pi R  } } \left[  - \psi_{1}
+ \frac{ 1 } {\sqrt{\kappa^{-1} \pi R } } \psi_{2} \right] \, \, \sim \, \, -
\frac{1}{ \sqrt{ \kappa^{-1} \pi R }}  \, \, \psi_{1}
\nonumber \\
\chi_{0} &\simeq&
\frac{1}{ \sqrt{\kappa^{-1} \pi R  } } \left[ \psi_{1}
+ \sqrt{\kappa^{-1} \pi R } \psi_{2} \right] \, \, \sim \, \, \psi_{2}
\nonumber \\
\chi_{\pi} &\simeq&
\frac{1}{ \sqrt{ \kappa^{-1} \pi R  } } \left[ \psi_{1}
- \frac{ 1 } {\sqrt{\kappa^{-1} \pi R } } \psi_{2} \right]\, \, \sim \, \,
\frac{1}{
\sqrt{ \kappa^{-1} \pi R  } } \, \, \psi_{1}
\label{5bMpfty2}
\end{eqnarray}

\item Case $R \kappa^{-1} \rightarrow 0$:

Another simple limit is obtained by combining both $\omega=0$ and  $R
\kappa^{-1}
\rightarrow 0$, 
in which case (\ref{Mass8}) and (\ref{Fields8}) lead to: 
\begin{eqnarray}
    m_{1} & = & 0
\nonumber \\
    m_{2} & \simeq & \frac{ 4 M_{0} M_{\pi} }{ M_{0} + M_{\pi} }
\nonumber \\
\nonumber \\
\psi_{5 1}(y) &\simeq& - \frac{\sqrt{2} \kappa M_{0} M_{\pi}}{M_{0} + M_{\pi}} 
\psi_{2}
\nonumber \\
\psi_{5 2}(y) &\simeq& - \frac{1}{ \sqrt{2} } \psi_{1}
+  \frac{ M_{\pi} - M_{0} }{ \sqrt{2} \left( M_{0} + M_{\pi} \right) } \psi_{2}
\nonumber \\
\chi_0 &\simeq& \frac{1}{ \sqrt{2} } \psi_{1} + \frac{1}{ \sqrt{2} } \psi_{2} 
\nonumber \\
\chi_{\pi} &\simeq& \frac{1}{ \sqrt{2} } \psi_{1} - \frac{1}{ \sqrt{2} }
\psi_{2}
\label{5bR02}
\end{eqnarray}
It is easy to check the agreement of (\ref{5bR02}) with the results presented in
section 
\ref{subsecSmallRadius} in the limit $\omega = 0$. If we add $M_\pi >> M_0$,
they
become:

\begin{eqnarray}
    m_{1} & = & 0
\nonumber \\
    m_{2} & \simeq & 4 M_{0}
\nonumber \\
\nonumber \\
\psi_{5 1}(y) &\simeq& - \sqrt{2} \kappa M_{0}  \psi_{2}
\nonumber \\
\psi_{5 2}(y) &\simeq& \frac{1}{ \sqrt{2} } \left( - \psi_{1}+  \psi_{2} \right)
\nonumber \\
\chi_0 &\simeq& \frac{1}{ \sqrt{2} } \left( \psi_{1} +  \psi_{2} \right)
\nonumber \\
\chi_{\pi} &\simeq& \frac{1}{ \sqrt{2} } \left( \psi_{1} -  \psi_{2}\right) .
\label{5bR023}
\end{eqnarray}

\end{itemize}

\section*{Acknowledgments}

Work supported in part by the French ANR contracts 
BLAN05-0079-01 and 
PHYS@COL\&COS, 
and in part by the EU contract 
MRTN-CT-2004-005104.

\appendix
\section{Conventions}
\label{apeConventions}

We use lower case letters from the middle of the Greek alphabet 
($\mu, \nu, \rho, \lambda$)
for the four-dimensional Minkowski indices  $(\mu=0, \cdots, 3)$ 
and lower case letters from the beginning of the Latin alphabet for the 
four-dimensional Lorentz indices $(a=\hat{0},\cdots,\hat{3})$. 
Capital indices are five dimensional space indices: 
$M, N, P, Q, R$ are five-dimensional coordinate space indices $(M=0,\cdots,3,5)$
and $A, B, C, D, E$ are five-dimensional tangent space indices
$(A=\hat{0},\cdots,\hat{3},\hat{5})$.
Hated numbers $(\hat{0},\hat{1},\hat{2},\hat{3},\hat{5})$ are used for tangent
space indices. 
Indices $I,J$ are used as $SU(2)$ indices ($I= 1,2 $) and $i, j, k, l, i^*, j^*,
k^*, l^*$ are K\"{a}hler manifold indices
($i = 1,\cdots,N$ for $N$ chiral multiplets).

The f\"{u}nfbein $e^{A}_{M}$ and the vierbein $e^{a}_{\mu}$ allow to convert
between 
coordinate space and tangent space indices:
\begin{equation}
g_{MN} = e^{A}_{M}e^{B}_{N}\eta_{AB} , \quad
    \Gamma_{M} = e^{A}_{M} \Gamma_{A} , \quad
g_{\mu \nu} = e^{a}_{\mu}e^{b}_{\nu}\eta_{ab} , \quad
    \Gamma_{\mu} = e^{a}_{\mu}\Gamma_{a},
\label{IndexConv}
\end{equation}
their determinant is denoted:
\begin{equation}
    e_{5} = det\left(  e^{A}_{M}\right) , \quad
    e_{4} = det\left(  e^{a}_{\mu}\right).
\label{VielBeinDeterminants}
\end{equation}

The five-dimensional gamma matrices obey the relations:
\begin{equation}
\left\lbrace \Gamma_{A},\Gamma_{B} \right\rbrace  = -2 \eta_{AB}  ,  \quad
\eta_{AB} = diag(-1,1,1,1,1)
\label{GammaComutator}
\end{equation}

Our conventions follow closely those \cite{Bagger:2002rw}
(see also \cite{WessAndBagger}).
We use the following representation for the gamma matrices:
\begin{equation}
\Gamma^{a} = 
\left(\begin{matrix}
0  & \sigma^{a} \\
\overline{\sigma}^{a} & 0
\end{matrix}\right) 
,  \quad
\Gamma^{\hat{5}} = 
\left(\begin{matrix}
-i  & 0 \\
0 & i
\end{matrix}\right) 
\label{GammaRepresentation}
\end{equation}
where the Pauli matrices are:
\begin{eqnarray}
& \sigma^{\hat{0}} = 
\left(\begin{matrix}
-1 & 0 \\
0 & -1
\end{matrix}\right) 
,  \quad
\sigma^{\hat{1}} = 
\left(\begin{matrix}
0 & 1 \\
1 & 0
\end{matrix}\right) 
,  \quad
\sigma^{\hat{2}} = 
\left(\begin{matrix}
0 & -i \\
i & 0
\end{matrix}\right) 
,  \quad
\sigma^{\hat{3}} = 
\left(\begin{matrix}
1 & 0 \\
0 & -1
\end{matrix}\right)
\nonumber \\ &
\overline{\sigma}^{\hat{0}} = \sigma^{\hat{0}} ,  \quad
\overline{\sigma}^{\hat{1}} = -\sigma^{\hat{1}} ,  \quad
\overline{\sigma}^{\hat{2}} = -\sigma^{\hat{2}} ,  \quad
\overline{\sigma}^{\hat{3}} = -\sigma^{\hat{3}}.
\label{SigmaMatrices}
\end{eqnarray}

The gamma matrices obey the following properties
\begin{equation}
\Gamma^{ABCD} = \epsilon^{ABCDE} \Gamma_{E} ,\quad 
\Gamma^{ABC} = \epsilon^{ABCDE} \Sigma_{DE} ,\quad 
\Gamma^{ABCDE} = - \epsilon^{ABCDE}
\label{GammaProperties}
\end{equation}
where $\epsilon^{ABCDE}$ is the completely antisymmetric tensor 
\begin{equation}
\epsilon^{\hat{0}\hat{1}\hat{2}\hat{3}\hat{5}} = + 1 ,\quad 
\epsilon^{MNPQR} = e^{M}_{A} e^{N}_{B} e^{P}_{C} e^{Q}_{D} e^{R}_{E}
\epsilon^{ABCDE}
\label{EpsilonDefinition}
\end{equation}
and 
\begin{equation}
\Sigma^{AB} = \frac{1}{2}\Gamma^{AB} = \frac{1}{4} \left[ \Gamma^{A} ,
\Gamma^{B} \right] .
\label{SigmaDefinition}
\end{equation}

From the representation (\ref{GammaRepresentation}) we find
\begin{equation}
\Sigma^{ab} = 
\left(\begin{matrix}
\sigma^{ab} & 0 \\
0 & \overline{\sigma}^{ab}
\end{matrix}\right) 
,  \quad
\Sigma^{a\hat{5}} = \frac{i}{2}
\left(\begin{matrix}
0  &  \sigma^{a} \\
-\overline{\sigma}^{a} & 0
\end{matrix}\right) .
\label{SigmaRepresentation}
\end{equation}

The charge conjugation matrix is:
\begin{equation}
C = 
\left(\begin{matrix}
i \sigma^{\hat{2}} & 0 \\
0 & i \sigma^{\hat{2}}
\end{matrix}\right) 
\label{ChargeConjMatrix}
\end{equation}
and obeys:
\begin{equation}
C^{T} = - C , \quad 
\left( \Gamma^{a}\right) ^{T} = C \Gamma^{a} C^{-1}
\label{ChargeConjProperties}
\end{equation}

In the five-dimensional Lagrangians we use symplectic Majorana spinors
$\Psi_{I}$. 
We define:
\begin{equation}
\Psi^{I} = \epsilon^{I J} \Psi_{J} , \quad 
\Psi_{I} = \epsilon_{I J} \Psi^{J}
\label{SimplMajUpAndLowIndices}
\end{equation}
where $\epsilon^{I J}$ is the completely antisymmetric tensor: $\epsilon^{1 2} =
\epsilon_{2 1} = 1$.
The symplectic Majorana spinors $\Psi_{J}$ obey the reality condition
\cite{VanProeyen:1999ni}:
\begin{equation}
\overline{\Psi}_{J} = \check{\Psi}^{J}
\label{SimplMajCondition}
\end{equation}
where
\begin{equation}
\overline{\Psi}_{J} = \Psi^{\dagger}_{J} \Gamma_{\hat{0}} , \quad 
\check{\Psi}_{J} =  \Psi^{T}_{J}C .
\label{SpinorConjugated}
\end{equation}

We can express then in the two-component spinor notation 
as follows,
\begin{equation}
\Psi_{1} = 
\left(\begin{matrix}
\psi_{1}\\
\overline{\psi}_{2}
\end{matrix}\right) 
\label{SimplMajInWeylNotation}
\end{equation}
where $\psi_{1}$ and $\psi_{2}$ are two-component Weyl spinors.
Equation (\ref{SpinorConjugated}) implies
\begin{equation}
\overline{\Psi}_{1} = 
\left(\begin{matrix}
\psi_{2}, & \overline{\psi}_{1}
\end{matrix}\right)
    , \quad 
\check{\Psi}_{1} = 
\left(\begin{matrix}
- \psi_{1}, & \overline{\psi}_{2}
\end{matrix}\right) 
\label{ConjgatedSpinorInWeylNotation}
\end{equation}
and the reality condition (\ref{SimplMajCondition}) gives :
\begin{eqnarray}
\Psi_{1} = - \Psi^{2} = 
\left(\begin{matrix}
\psi_{1}\\
\overline{\psi}_{2}
\end{matrix}\right) 
, & \quad &
\Psi_{2} = \Psi^{1} =
\left(\begin{matrix}
- \psi_{2} \\
\overline{\psi}_{1}
\end{matrix}\right) 
\nonumber \\
\overline{\Psi}_{1} = - \overline{\Psi}^{2} =
\left(\begin{matrix}
\psi_{2}, & \overline{\psi}_{1}
\end{matrix}\right) 
, & \quad &
\overline{\Psi}_{2} = \overline{\Psi}^{1} =
\left(\begin{matrix}
\psi_{1} ,& - \overline{\psi}_{2}
\end{matrix}\right) 
\nonumber \\
\check{\Psi}_{1} = - \check{\Psi}^{2} =
\left(\begin{matrix}
- \psi_{1} ,& \overline{\psi}_{2}
\end{matrix}\right) 
, & \quad &
\check{\Psi}_{2} = \check{\Psi}^{1} =
\left(\begin{matrix}
\psi_{2} ,& \overline{\psi}_{1}
\end{matrix}\right) .
\label{SimpMajSpinors}
\end{eqnarray}

The five-dimensional covariant derivative of a spinor is given by
\begin{equation}
D_{M}\Psi_{J} = \partial_{M} \Psi_{J} +\frac{1}{2} \omega_{MAB}\Sigma^{AB}
\Psi_{J} ,
\label{CovDerivOfSpinor}
\end{equation}
the five-dimensional connection and curvature tensors are
\begin{eqnarray}
\omega_{MAB} &=& \frac{1}{2} e^{P}_{A}e^{N}_{B} \left(
e_{MC} \partial_{[ P} e^{C}_{N ]} - e_{PC} \partial_{[ N} e^{C}_{M ]} 
- e_{NC} \partial_{[ M} e^{C}_{P ] }
\right) 
\nonumber \\
R_{MNAB} &=& \partial_{M} \omega_{NAB} - \partial_{N} \omega_{MAB} +
{\omega_{MA}}^{C}\omega_{NBC} - {\omega_{NA}}^{C}\omega_{MBC}
\nonumber \\
R_{MA} &=& R_{MNAB}e^{NB}  ,  \quad R(\omega) = e^{MA}R_{MA}.
\label{ConnectionDefinition}
\end{eqnarray}

The five-dimensional covariant derivatives (equation \ref{CovDerivOfSpinor})
expressed in the two-component spinors notation are:
\begin{eqnarray}
D_{M}\psi_{1} &=& \partial_{M} \psi_{1} +\frac{1}{2} \omega_{Mab}\sigma^{ab}
\psi_{1}
+ i \frac{1}{2} \omega_{Ma\hat{5}}\sigma^{a} \overline{\psi}_{2}
\nonumber \\
D_{M}\psi_{2} &=& \partial_{M} \psi_{2} +\frac{1}{2} \omega_{Mab}\sigma^{ab}
\psi_{2}
- i \frac{1}{2} \omega_{Ma\hat{5}}\sigma^{a} \overline{\psi}_{1} .
\label{CovDerivWeylSpinor}
\end{eqnarray}

The four-dimensional connection and curvature tensors are denoted:
\begin{eqnarray}
\hat{\omega}_{\mu a b} &=& \frac{1}{2} e^{\rho}_{a}e^{\nu}_{b} \left(
e_{\mu c} \partial_{[ \rho} e^{c}_{\nu ]} - e_{\rho c} \partial_{[ \nu}
e^{c}_{\mu ]} 
- e_{\nu c} \partial_{[ \mu} e^{c}_{\rho ] }
\right) 
\nonumber \\
\hat{R}_{\mu \nu a b} &=& \partial_{\mu} \hat{\omega}_{\nu a b} - \partial_{\nu}
\hat{\omega}_{\mu a b} +
{ \left. \hat{\omega}_{\mu a} \right. }^{c}\hat{\omega}_{\nu b c} 
- { \left. \hat{\omega}_{\nu a} \right. }^{c}\hat{\omega}_{\mu b c}
\nonumber \\
\hat{R}_{\mu a} &=& \hat{R}_{\mu \nu a b}e^{\nu b} , \quad 
\hat{R}(\hat{\omega}) = e^{\mu a}\hat{R}_{\mu a},
\label{R4dDefinition}
\end{eqnarray}
and the four-dimensional covariant derivative of a spinor is denoted:
\begin{equation}
\hat{D}_{\mu}\psi = \partial_{\mu} \psi +\frac{1}{2} \hat{\omega}_{\mu a
b}\sigma^{ab} \psi .
\label{CovDer4dDefinition}
\end{equation}

\section{Supersymmetric action}
\label{apeTotalSusyAction}

The space onsidered here is five-dimensional with the fifth dimension
compactified on 
the $S^1/\mathbb{Z}_2$ orbifold through the
identification $y \sim - y$. 
Matter fields live on branes localized in the
boundaries $y_n=y_b\in \{ 0, \pi R\}$. 
The total action is:
\begin{equation}
S = \int^{2 \pi R}_{0} dy \int d^{4}x \left\{ \frac{1}{2} {\cal L}_{BULK} + 
{\cal L}_{0} \delta(y)
    +  {\cal L}_{\pi} \delta(y-\pi R) \right\}.
\label{TotalAction}
\end{equation}

For simplicity we fix $e^{a}_{5} = 0$ and $e^{\hat{5}}_{\mu} = 0$ on the
following formulas.

The bulk fields are composed by the five-dimensional supergravity multiplet and
some auxiliary fields.
The on-shell supergravity multiplet contains the 
f\"{u}nfbein $e^{A}_{M}$, the gravitinos $\psi_{M I}$
and the graviphoton $B_{M}$.

The bulk Lagrangian density is given by:
\begin{equation}
{\cal L}_{BULK} = {\cal L}_{SUGRA} + {\cal L}_{AUX} .
\label{Lbulk3}
\end{equation}

Auxiliary fields are present in the off-shell part of the Lagrangian density,
\begin{equation}
{\cal L}_{AUX} = e_{5} \frac{1}{2} \left( u u + v_{M} v^{M} \right) .
\label{Laux2}
\end{equation}
Here $u$ is a real scalar and $v_{M}$ is a real five-dimensional vector field. 

The on-shell part of the bulk Lagrangian density reads 
\begin{equation}
{\cal L}_{SUGRA} = {\cal L}_{Boson} + {\cal L}_{Fermi} ,
\label{Lsugra2}
\end{equation}
with the on-shell bosonic Lagrangian in the bulk given by
\begin{equation}
{\cal L}_{Boson} = - e_{5} \left\lbrace  \frac{1}{2}R \left( \omega \right) 
+ \frac{1}{4}F_{MN}F^{MN} + \frac{1}{6\sqrt{6}}
\epsilon^{ABCDE}F_{AB}F_{CD}B_{E} \right\rbrace .
\label{Lboson}
\end{equation}
The fermionic part of the bulk Lagrangian expressed in two-component spinor
notation reads
\footnote{We recall that in this paper we use the following approximation, in
the Lagrangians 
we drop the four-fermions terms and in the supersymmetry transformations we
drop 
the three and four-fermions terms.}
:
\begin{eqnarray}
{\cal L}_{Fermi} &=& e_{5} \Bigg \lbrace  \frac{1}{2} \epsilon^{\mu \nu \rho
\lambda} \left( 
\overline{\psi}_{\mu 1}\overline{\sigma}_{\nu}D_{\rho}\psi_{\lambda 1}
+ \overline{\psi}_{\mu 2}\overline{\sigma}_{\nu}D_{\rho}\psi_{\lambda 2}
\right)
    + e^{5}_{\hat{5}} \left(   \psi_{\mu 1} \sigma^{\mu \nu}D_{5}\psi_{\nu 2} 
- \psi_{\mu2} \sigma^{\mu \nu}D_{5}\psi_{\nu1}  \right) 
\nonumber    \\
&&
- e^{5}_{\hat{5}} \left( \psi_{51} \sigma^{\mu \nu}D_{\mu}\psi_{\nu2} 
- \psi_{52} \sigma^{\mu \nu}D_{\mu}\psi_{\nu 1} 
+ \psi_{\mu 1} \sigma^{\mu \nu}D_{\nu}\psi_{52} 
- \psi_{\mu 2} \sigma^{\mu \nu}D_{\nu}\psi_{51} \right) 
\nonumber    \\
&&
- i\frac{\sqrt{6}}{8} e^{5}_{\hat{5}} \epsilon^{\mu \nu \rho \lambda} F_{\mu
\nu} 
\left(  \psi_{\lambda 1} \sigma_{\rho} \overline{\psi}_{51} 
+ \psi_{\lambda2} \sigma_{\rho} \overline{\psi}_{52}
+i \psi_{\rho 1} \psi_{\lambda 2} \right) 
\nonumber    \\
&&
+ i\frac{\sqrt{6}}{4} \left[  F^{\mu \nu} \psi_{\mu1} \psi_{\nu2} 
+ F^{\mu 5} \left( \psi_{\mu 1} \psi_{52}  - \psi_{\mu 2} \psi_{51} 
\right) \right]
\nonumber    \\
&&
+ i\frac{\sqrt{6}}{8} e^{5}_{\hat{5}}  \epsilon^{\mu \nu \rho \lambda} F_{\mu5} 
\left( \psi_{\rho 1} \sigma_{\nu} \overline{\psi}_{\lambda 1} 
+ \psi_{\rho 2} \sigma_{\nu} \overline{\psi}_{\lambda2} \right)  + h.c. \Bigg
\rbrace .
\label{Lferm}
\end{eqnarray}
The covariant derivatives employed here are defined in equation
(\ref{CovDerivWeylSpinor}).

In order to express the supersymmetry transformations (\ref{SusyTransf0}) in 
two-component spinor notation we adopt, in parallel to equation
(\ref{SimpMajSpinors}), the 
following notation for the supersymmetry transformation parameter:
\begin{eqnarray}
\Xi_{1} = - \Xi^{2} =
\left(\begin{matrix}
\xi_{1}\\
\overline{\xi}_{2}
\end{matrix}\right) 
, & \quad &
\Xi_{2} = \Xi^{1} =
\left(\begin{matrix}
- \xi_{2}\\
\overline{\xi}_{1}
\end{matrix}\right) 
\nonumber \\
\check{\Xi}_{1} = - \check{\Xi}^{2} =
\left(\begin{matrix}
- \xi_{1} ,& \overline{\xi}_{2}
\end{matrix}\right) 
, & \quad &
\check{\Xi}_{2} = \check{\Xi}^{1} =
\left(\begin{matrix}
\xi_{2} ,& \overline{\xi}_{1}
\end{matrix}\right) .
\label{SusyParam}
\end{eqnarray}
With these definitions, the on-shell supersymmetry transformations in
two-component spinor notation are given by:
\begin{eqnarray}
\delta_{\prime} e^{a}_{M}  &=& i \left( \xi_{1} \sigma^{a}
\overline{\psi}_{M 1}
+ \xi_{2} \sigma^{a} \overline{\psi}_{M 2} \right) + h.c.
\nonumber   \\
\delta_{\prime} e^{\hat{5}}_{M}  &=& \xi_{2}\psi_{M 1} - \xi_{1} \psi_{M 2} 
+ h.c.
\nonumber   \\
\delta_{\prime} B_{M}  &=& i \frac{\sqrt{6}}{2} \left( \xi_{1}\psi_{M 2}
-\xi_{2}\psi_{M 1} \right) +h.c.
\nonumber   \\
\delta_{\prime} \psi_{\mu 1}  &=& 2 D_{\mu}\xi_{1} 
+ \frac{1}{2\sqrt{6}} F^{\nu \rho} \left( i \epsilon_{\mu \nu \rho \lambda}
\sigma^{\lambda} - 4 g_{\mu \rho} \sigma_{\nu} \right) \overline{\xi}_{2}
- i\frac{2}{\sqrt{6}} e^{\hat{5}}_{5} F^{\nu5} \left( \sigma_{\mu \nu} + g_{\mu
\nu} \right) \xi_{1}
\nonumber   \\
\delta_{\prime} \psi_{\mu 2}  &=& 2 D_{\mu}\xi_{2} 
- \frac{1}{2\sqrt{6}} F^{\nu \rho} \left( i \epsilon_{\mu \nu \rho \lambda}
\sigma^{\lambda} - 4 g_{\mu \rho} \sigma_{\nu} \right) \overline{\xi}_{1}
- i\frac{2}{\sqrt{6}} e^{\hat{5}}_{5} F^{\nu 5} \left( \sigma_{\mu \nu} + g_{\mu
\nu} \right) \xi_{2}
\nonumber   \\
\delta_{\prime} \psi_{5 1}  &=& 2 D_{5}\xi_{1} 
- i\frac{1}{\sqrt{6}} e^{\hat{5}}_{5} F_{\mu \nu} \sigma^{\mu \nu} \xi_{1}
- \frac{2}{\sqrt{6}} F_{\mu 5} \sigma^{\mu} \overline{\xi}_{2}
\nonumber   \\
\delta_{\prime} \psi_{5 2}  &=& 2 D_{5}\xi_{2} 
- i\frac{1}{\sqrt{6}} e^{\hat{5}}_{5} F_{\mu \nu} \sigma^{\mu \nu} \xi_{2}
+ \frac{2}{\sqrt{6}} F_{\mu 5} \sigma^{\mu} \overline{\xi}_{1} .
\label{SusyWeylTransf0}
\end{eqnarray}

The bulk fields have well defined $\mathbb{Z}_{2}$ parities as described in
tables
\ref{FieldParities0} and \ref{FieldParitiesPi}.
\begin{table}[htb]
      \begin{center}
         \begin{tabular}{|l|c|c|c|c|c|c|c|c|}
         \hline
         $ \text{Even} $ & $e^{a}_{\mu}$ & $e^{\hat{5}}_{5}$ & $B_{5}$ &
$\psi_{\mu 1}$ & $\psi_{52}$ & $\xi_{1}$
           & $v_{\mu}$ & $u$ \\
         \hline
         $ \text{Odd} $ & $e^{a}_{5}$ & $e^{\hat{5}}_{\mu}$ & $B_{\mu}$ &
$\psi_{\mu 2}$ & $\psi_{51}$ & $\xi_{2}$
           & $v_{5}$ & \\
         \hline
         \end{tabular}
      \caption{Parity assignments for bulk fields at $y=0$.}
\label{FieldParities0}
      \end{center}
\end{table}
\begin{table}[htb]
      \begin{center}
         \begin{tabular}{|l|c|c|c|c|c|c|c|c|}
         \hline
         $ \text{Even} $ & $e^{a}_{\mu}$ & $e^{\hat{5}}_{5}$ & $B_{5}$ &
$\psi_{\mu +}$ & $\psi_{5 +}$ & $\xi_{+}$
           & $v_{\mu}$ & $u$ \\
         \hline
         $ \text{Odd} $ & $e^{a}_{5}$ & $e^{\hat{5}}_{\mu}$ & $B_{\mu}$ &
$\psi_{\mu -}$ & $\psi_{5 -}$ & $\xi_{-}$
           & $v_{5}$ & \\
         \hline
         \end{tabular}
      \caption{Parity assignments for bulk fields at $y=\pi R$.}
\label{FieldParitiesPi}
      \end{center}
\end{table}

Here the following definitions are used:
\begin{eqnarray}
\psi_{\mu +} & = &
\cos( \pi \omega) \psi_{\mu 1} - \sin( \pi \omega) \psi_{\mu 2} 
\nonumber \\
\psi_{\mu -} & = &
\sin( \pi \omega) \psi_{\mu 1} + \cos( \pi \omega) \psi_{\mu 2} 
\nonumber \\
\psi_{5 +} & = &
\sin( \pi \omega) \psi_{5 1} + \cos( \pi \omega) \psi_{5 2} 
\nonumber \\
\psi_{5 -} & = &
\cos( \pi \omega) \psi_{5 1} - \sin( \pi \omega) \psi_{5 2} 
\nonumber \\
\xi_{+} & = &
\cos( \pi \omega) \xi_{1} - \sin( \pi \omega) \xi_{2} 
\nonumber \\
\xi_{-} & = &
\sin( \pi \omega) \xi_{1} + \cos( \pi \omega) \xi_{2} 
\label{ParityEigenvectors}
\end{eqnarray}

At each boundary $y_b$, $y_b \in \{0 , \pi R \}$, $N_b$ chiral multiplets are
placed, 
each containing one sacalar $\phi^{i}_{b} $ and one fermionic field
$\chi^{i}_{b}$ 
($i = 1 , \cdots N_b$). The Lagrangian density for the brane $b$, $b \in \{0 ,
\pi\}$, is given by:
\begin{eqnarray}
{\cal L}_{b} &=& e_{4} \Bigg \lbrace - \frac{1}{2} g_{i j^*} \partial_{\mu}
\phi_{b}^{i} \partial^{\mu} \phi_{b}^{* j} 
- i \frac{1}{2} g_{i j^*} \overline{\chi}_{b}^{j}
\overline{\sigma}^{\mu}\tilde{D}_{\mu}\chi_{b}^{i}
+ \frac{1}{8}\left( {\cal G}_{b j} \partial_{\mu} \phi_{b}^{j} - {\cal G}_{b
j^*} 
\partial_{\mu} \phi_{b}^{* j} \right) 
\epsilon^{\mu \nu \rho \lambda} \overline{\psi}_{\rho 1}
\overline{\sigma}_{\lambda} \psi_{\nu 1}
\nonumber \\ &&
- e^{{\cal G}_{b}/2} \left[ \psi_{\mu 1} \sigma^{\mu \nu} \psi_{\nu 1} 
+ i \frac{\sqrt{2}}{2} {\cal G}_{b j^*}\overline{\chi}_{b}^{j}
\overline{\sigma}^{\mu} \psi_{\mu 1}
+ \frac{1}{2} \left( {\cal G}_{b i j} + {\cal G}_{b i}{\cal G}_{b j} 
- \Gamma^{k}_{i j} {\cal G}_{b k} \right)^{*} 
\overline{\chi}_{b}^{i} \overline{\chi}_{b}^{j} \right] 
\nonumber \\ &&
- \frac{\sqrt{2}}{2} g_{i j^*}  \partial_{\nu} \phi_{b}^{* j} \chi_{b}^{i}
\sigma^{\mu} \overline{\sigma}^{\nu} \psi_{\mu 1}
- \frac{1}{2} e^{{\cal G}_{b}} \left( g^{i j^*} {\cal G}_{b i} {\cal G}_{b j^*}
- 3 \right)
+ h.c. \Bigg \rbrace ,
\label{BraneLagrangian}
\end{eqnarray}
where
\begin{equation}
\tilde{D}_{\mu}\chi_{b}^{i} = \partial_{\mu}\chi_{b}^{i} + \frac{1}{2}
\omega_{\mu ab}\sigma^{ab}\chi_{b}^{i} 
- \frac{1}{4} \left( {\cal G}_{b j} \partial_{\mu} \phi_{b}^{j} - {\cal G}_{b
j^*} 
\partial_{\mu} \phi_{b}^{* j} \right)  \chi_{b}^{i}
\label{CovDerivChiB}
\end{equation}
and the hermitian function ${\mathcal G}_{b}(\phi_{b} , \phi_{b}^{*})$ 
is given in terms of the K\"ahler potential and
superpotential by 
\begin{equation}
{\mathcal G}_{b}(\phi_{b} , \phi_{b}^{*}) = K_{b}(\phi_{b} , \phi_{b}^{*}) +
\ln\left[ W_{b}(\phi_{b})\right]
    + \ln\left[ W_{b}(\phi_{b}) \right] ^{*}.
\label{KahlerFunctionB}
\end{equation}

We impose also the following boundary conditions at $y = 0$ and $y = \pi R$:
\begin{equation}
\left. u \right|_{y = 0} = e^{{\cal G}_{0}/2} , \quad 
i \left. v_{\mu} \right|_{y = 0} = \frac{1}{2}  \left( {\cal G}_{0 j}
\partial_{\mu} \phi_{0}^{j} 
- {\cal G}_{0 j^*} \partial_{\mu} \phi_{0}^{* j} \right) , \quad
\left. F^{\mu 5} \right|_{y = 0} = 0 ,
\label{CDFBoundCond0}
\end{equation}
\begin{equation}
\left. u \right|_{y = \pi R}  = e^{{\cal G}_{\pi}/2} , \quad 
i \left. v_{\mu} \right|_{y = \pi R} =  \frac{1}{2} \left( {\cal G}_{\pi j}
\partial_{\mu} \phi_{\pi}^{j} 
- {\cal G}_{\pi j^*} \partial_{\mu} \phi_{\pi}^{* j} \right) , \quad
\left. F^{\mu 5} \right|_{y = \pi R} = 0 .
\label{CDFBoundCondPi}
\end{equation}

The modified supersymmetry transformations for the bulk fields are given by:
\begin{eqnarray}
\delta e^{A}_{M} &=& \delta_{\prime} e^{A}_{M}
\nonumber \\
\delta B_{M} &=& \delta_{\prime} B_{M}
\nonumber \\
\delta \psi_{\mu 1}  &=& \delta_{\prime} \psi_{\mu 1} + i v_{\mu} \xi_{1} + i u
\sigma_{\mu} \overline{\xi}_{1}
\nonumber   \\
\delta \psi_{\mu 2}  &=& \delta_{\prime} \psi_{\mu 2} + i v_{\mu} \xi_{2} + i u
\sigma_{\mu} \overline{\xi}_{2}
\nonumber \\
\delta \psi_{5 1} &=& \delta_{\prime} \psi_{5 1} 
- 4 e^{{\cal G}_{\pi}/2} \sin(\omega \pi)\xi_{+} \delta(y-\pi R)
\nonumber \\
\delta \psi_{5 2} &=& \delta_{\prime} \psi_{5 2} 
- 4 e^{{\cal G}_{0}/2}\xi_{1} \delta(y) 
- 4 e^{{\cal G}_{\pi}/2} \cos(\omega \pi)\xi_{+} \delta(y-\pi R)
\nonumber   \\
\delta u &=& - \frac{1}{2} u \left( i\xi_{1} \sigma^{\nu} \overline{\psi}_{\nu1}
+ i\xi_{2} \sigma^{\nu} \overline{\psi}_{\nu2} 
+ \xi_{2}\psi_{51} - \xi_{1} \psi_{52} \right) 
\nonumber \\ &&
+ \frac{i}{e_{5}} \left[
    \overline{\xi}_{J} \overline{\sigma}^{\mu} \frac{\partial {\cal
L}_{SUGRA}}{\partial \psi^{\mu}_{J}}
- \overline{\xi}_{J} \overline{\sigma}^{\mu} D_{N} \frac{\partial {\cal
L}_{SUGRA}}{\partial \left( D_{N}\psi^{\mu}_{J} \right) }
\right] + h.c.
\nonumber \\
\delta v_{\mu} &=& - \frac{1}{2} v_{\mu} \left( i\xi_{1} \sigma^{\nu}
\overline{\psi}_{\nu1}
+ i\xi_{2} \sigma^{\nu} \overline{\psi}_{\nu2} 
+ \xi_{2}\psi_{51} - \xi_{1} \psi_{52} \right) 
- \frac{i}{e_{5}} \left[
    \xi_{J} \frac{\partial {\cal L}_{SUGRA}}{\partial \psi^{\mu}_{J}}
+ \xi_{J} D_{N} \frac{\partial {\cal L}_{SUGRA}}{\partial \left(
D_{N}\psi^{\mu}_{J} \right) }
\right] 
\nonumber \\ &&
    - 2 i e^{5}_{\hat{5}} \epsilon^{\mu \nu \rho \lambda} \overline{\xi}_{1}
\overline{\sigma}_{\nu} D_{\rho}\psi_{\lambda 1} \delta(y) 
- 2 i e^{5}_{\hat{5}} \epsilon^{\mu \nu \rho \lambda} \overline{\xi}_{+}
\overline{\sigma}_{\nu} D_{\rho}\psi_{\lambda +} \delta(y - \pi R) + h.c.
\nonumber \\
\delta v_{5} &=& - \frac{1}{2} v_{5} \left( i\xi_{1} \sigma^{\nu}
\overline{\psi}_{\nu1}
+ i\xi_{2} \sigma^{\nu} \overline{\psi}_{\nu2} 
+ \xi_{2}\psi_{51} - \xi_{1} \psi_{52} \right) + h.c.
\label{BulkFieldsSusyTransf}
\end{eqnarray}

In the branes at $y = 0$ and $y = \pi R$ the supersymmetry transformations of 
the fields $e^{a}_{\mu}$ and $\psi_{\mu I}$ are those
induced by the bulk (given by equation (\ref{BulkFieldsSusyTransf}) calculated
at $y = 0$ and $y=\pi R$). 
Together with the supersymmetry transformations of the brane matter fields 
($\phi_{0}$, $\chi_{0}$ , $\phi_{\pi}$ and $\chi_{\pi}$) they read at the brane
sitting 
on the boundary $y_b$, $y_b \in \{0 , \pi R \}$:
\begin{eqnarray}
\left. \delta e^{a}_{\mu} \right|_{y = y_{b}}  &=& i \left(
\left.\xi_{1}\right|_{y
= y_{b} } \sigma^{a} \left. \overline{\psi}_{\mu 1} \right|_{y = y_{b}} \right)
+ h.c.
\nonumber   \\
\delta \phi_{b}^{i}  &=& \sqrt{2} \left.\xi_{1}\right|_{y = y_{b}}\chi_{b}^{i}
\nonumber   \\
\delta \chi_{b}^{i}  &=& i \sqrt{2} \sigma^{\mu} \left.\xi_{1}\right|_{y =
y_{b}}
\partial_{\mu} \phi_{b}^{i}
- \sqrt{2} e^{{\cal G}_{b}/2} g^{i j^*} {\cal G}_{b j^*} \left.\xi_{1}\right|_{y
= b}
\nonumber   \\
\left. \delta \psi_{\mu 1} \right|_{y = y_{b}}  &=& 2
\hat{D}_{\mu}\left.\xi_{1}\right|_{y = y_{b}}
+ \frac{1}{2} \left( {\cal G}_{b j} \partial_{\mu} \phi_{b}^{j} 
- {\cal G}_{b j^*} \partial_{\mu} \phi_{b}^{* j} \right) 
\left.\xi_{1}\right|_{y = y_{b}}
+ i e^{{\cal G}_{b}/2} \sigma_{\mu} \left.\overline{\xi_{1}}\right|_{y = y_{b}}
\label{BraneBSusyTrans}
\end{eqnarray}
where $\hat{D}_{\mu}\xi_{1}$ and $\hat{D}_{\mu}\xi_{+}$ are given by equation
(\ref{CovDer4dDefinition}).

With the parity assignments of tables \ref{FieldParities0} and
\ref{FieldParitiesPi} and the boundary conditions (\ref{CDFBoundCond0}) and
(\ref{CDFBoundCondPi}) the action (\ref{TotalAction}) is invariant under the
supersymmetry transformations (\ref{BulkFieldsSusyTransf}) and 
(\ref{BraneBSusyTrans}) up to four-fermions terms
(which is the approximation we use in this paper).

\section{A simple example of bulk-brane supersymmetry breaking}
\label{apeSimpleExample}

In this appendix we provide a simple example of supersymmetry breaking in both
sectors (bulk and brane) of 
the 5d space-time. We consider only one chiral multiplet living in 
a brane placed at $y = 0$. Supersymmetry is broken in the bulk by a non trivial
Scherk-Schwarz 
twist described by the angle $\omega \ne 0 $.

To keep things as simple as possible, in the brane the K\"ahler potential is the
canonical 
one, $K = \phi \phi^*$ and the superpotential considered here 
is $ W = e^{ - \phi^{2}/2 + \sqrt{3} \phi }$. This implies the following 
K\"ahler function for the brane:
\begin{equation}
{\cal G} = \phi \phi^* - \frac{\phi^{2}}{2} + \sqrt{3} \phi - \frac{\phi^{*
2}}{2} + \sqrt{3} \phi^* .
\label{SimpleKalerPot}
\end{equation}
We will now show that this choice for the K\"ahler function provide
supersymmetry
breaking with a 
vanishing cosmological constant in the brane.
The Lagrangian density \ref{Lbrane0} gives the following brane potential:
\begin{equation}
{\cal V} = e^{\cal G} \left( \frac{\partial {\cal G}}{\partial \phi} 
\frac{\partial {\cal G}}{\partial \phi^*}  -3 \right) 
\label{SimpleBranePot}
\end{equation}
From (\ref{SimpleKalerPot}) it is easy to obtain that 
$ {\cal V} = e^{\cal G} \left| \phi - \phi^* \right|^{2} $.
Then at the extremun of the potential $ \, \left\langle Im(\phi) \,
\right\rangle = 0$ 
and $ \left\langle \, {\cal V} \, \right\rangle = 0$, giving a vanishing brane
cosmological 
constant, as claimed.
It is useful to parametrize the complex field $\phi$ by two real-valued fields
$\varphi$ 
and $\sigma$:
\begin{equation}
\phi = \frac{1}{\sqrt{2}} \left[ \, \varphi + i \sigma \, \right] 
\label{SimplePhi}
\end{equation}
From potential \ref{SimpleBranePot} it is clear that $\varphi$ is massless and
$\sigma$ has mass squared
$m^{2}_{\sigma} = 4 \left\langle \, e^{\sqrt{6} \varphi } \, \right\rangle $.

The fermionic spectrum is easily calculated with help of 
formulae (\ref{PsiMass}), (\ref{Mass1}) and the $F$-terms values 
$M_{0} = \left\langle e ^{{\cal G} / 2} \right\rangle = \left\langle \, 
e^{\sqrt{6} \varphi / 2 } \, \right\rangle $ and 
$M_{\pi} = 0$. Taking the v.e.v. $\left\langle \varphi \right\rangle = A$, the
fermion masses are 
given as follows, the gravitino tower of Kaluza-Klein masses are:
\begin{equation}
m_{3/2} = \frac{\omega}{R} + \frac{1}{\pi R} \arctan \left( e^{\sqrt{6} A / 2 }
\right) 
+ \frac{n}{R} ,  \quad n \in \mathbb{Z}
\label{SimpleMassGravitino}
\end{equation}
and the pseudo Goldstino mass is:
\begin{equation}
m_{PG} = \frac{2 e^{\sqrt{6} A / 2 } \sin(\omega \pi) \left[ e^{\sqrt{6} A / 2
} 
\cos(\omega \pi) + \sin(\omega \pi) \right]}
{\pi R e^{\sqrt{6} A } + \left[ e^{\sqrt{6} A / 2 } \cos(\omega \pi) 
+ \sin(\omega \pi) \right]^{2}}
\label{SimpleMassPseudoGold}
\end{equation}

\section{Pseudo-Goldstino mass eigenstates}
\label{apePGMassEigenstate}

In this appendix we wish to present the eigenstates and masses of the pseudo
Goldstinos for 
general $M_{0}$, $M_{\pi}$ and $\omega$. As said at the end of section
\ref{secUnitGauge}, the 
procedure to identify the mass eigenstates of the pseudo Goldstinos is long but
straight forward :
one must plug (\ref{Psi5UG}) and (\ref{ChiUG2}) in the Lagrangian (\ref{LkmGF}),
integrate over the $y$ 
dimension, diagonalize and canonically normalize the kinetic terms of the 
fields $\chi_{1}$ and $\chi_{2}$ and finally diagonalize their 
mass matrix. To do this we set $\theta = - \omega \pi / 2$ in (\ref{Psi5UG}) and
do all the diagonalizations 
described above. We now present the final results.

We call the mass eigenstates $\psi_{1}$ and $\psi_{2}$, their masses are
respectively:
\begin{eqnarray}
m_{1}  &=& \frac{m_{11} a_{22} + m_{22} a_{11} -2 a_{12} m_{12} + d
\sqrt{\Delta}} 
{2 (a_{11} a_{22} - a_{12}^{2})}
\nonumber   \\
m_{2}  &=& \frac{m_{11} a_{22} + m_{22} a_{11} -2 a_{12} m_{12} - d
\sqrt{\Delta}} 
{2 (a_{11} a_{22} - a_{12}^{2})}
\label{PGMasses}
\end{eqnarray}
where we defined
\begin{eqnarray}
a_{11}  &=& 1 + \frac{\kappa}{\pi R} \left[ 2 \sin\left( \frac{\omega \pi}{2}
\right)^{2} 
- \left( \frac{1}{\kappa M_{0}} + \frac{1}{\kappa M_{\pi}} \right) \sin(\omega
\pi) 
+ \left( \frac{1}{ \kappa^{2} M_{0}^{2}} + \frac{1}{ \kappa^{2} M_{\pi}^{2}}
\right) 
\cos\left( \frac{\omega \pi}{2} \right)^{2} \right] 
\nonumber   \\
a_{22}  &=& 1 + \frac{\kappa}{\pi R} \left[ 2 \cos\left( \frac{\omega \pi}{2}
\right)^{2} 
+ \left( \frac{1}{\kappa M_{0}} + \frac{1}{\kappa M_{\pi}} \right) \sin(\omega
\pi) 
+ \left( \frac{1}{ \kappa^{2} M_{0}^{2}} + \frac{1}{ \kappa^{2} M_{\pi}^{2}}
\right) 
\sin\left( \frac{\omega \pi}{2} \right)^{2} \right]
\nonumber   \\
a_{12}  &=& \frac{\kappa}{\pi R} \left[
    \frac{1}{2} \left( \frac{1}{\kappa^{2} M_{\pi}^{2}} - \frac{1}{\kappa^{2}
M_{0}^{2}} \right)  \sin(\omega \pi) 
+ \left( \frac{1}{\kappa M_{\pi}} - \frac{1}{\kappa M_{0}} \right) \cos(\omega
\pi) \right]
\nonumber   \\
m_{11}  &=& \frac{1}{\pi R} \left[
    2 \left( \frac{1}{\kappa M_{0}} + \frac{1}{\kappa M_{\pi}} \right)  \cos
\left(
\frac{\omega \pi}{2} \right)^{2} 
- 2 \sin(\omega \pi) \right]
\nonumber   \\
m_{22}  &=& \frac{1}{\pi R} \left[
    2 \left( \frac{1}{\kappa M_{0}} + \frac{1}{\kappa M_{\pi}} \right)  \sin
\left(
\frac{\omega \pi}{2} \right)^{2} 
+ 2 \sin(\omega \pi) \right]
\nonumber   \\
m_{12}  &=& \frac{1}{\pi R} \left( \frac{1}{\kappa M_{\pi}} - \frac{1}{\kappa
M_{0}} \right) \sin(\omega \pi) 
\label{AMDef}
\end{eqnarray}
and
\begin{eqnarray}
d  &=& \frac{a_{11} a_{22} -a_{12}^{2}}{ \left| a_{11} a_{22} - a_{12}^{2}
\right| } 
\nonumber   \\
\Delta  &=& 2 a_{11} a_{22} \left( 2 m_{12}^{2} - m_{11} m_{22} \right) +
a_{11}^{2} m_{22}^{2} + a_{22}^{2} m_{11}^{2} + 4 m_{11} m_{22} a_{12}^{2} 
\nonumber   \\ &&
- 4 a_{12} m_{12} \left( a_{11} m_{22} + m_{11} a_{22} \right) 
\label{DDDef}
\end{eqnarray}

The canonically normalized mass eigenstates are:
\begin{eqnarray}
\psi_{1}  &=& \frac{1}{\sqrt{(a^{2} + b^{2}) r_{2} }} 
\{ \left[ a \sqrt{r_{3}} ( a_{11} - a_{22} + \sqrt{r_{1}}) 
- 2 b a_{12} \sqrt{r_{4}} \right] \chi_{1} 
\nonumber   \\ &&
+ \left[ b \sqrt{r_{4}} ( a_{11} - a_{22} + \sqrt{r_{1}} )
+ 2 a a_{12} \sqrt{r_{3}}  \right] \chi_{2} \} 
\nonumber   \\
\psi_{2}  &=& \frac{1}{\sqrt{(a^{2} + b^{2}) r_{2} }} 
\{ - \left[ b \sqrt{r_{3}} ( a_{11} - a_{22} + \sqrt{r_{1}}) 
+ 2 a a_{12} \sqrt{r_{4}} \right] \chi_{1} 
\nonumber   \\ &&
+ \left[ a \sqrt{r_{4}} ( a_{11} - a_{22} + \sqrt{r_{1}} )
- 2 b a_{12} \sqrt{r_{3}}  \right] \chi_{2} \} 
\label{PGStates}
\end{eqnarray}
where we defined
\begin{eqnarray}
a &=& \left( m_{11} + m_{22} \right) \left( a_{11} a_{22} - 2a_{12}^{2} \right) 
-m_{11} a_{22}^{2} -m_{22} a_{11}^{2} 
\nonumber   \\ &&
+ 2 \left( a_{11} + a_{22} \right) a_{12} m_{12} + d \sqrt{r_{1} \Delta }
\nonumber   \\
b &=& 2 d \left[ a_{12} \left( m_{11} - m_{22} \right) - m_{12} \left( a_{11} -
a_{12} \right) 
\right] t \sqrt{a_{11} a_{22} - a_{12}^{2}}
\nonumber   \\
r_{1}  &=& \left( a_{11} - a_{22} \right) ^{2} + 4 a_{12}^{2}
\nonumber   \\
r_{2}  &=& \left( a_{11} - a_{22} + \sqrt{r_{1}} \right) ^{2} + 4 a_{12}^{2}
\nonumber   \\
r_{3}  &=& ( a_{11} + a_{22} + \sqrt{r_{1}} ) / 2 
\nonumber   \\
r_{4}  &=& ( a_{11} + a_{22} - \sqrt{r_{1}} ) / 2
\nonumber   \\
t &=& \frac{ a_{11} - a_{22} - \sqrt{r_{1}} } { \left| a_{11} - a_{22} -
\sqrt{r_{1}} \right| } .
\label{ABDef}
\end{eqnarray}

The fields $\psi_{5 1}(y)$, $\psi_{5 2}(y)$, $\chi_{0}$ and $\chi_{\pi}$ can be 
written in terms of the mass eigenstates $\psi_{1}$ and $\psi_{2}$ with help 
of (\ref{Psi5UG}), (\ref{ChiUG2}) and
\begin{eqnarray}
\chi_{1}  &=& \frac{1}{ \sqrt{(a^{2} + b^{2}) r_{2} r_{3} r_{4} } } 
\{ \left[ a  \sqrt{r_{4}} (a_{11} - a_{22} + \sqrt{r_{1}} ) 
- 2 b a_{12} \sqrt{r_{3}} \right] \psi_{1} 
\nonumber   \\ &&
    - \left[ b \sqrt{r_{4}} (a_{11} - a_{22} + \sqrt{r_{1}} ) 
+ 2 a a_{12} \sqrt{r_{3}} \right] \psi_{2} \} 
\nonumber   \\
\chi_{2}  &=& \frac{1}{ \sqrt{(a^{2} + b^{2}) r_{2} r_{3} r_{4} } } 
\{ \left[ b \sqrt{r_{3}} (a_{11} - a_{22} +  \sqrt{r_{1}} ) 
+ 2 a a_{12} \sqrt{r_{4}} \right] \psi_{1} 
\nonumber   \\ &&
    + \left[ a \sqrt{r_{3}} (a_{11} - a_{22} + \sqrt{r_{1}} ) 
- 2 b a_{12} \sqrt{r_{4}} \right] \psi_{2} \} 
\label{PGChi}
\end{eqnarray}


\begin{thebibliography}{99}

\bibitem{Fayet:1974jb}
     P.~Fayet and J.~Iliopoulos,
     Phys.\ Lett.\  B {\bf 51} (1974) 461.

\bibitem{Volkov:1973jd}
     D.~V.~Volkov and V.~A.~Soroka,
     JETP Lett.\  {\bf 18} (1973) 312
     [Pisma Zh.\ Eksp.\ Teor.\ Fiz.\  {\bf 18} (1973) 529];
     S.~Deser and B.~Zumino,
     Phys.\ Rev.\ Lett.\  {\bf 38} (1977) 1433;
     E.~Cremmer, B.~Julia, J.~Scherk, S.~Ferrara, L.~Girardello and P.~van
Nieuwenhuizen,
     Nucl.\ Phys.\  B {\bf 147} (1979) 105.


\bibitem{Fayet:1977vd}
     P.~Fayet,
     Phys.\ Lett.\  B {\bf 70} (1977) 461.

\bibitem{Casalbuoni:1988qd}
     R.~Casalbuoni, S.~De Curtis, D.~Dominici, F.~Feruglio and R.~Gatto,
     Phys.\ Rev.\  D {\bf 39} (1989) 2281;
     Phys.\ Lett.\  B {\bf 215} (1988) 313.


\bibitem{Antoniadis:1990ew}
     I.~Antoniadis,
     Phys.\ Lett.\  B {\bf 246} (1990) 377;
     I.~Antoniadis, C.~Munoz and M.~Quiros,
     Nucl.\ Phys.\  B {\bf 397} (1993) 515
     [arXiv:hep-ph/9211309];
     I.~Antoniadis and K.~Benakli,
     Phys.\ Lett.\  B {\bf 326} (1994) 69
     [arXiv:hep-th/9310151];
     K.~Benakli,
     Phys.\ Lett.\  B {\bf 386} (1996) 106
     [arXiv:hep-th/9509115].



\bibitem{Horava:1996ma}
     P.~Horava and E.~Witten,
     Nucl.\ Phys.\  B {\bf 475} (1996) 94
     [arXiv:hep-th/9603142];
     E.~Witten,
     Nucl.\ Phys.\  B {\bf 471}, 135 (1996)
     [arXiv:hep-th/9602070];
     K.~Benakli,
     Phys.\ Lett.\  B {\bf 447} (1999) 51
     [arXiv:hep-th/9805181];
     S.~Stieberger,
     Nucl.\ Phys.\  B {\bf 541} (1999) 109
     [arXiv:hep-th/9807124];
     Z.~Lalak, S.~Pokorski and S.~Thomas,
     Nucl.\ Phys.\  B {\bf 549} (1999) 63
     [arXiv:hep-ph/9807503].

\bibitem{Lykken:1996fj}
     J.~D.~Lykken,
     Phys.\ Rev.\  D {\bf 54} (1996) 3693
     [arXiv:hep-th/9603133].

\bibitem{Arkani-Hamed:1998rs}
     N.~Arkani-Hamed, S.~Dimopoulos and G.~R.~Dvali,
     Phys.\ Lett.\  B {\bf 429} (1998) 263
     [arXiv:hep-ph/9803315];
     I.~Antoniadis, N.~Arkani-Hamed, S.~Dimopoulos and G.~R.~Dvali,
     Phys.\ Lett.\  B {\bf 436} (1998) 257
     [arXiv:hep-ph/9804398].

\bibitem{Dienes:1998vh}
     K.~R.~Dienes, E.~Dudas and T.~Gherghetta,
     Phys.\ Lett.\  B {\bf 436}, 55 (1998)
     [arXiv:hep-ph/9803466];
     Nucl.\ Phys.\  B {\bf 537}, 47 (1999)
     [arXiv:hep-ph/9806292].

\bibitem{Benakli:1998pw}
     K.~Benakli,
     Phys.\ Rev.\  D {\bf 60}, 104002 (1999)
     [arXiv:hep-ph/9809582];
     C.~P.~Burgess, L.~E.~Ibanez and F.~Quevedo,
     Phys.\ Lett.\  B {\bf 447}, 257 (1999)
     [arXiv:hep-ph/9810535].



\bibitem{Randall:1999ee}
     L.~Randall and R.~Sundrum,
     Phys.\ Rev.\ Lett.\  {\bf 83} (1999) 3370
     [arXiv:hep-ph/9905221];
     Phys.\ Rev.\ Lett.\  {\bf 83} (1999) 4690
     [arXiv:hep-th/9906064].


\bibitem{Witten:1981nf}
     E.~Witten,
     Nucl.\ Phys.\  B {\bf 188} (1981) 513.


\bibitem{Horava:1996vs}
     P.~Horava,
     Phys.\ Rev.\  D {\bf 54} (1996) 7561
     [arXiv:hep-th/9608019].

\bibitem{Veneziano:1982ah}
     G.~Veneziano and S.~Yankielowicz,
     Phys.\ Lett.\  B {\bf 113} (1982) 231;
     H.~P.~Nilles,
     Phys.\ Lett.\  B {\bf 115} (1982) 193;
     Nucl.\ Phys.\  B {\bf 217} (1983) 366;
     T.~R.~Taylor, G.~Veneziano and S.~Yankielowicz,
     Nucl.\ Phys.\  B {\bf 218} (1983) 493;
     S.~Ferrara, L.~Girardello and H.~P.~Nilles,
     Phys.\ Lett.\  B {\bf 125} (1983) 457;
     I.~Affleck, M.~Dine and N.~Seiberg,
     Nucl.\ Phys.\  B {\bf 241} (1984) 493;
     Nucl.\ Phys.\  B {\bf 256} (1985) 557;
     J.~P.~Derendinger, L.~E.~Ibanez and H.~P.~Nilles,
     Phys.\ Lett.\  B {\bf 155} (1985) 65;
     M.~Dine, R.~Rohm, N.~Seiberg and E.~Witten,
     Phys.\ Lett.\  B {\bf 156} (1985) 55;
     C.~Kounnas and M.~Porrati,
     Phys.\ Lett.\  B {\bf 191} (1987) 91;
     K.~A.~Intriligator and S.~D.~Thomas,
     Nucl.\ Phys.\  B {\bf 473} (1996) 121
     [arXiv:hep-th/9603158];
     K.~I.~Izawa and T.~Yanagida,
     Prog.\ Theor.\ Phys.\  {\bf 95} (1996) 829
     [arXiv:hep-th/9602180].




\bibitem{Intriligator:2007cp}
    Excellent guides through recent developpements are  K.~Intriligator and
N.~Seiberg,
     arXiv:hep-ph/0702069;
     Y.~Shadmi and Y.~Shirman,
     Rev.\ Mod.\ Phys.\  {\bf 72} (2000) 25
     [arXiv:hep-th/9907225];
     G.~F.~Giudice and R.~Rattazzi,
     Phys.\ Rept.\  {\bf 322} (1999) 419
     [arXiv:hep-ph/9801271], and references therein.



\bibitem{Scherk:1978ta}
     J.~Scherk and J.~H.~Schwarz,
     Phys.\ Lett.\  B {\bf 82} (1979) 60;
     Nucl.\ Phys.\  B {\bf 153}, 61 (1979);
     P.~Fayet,
     Phys.\ Lett.\  B {\bf 159}, 121 (1985);
     Nucl.\ Phys.\  B {\bf 263}, 649 (1986).


\bibitem{Mirabelli:1997aj}
     E.~A.~Mirabelli and M.~E.~Peskin,
     Phys.\ Rev.\  D {\bf 58} (1998) 065002
     [arXiv:hep-th/9712214].

\bibitem{Randall:1998uk}
     L.~Randall and R.~Sundrum,
     Nucl.\ Phys.\  B {\bf 557}, 79 (1999)
     [arXiv:hep-th/9810155];
     G.~F.~Giudice, M.~A.~Luty, H.~Murayama and R.~Rattazzi,
     JHEP {\bf 9812}, 027 (1998)
     [arXiv:hep-ph/9810442].



\bibitem{Antoniadis:1997ic}
     I.~Antoniadis and M.~Quiros,
     Nucl.\ Phys.\  B {\bf 505} (1997) 109
     [arXiv:hep-th/9705037];
     E.~Dudas and C.~Grojean,
     Nucl.\ Phys.\  B {\bf 507} (1997) 553
     [arXiv:hep-th/9704177];
     H.~P.~Nilles, M.~Olechowski and M.~Yamaguchi,
     Phys.\ Lett.\  B {\bf 415} (1997) 24
     [arXiv:hep-th/9707143];
     A.~Lukas, B.~A.~Ovrut and D.~Waldram,
     Phys.\ Rev.\  D {\bf 57}, 7529 (1998)
     [arXiv:hep-th/9711197];
     K.~Choi, H.~B.~Kim and C.~Munoz,
     Phys.\ Rev.\  D {\bf 57} (1998) 7521
     [arXiv:hep-th/9711158];
     J.~R.~Ellis, Z.~Lalak, S.~Pokorski and W.~Pokorski,
     Nucl.\ Phys.\  B {\bf 540} (1999) 149
     [arXiv:hep-ph/9805377];

\bibitem{Pomarol:1998sd}
     A.~Pomarol and M.~Quiros,
     Phys.\ Lett.\  B {\bf 438} (1998) 255
     [arXiv:hep-ph/9806263];
     I.~Antoniadis, S.~Dimopoulos, A.~Pomarol and M.~Quiros,
     Nucl.\ Phys.\  B {\bf 544}, 503 (1999)
     [arXiv:hep-ph/9810410];
     R.~Barbieri, L.~J.~Hall and Y.~Nomura,
     Phys.\ Rev.\  D {\bf 63} (2001) 105007
     [arXiv:hep-ph/0011311];
     D.~Marti and A.~Pomarol,
     Phys.\ Rev.\  D {\bf 64} (2001) 105025
     [arXiv:hep-th/0106256];
     Phys.\ Rev.\  D {\bf 66} (2002) 125005
     [arXiv:hep-ph/0205034];
     D.~E.~Kaplan, G.~D.~Kribs and M.~Schmaltz,
     Phys.\ Rev.\  D {\bf 62}, 035010 (2000)
     [arXiv:hep-ph/9911293];
     Z.~Chacko, M.~A.~Luty, A.~E.~Nelson and E.~Ponton,
     JHEP {\bf 0001}, 003 (2000)
     [arXiv:hep-ph/9911323];
     K.~Benakli,
     Phys.\ Lett.\  B {\bf 475} (2000) 77
     [arXiv:hep-ph/9911517];
     N.~Arkani-Hamed, L.~J.~Hall, Y.~Nomura, D.~R.~Smith and N.~Weiner,
     Nucl.\ Phys.\  B {\bf 605}, 81 (2001)
     [arXiv:hep-ph/0102090].


\bibitem{Gherghetta:2000qt}
     T.~Gherghetta and A.~Pomarol,
     Nucl.\ Phys.\  B {\bf 586} (2000) 141
     [arXiv:hep-ph/0003129];
     Nucl.\ Phys.\  B {\bf 602} (2001) 3
     [arXiv:hep-ph/0012378];
     A.~Falkowski, Z.~Lalak and S.~Pokorski,
     Phys.\ Lett.\  B {\bf 491} (2000) 172
     [arXiv:hep-th/0004093];
     R.~Altendorfer, J.~Bagger and D.~Nemeschansky,
     Phys.\ Rev.\  D {\bf 63} (2001) 125025
     [arXiv:hep-th/0003117];
     E.~Bergshoeff, R.~Kallosh and A.~Van Proeyen,
     JHEP {\bf 0010} (2000) 033
     [arXiv:hep-th/0007044];
     G.~von Gersdorff and M.~Quiros,
     Phys.\ Rev.\  D {\bf 65} (2002) 064016
     [arXiv:hep-th/0110132];
  J.~L.~Lehners, P.~Smyth and K.~S.~Stelle,
  arXiv:0704.3343 [hep-th].


\bibitem{Zucker:2003qv}
     M.~Zucker,
     Fortsch.\ Phys.\  {\bf 51} (2003) 899;
     Phys.\ Rev.\  D {\bf 64} (2001) 024024
     [arXiv:hep-th/0009083];
     JHEP {\bf 0008} (2000) 016
     [arXiv:hep-th/9909144];
     Nucl.\ Phys.\  B {\bf 570} (2000) 267
     [arXiv:hep-th/9907082].


\bibitem{PaccettiCorreia:2004ri}
  F.~Paccetti Correia, M.~G.~Schmidt and Z.~Tavartkiladze,
  Nucl.\ Phys.\  B {\bf 709} (2005) 141
  [arXiv:hep-th/0408138];
  Phys.\ Lett.\  B {\bf 613} (2005) 83
  [arXiv:hep-th/0410281];
  Nucl.\ Phys.\  B {\bf 751} (2006) 222
  [arXiv:hep-th/0602173].


\bibitem{Kugo:2000hn}
     T.~Kugo and K.~Ohashi,
     Prog.\ Theor.\ Phys.\  {\bf 104} (2000) 835
     [arXiv:hep-ph/0006231];
     Prog.\ Theor.\ Phys.\  {\bf 105} (2001) 323
     [arXiv:hep-ph/0010288];
     T.~Fujita and K.~Ohashi,
     Prog.\ Theor.\ Phys.\  {\bf 106} (2001) 221
     [arXiv:hep-th/0104130];
     H.~Abe and Y.~Sakamura,
     JHEP {\bf 0410} (2004) 013
     [arXiv:hep-th/0408224];
     JHEP {\bf 0602} (2006) 014
     [arXiv:hep-th/0512326].


\bibitem{Gherghetta:2001sa}
     T.~Gherghetta and A.~Riotto,
     Nucl.\ Phys.\  B {\bf 623}, 97 (2002)
     [arXiv:hep-th/0110022].

\bibitem{Rattazzi:2003rj}
     R.~Rattazzi, C.~A.~Scrucca and A.~Strumia,
     Nucl.\ Phys.\  B {\bf 674} (2003) 171
     [arXiv:hep-th/0305184].

\bibitem{Buchbinder:2003qu}
     I.~L.~Buchbinder, S.~J.~J.~Gates, H.~S.~J.~Goh, W.~D.~I.~Linch, 
M.~A.~Luty, S.~P.~Ng and J.~Phillips,
     Phys.\ Rev.\  D {\bf 70} (2004) 025008
     [arXiv:hep-th/0305169].

\bibitem{Bagger:2001qi}
     J.~A.~Bagger, F.~Feruglio and F.~Zwirner,
     Phys.\ Rev.\ Lett.\  {\bf 88} (2002) 101601
     [arXiv:hep-th/0107128];
     JHEP {\bf 0202} (2002) 010
     [arXiv:hep-th/0108010];
     C.~Biggio, F.~Feruglio, A.~Wulzer and F.~Zwirner,
     JHEP {\bf 0211} (2002) 013
     [arXiv:hep-th/0209046].


\bibitem{vonGersdorff:2002tj}
     G.~von Gersdorff, M.~Quiros and A.~Riotto,
     Nucl.\ Phys.\  B {\bf 634} (2002) 90
     [arXiv:hep-th/0204041];
     G.~von Gersdorff, L.~Pilo, M.~Quiros, A.~Riotto and V.~Sanz,
     Phys.\ Lett.\  B {\bf 598}, 106 (2004)
     [arXiv:hep-th/0404091];
     K.~A.~Meissner, H.~P.~Nilles and M.~Olechowski,
     Acta Phys.\ Polon.\  B {\bf 33} (2002) 2435
     [arXiv:hep-th/0205166];
     A.~Delgado, G.~von Gersdorff and M.~Quiros,
     JHEP {\bf 0212} (2002) 002
     [arXiv:hep-th/0210181];
     Z.~Lalak and R.~Matyszkiewicz,
     Nucl.\ Phys.\  B {\bf 730} (2005) 37
     [arXiv:hep-ph/0506223].



\bibitem{Meissner:1999ja}
     K.~A.~Meissner, H.~P.~Nilles and M.~Olechowski,
     Nucl.\ Phys.\  B {\bf 561} (1999) 30
     [arXiv:hep-th/9905139].


\bibitem{DeCurtis:2003hs}
     S.~De Curtis, D.~Dominici and J.~R.~Pelaez,
     JHEP {\bf 0401} (2004) 052
     [arXiv:hep-th/0311226].

\bibitem{Bagger:2004rr}
     J.~A.~Bagger and D.~V.~Belyaev,
     Phys.\ Rev.\  D {\bf 72} (2005) 065007
     [arXiv:hep-th/0406126].

\bibitem{Weinberg:1972fn}
     S.~Weinberg,
     Phys.\ Rev.\ Lett.\  {\bf 29} (1972) 1698.


\bibitem{Georgi:1974yw}
     H.~Georgi and A.~Pais,
     Phys.\ Rev.\  D {\bf 10} (1974) 539;
     Phys.\ Rev.\  D {\bf 12} (1975) 508.


\bibitem{Arkani-Hamed:2001nc}
     N.~Arkani-Hamed, A.~G.~Cohen and H.~Georgi,
     Phys.\ Lett.\  B {\bf 513}, 232 (2001)
     [arXiv:hep-ph/0105239].

\bibitem{Hatanaka:1998yp}
     H.~Hatanaka, T.~Inami and C.~S.~Lim,
     Mod.\ Phys.\ Lett.\  A {\bf 13} (1998) 2601
     [arXiv:hep-th/9805067];
     A.~Masiero, C.~A.~Scrucca, M.~Serone and L.~Silvestrini,
     Phys.\ Rev.\ Lett.\  {\bf 87}, 251601 (2001)
     [arXiv:hep-ph/0107201];
     C.~P.~Bachas,
     arXiv:hep-th/9509067;
     G.~R.~Dvali, S.~Randjbar-Daemi and R.~Tabbash,
     Phys.\ Rev.\  D {\bf 65}, 064021 (2002)
     [arXiv:hep-ph/0102307];
     I.~Antoniadis, K.~Benakli and M.~Quiros,
     New J.\ Phys.\  {\bf 3} (2001) 20
     [arXiv:hep-th/0108005];
     C.~Csaki, C.~Grojean and H.~Murayama,
     Phys.\ Rev.\  D {\bf 67} (2003) 085012
     [arXiv:hep-ph/0210133];
     C.~A.~Scrucca, M.~Serone and L.~Silvestrini,
     Nucl.\ Phys.\  B {\bf 669} (2003) 128
     [arXiv:hep-ph/0304220].


\bibitem{Cacciapaglia:2005pa}
     G.~Cacciapaglia, C.~Csaki, C.~Grojean, M.~Reece and J.~Terning,
     Phys.\ Rev.\  D {\bf 72} (2005) 095018
     [arXiv:hep-ph/0505001].


\bibitem{Cremmer:1980gs}
     E.~Cremmer,
{\it Invited paper at the Nuffield Gravity Workshop, 
Cambridge, Eng., Jun 22 - Jul 12, 1980}; 
     A.~H.~Chamseddine and H.~Nicolai,
     Phys.\ Lett.\ B {\bf 96} (1980) 89.


\bibitem{WessAndBagger}
     J.~Wess and J.~Bagger,
     \textit{Supersymmetry and supergravity,}
     2nd edition, Princeton University Press, 1992.


\bibitem{Baulieu:1985wa}
     L.~Baulieu, A.~Georges and S.~Ouvry,
     Nucl.\ Phys.\  B {\bf 273} (1986) 366.


\bibitem{Bagger:2002rw}
  J.~Bagger and D.~V.~Belyaev,
  Phys.\ Rev.\  D {\bf 67} (2003) 025004
  [arXiv:hep-th/0206024].


\bibitem{VanProeyen:1999ni}
     A.~Van Proeyen,
     arXiv:hep-th/9910030.


\end{thebibliography}
\end{document}